\documentclass{aastex6}

\usepackage{natbib}
\usepackage{graphicx}
\usepackage{amssymb}
\usepackage{float}
\usepackage{url}

\newcommand{\Lcrit}{L$_{crit}$}
\newcommand{\Lcrits}{L$_{crit}$ }
\newcommand{\microns}{$\mu$m }
\newcommand{\get}{{\it getsources}}
\newcommand{\fell}{{\it FellWalker}}

\defcitealias{gutermuth09}{G09}
\defcitealias{kirk11}{KM11}
\defcitealias{Kirk16b}{K16}

\begin{document}

\title{The JCMT Gould Belt Survey: Dense Core Clusters in Orion~A}
\author{
J. Lane\altaffilmark{1,2},
H. Kirk\altaffilmark{2},
D. Johnstone\altaffilmark{1, 2, 3},
S. Mairs\altaffilmark{1,2},
J. Di Francesco\altaffilmark{1, 2}, 
S. Sadavoy\altaffilmark{4}
J. Hatchell\altaffilmark{5}, 
D.S. Berry\altaffilmark{3}, 
T. Jenness\altaffilmark{6,7},
M.R. Hogerheijde\altaffilmark{8}, 
D. Ward-Thompson\altaffilmark{9}, 
and the JCMT Gould Belt Survey team\footnote{The full members of the JCMT Gould Belt Survey
Consortium are
P. Bastien, D. S. Berry, D. Bresnahan, H. Broekhoven-Fiene, J. Buckle, H. Butner,
M. Chen, A. Chrysostomou, S. Coude, M. J. Currie, C. J. Davis, E. Drabek-Maunder, A. Duarte-Cabral,
J. Di Francesco, M. Fich, J. Fiege, P. Friberg, R. Friesen, G.A. Fuller,
S. Graves, J. Greaves, J. Gregson, J. Hatchell, M.R. Hogerheijde, W. Holland, T. Jenness,
D. Johnstone, G. Joncas, J.M. Kirk ,H. Kirk, L.B.G. Knee, S. Mairs, K. Marsh, B.C. Matthews,
G. Moriarty-Schieven, J.C. Mottram, C. Mowat, K. Pattle, J. Rawlings, J. Richer, D. Robertson, 
E. Rosolowsky, D. Rumble, S. Sadavoy, N. Tothill, H. Thomas, S. Viti, D. Ward-Thompson,
G.J. White, J. Wouterloot, J. Yates, and M. Zhu}
}
\altaffiltext{1}{NRC Herzberg Astronomy and Astrophysics, 5071 West Saanich Rd, Victoria, BC, V9E 2E7, Canada}
\altaffiltext{2}{Department of Physics and Astronomy, University of Victoria, Victoria, BC, V8P 1A1, Canada}
\altaffiltext{3}{Joint Astronomy Centre, 660 N. A`oh\={o}k\={u} Place, University Park, Hilo, Hawaii 96720, USA}
\altaffiltext{4}{Max Planck Institute for Astronomy, K\"{o}nigstuhl 17, D-69117 Heidelberg, Germany
}
\altaffiltext{5}{Physics and Astronomy, University of Exeter, Stocker Road, Exeter EX4 4QL, UK}
\altaffiltext{6}{Joint Astronomy Centre, 660 N. A`oh\={o}k\={u} Place, University Park, Hilo, Hawaii 96720, USA}
\altaffiltext{7}{LSST Project Office, 933 N. Cherry Ave, Tucson, AZ 85719, USA}
\altaffiltext{8}{Leiden Observatory, Leiden University, PO Box 9513, 2300 RA Leiden, The Netherlands}
\altaffiltext{9}{Jeremiah Horrocks Institute, University of Central Lancashire, Preston, Lancashire, PR1 2HE, UK}

\begin{abstract}
The Orion~A molecular cloud is one of the most well-studied nearby star-forming regions,
and includes regions of both highly clustered and more dispersed star formation across
its full extent.  Here, we analyze dense, star-forming cores identified in the 
850~$\mu$m and 450~$\mu$m SCUBA-2 maps from the JCMT Gould Belt Legacy Survey.  
We identify dense cores
in a uniform manner across the Orion~A cloud and analyze their clustering
properties.  Using two independent lines of analysis, we find evidence that clusters
of dense cores tend to be mass segregated, suggesting that stellar clusters may 
have some amount of primordial mass segregation already imprinted in them 
at an early
stage.  We also demonstrate that the dense core clusters have a tendency to
be elongated, perhaps indicating a formation mechanism
linked to the filamentary structure within molecular clouds.
\end{abstract}

\section{INTRODUCTION}

While the \citet{shu87} model provides a framework to understand many aspects of the
formation of a single, low-mass, isolated star, much remains to be done to expand a similar
understanding to a wider range of systems.
Most low-mass stars and nearly all high-mass stars form in clustered
environments \citep{ladalada03, zinnecker07}, making understanding the clustered mode of
star formation important.  Nevertheless, there is little consensus on the formation process
of either high-mass stars or stellar clusters.
One avenue which will help to guide models of cluster formation is a 
careful characterization of the basic properties of clustered systems at early times, which will
put constraints on the initial conditions of such systems.  Our goal in this paper is to present
one such characterization of dense core clusters in Orion~A, which can easily be reproduced in
other star-forming environments, allowing for a quantification of the range of typical
cluster properties held by an ensemble of clouds.

There have been many studies of the clustering properties of young stars, including 
\citet[hereafter G09]{gutermuth09} and \citet[hereafter KM11]{kirk11}, both of which 
analyzed samples of nearby star-forming regions (mostly within 1000~pc for the former and 
all within 500~pc for the latter), thereby ensuring that even low mass stars were included. 
One intriguing result from \citetalias{kirk11} was that young (1--2~Myr), small (20--40 member), and 
sparse (less than 100~stars~pc$^{-2}$) groupings of stars showed evidence of mass segregation 
through the presence of a centrally located, most massive cluster member. It would be 
difficult for such an arrangement to be fully caused by dynamical interactions after the 
stars had formed, given the young ages of the systems.
A later analysis of numerical simulations 
of the formation of small stellar groups also showed that 
mass segregation is present as early in 
the simulation as it was measurable \citep{kirk14}. These results raise the question of how 
the mass in a cluster-forming region is arranged {\it before} star formation begins. 
Mass segregation, particularly in clusters of dense cores, is therefore an important 
factor in understanding the initial conditions of star formation.

We study dense cores 
\citep[structures containing roughly 1~M$_{\odot}$ of material within $\sim$0.1~pc, as in][]{Bergin07}
in the Orion~A molecular cloud. 
Orion~A is a 
$\sim10^{5}~M_{\sun}$, nearby \citep[$\sim$415~pc,][]{menten07,kim08} cloud that is forming 
both high and low mass stars \citep[as reviewed in][]{bally08}. The brightest submillimetre feature 
in Orion~A is the Integral Shaped Filament (ISF), in the northern part of the cloud 
\citep[e.g.,][]{bally87,johnstone99}. The ISF contains the Orion Nebula Cluster (ONC), a large cluster 
of young stars including several O stars \citep{hillenbrand97}. The Orion~A cloud also contains many 
sites of more dispersed star formation, extending approximately 30~pc ($\sim4\arcdeg$) to the southeast 
from the center of the ISF, in an area known as Orion~A~South \citep{bally08}.
This southern portion of Orion~A, also known as L1641, 
was recently studied by \citet{Polychroni13} using
{\it Herschel} observations, showing that the dense cores, especially the more massive
dense cores, have a strong tendency to be associated with filamentary structures.
With its large number of 
dense cores and varied clustering envioronments, Orion~A provides the ideal laboratory to explore mass 
segregation in dense core clusters. To perform this analysis, we use SCUBA-2 \citep{holland13} 
observations at 450~$\mu$m and 850~$\mu$m 
taken as part of the JCMT Gould Belt Survey \citep{wardthompson07}. An analysis of dense cores and filamentary structure found in Orion~A~North were published by 
\citet{salji15a} and \citet{salji15b} 
respectively, while an analysis of the fragmentation stability of dense gas in southern Orion~A 
is given by \citet{Mairs16}.
The clustering analysis presented here uses a similar methodology to a recent clustering analysis
of Orion~B by our group \citep[][hereafter K16]{Kirk16b}, and we later discuss the implications
of finding similar results in both Orion clouds.


\begin{figure}[p!]
	\includegraphics[height=8in]{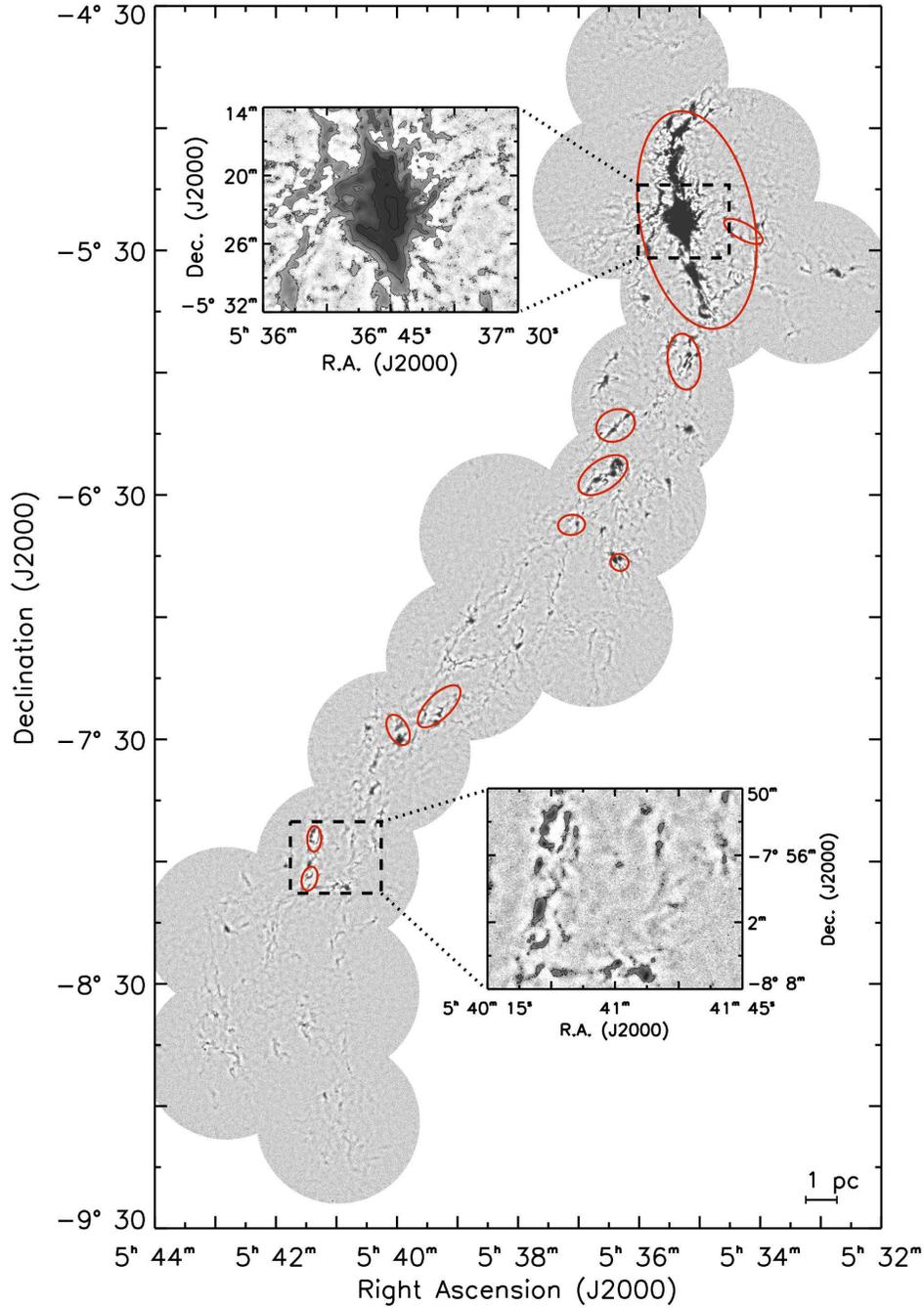}
	\caption{850~\microns flux density data for Orion~A, shown with logarithmic scaling. 
	The contour lines are placed at 0.01, 0.1, 0.5, and 2.5~Jy~arcsec$^{-2}$.
	The red circles show the approximate locations of the MST-based clusters
	identified in Section~\ref{sec_mst}.
	The two insets show selected 22.5\arcmin\ by 18\arcmin\ regions. The inset near the top 
	of the figure shows the bright ISF in the northern part of the cloud. This 
	structure is 
	contrasted by the sparser structure in the south, shown in the bottom inset.
	}
	\label{fig_overview}
\end{figure}


\section{OBSERVATIONS}
\label{sec_obs}

\citet{salji15a} and \citet{Mairs16} present a `first look' at the SCUBA-2 data
across the northern and southern portions of Orion~A respectively, and the data reduction
methods used in each of their analyses are given in their papers.  In short, the northern
Orion~A analysis by \citet{salji15a} used only the first portion of data available in
the region (i.e., a smaller area and shallower sensitivity than the final dataset).
These data were reduced using an earlier reduction method, Internal Release 1 (IR1), 
which had larger pixel sizes (6\arcsec versus the current 3\arcsec) and did not 
recover extended emission as well as the current reduction method.
The southern portion of Orion~A analyzed by \citet{Mairs16} includes the full
GBS data obtained in this region and applies the present best GBS reduction method,
Legacy Release 1 (LR1), for the data reduction.  Our present analysis of the entire Orion~A
complex also uses the full GBS dataset, but only uses a part of the full LR1 data reduction
procedure, due to availability at the time of our analysis.
We outline the differences between our reduction method and LR1 below, and highlight
the resultant differences in the final maps. 

Individual observations at 850~$\mu$m and 450~$\mu$m were made using fully sampled 30\arcmin\ 
diameter circular regions \citep[PONG 1800 mapping mode,][]{bintley14}; larger regions were 
mosaicked with overlapping scans.
The data were reduced as part of the GBS LR1 using an iterative 
map-making technique \citep[makemap in {\sc{smurf}},][]{chapin13}, and gridded to 3\arcsec\ 
pixels at 850~$\mu$m and 2\arcsec\ pixels at 450~$\mu$m. The iterations were halted when the map 
pixels, on average, changed by $<$0.1\% of the estimated map rms.  This reduction is 
referred to as the automask reduction. 
The final GBS LR1 data products are created through a second round of reductions,
wherein a signal-to-noise ratio mask is produced based on a mosaic of the automask-reduced data,
to define areas of probable emission \citep[see Mairs et al.~2016 for more details, or] 
[for a description of the same data reduction process in Orion~B]{Kirk16a}.
This second reduced map is referred to as the external mask reduction.
Both the automask and external mask reductions include a spatial filter of 600\arcsec,
limiting robust flux recovery to sources with a Gaussian FWHM less than 2.5\arcmin.
Sources between 2.5\arcmin\ and 7.5\arcmin\ will be detected, but both the flux density and the size 
are underestimated because Fourier components with scales greater than 5\arcmin\ are removed 
by the filtering process.  Use of an external mask allows for somewhat better source
recovery than the initial automask reduction, as demonstrated by \citet{Mairs15}. 
Testing with artificial sources in \citet{Mairs15}  shows that compact and brighter sources 
are reasonably well-recovered even with an automask-only reduction, whereas fainter and more 
diffuse sources tend to have underestimated sizes and flux 
densities.  Comparison of SCUBA-2 and {\it Herschel}~500~$\mu$m data in the L1495
region of Taurus also shows that SCUBA-2
is most sensitive to the coldest and densest gas structures (Ward-Thompson et al., submitted).

We use only the automask-reduced mosaics of Orion~A here.  This choice 
was motivated by the data available at the time
of analysis.  While the automask mosaic is insufficient for a detailed study of
core mass functions, since faint sources would have underestimated fluxes, this reduction
method is sufficient for clustering analyses.  Our requirements for
this analysis are to be able to identify dense cores, and, within sub-regions of the
map, rank the cores by flux (in particular, identify the highest
flux density dense core).  Accuracy on the absolute flux density values (or any size estimate) is 
not important for our analysis.  Furthermore, the small suppression of larger-scale, faint and
diffuse emission in complex regions such as the Integral Shaped Filament, likely
provides a slight advantage in simplifying the separation of this complex emission into
dense cores.  Figure~\ref{fig_overview} shows the 850~$\mu$m map that we analyzed,
highlighting the variety in emission structure along the cloud.

We include a copy of the automask mosaic used in our analysis at 
{\tt https://doi.org/10.11570/16.0008}.


\section{SOURCE EXTRACTION}
We applied two different source identification methods across our Orion~A map to create
independent uniform catalogues for our analysis.  
These two methods were the \get\ \citep{menshch12} and \fell\ \citep{berry15} algorithms.
In the \get\ algorithm\footnote{We used version 1.140127.}, 
cores are extracted by convolving an image on multiple 
spatial scales to find cores and filter out small-scale noise and large-scale structure.  
\fell\ identifies cores by 
following flux density gradients towards peaks to define the boundaries of cores. 
{\it Getsources} is a multi-wavelength extraction method and can therefore make use of both 850~\microns and 
450~\microns data simultaneously, while \fell\ was run only on the 850~\microns data. 
After removing sources that were likely noise artifacts, 
the final \get\ catalogue contained 919 cores and the final \fell\ catalogue 
contained 773 cores.  The details of \get, \fell, and the
automated procedure to eliminate spurious sources in each are discussed in more detail in
Appendix~\ref{sec_sourcefinding_gs} and \ref{sec_sourcefinding_fw} respectively.  
We found that \get\ did a 
better job of separating small, compact cores from the underlying large-scale cloud 
structure and we therefore use this catalogue for the analysis presented in this
paper.  In Appendices~\ref{sec_compare_gs_fell}, \ref{sec_fell_mst}, and
\ref{sec_fell_m_sigma}, we present comparable analyses using the \fell-based catalogue.

Creating our own dense core catalogue across Orion~A was necessary to ensure that
cores were identified uniformly across the entire cloud.  The differing sensitivities
of the maps and also core identification algorithms used by \citet{salji15a} and 
\citet{Mairs16} in the north and south could result in biases in our analysis
if we attempted to combine their two core catalogues.
Additionally, we note that the analysis of \citet{Mairs16} was not focussed
on identifying discrete dense cores, but instead on characterizing the fragmentation of
larger-scale features in the cloud, and hence their catalogue would not be directly applicable
to our science goals.
Source identification is particularly
difficult around the ISF in the northern part of Orion~A, where many compact peaks of emission
lie clustered within a larger zone of extended emission.  By running the two independent
core identification algorithms listed above, we can additionally test the
robustness of our analysis.
Despite large differences in the overall catalogues (total number
of cores and core boundaries, especially in the ISF), Appendix~\ref{sec_sourcefinding}
demonstrates that we
find similar results for our clustering analysis using either catalogue.

\subsection{Protostars}

After creating our \get-based dense core catalogue, we classified each core as being starless or
protostellar based on the {\it Spitzer} catalogue of \citet{megeath12} and
the {\it Herschel} catalogue of \citet{Stutz13}.  From the \citet{megeath12} catalogue,
we used only the protostars deemed `reliable', i.e., 
sources that were too faint at 24~\microns or went undetected in shorter wavebands were not 
included.  We defined a dense core as protostellar if one or more {\it Spitzer} 
or {\it Herschel} protostars 
were found within one beam radius (7.25\arcsec) of the central core peak.
The {\it Spitzer} catalogue was sufficient to identify most of the protostellar cores.
Indeed, only five cores were deemed protostellar based solely on the {\it Herschel} catalogue.
Our protostar -core association criterion is more stringent than the one applied in 
\citetalias{Kirk16b}, as
Orion A has a much higher surface density of protostars.
With our criterion, 814 of the cores are deemed starless, while 105 are
classified as protostellar, i.e., roughly 11\% of the cores are protostellar.
For comparison, \citet{Mairs16} found 75 of 359 islands and 103 of 431 fragments
contained at least one protostar (i.e., 21\% to 24\% protostellar) in the southern portion
of Orion~A, while \citet{salji15a} found 48 protostellar cores and 432 starless and 
prestellar cores around the ISF (i.e., 10\% protostellar).
We note that the \citet{Mairs16} protostellar fraction is significantly higher
than the other two cases because any structure which had a protostar located somewhere within
its boundary was classified as protostellar, rather than restricting the protostellar 
classification to those objects which have a protostar near the brightest peak within
the object.
We also note that additional candidate YSOs based on recent radio \citep{Forbrich16} or
near-infrared \citep{Meingast16} detections were not included in our analysis, nor in
that of \citet{salji15a} or \citet{Mairs16}, implying that all of the 
protostellar fractions listed here are lower limits.

\subsection{Core Masses}
\label{sec_core_mass}
Using the total flux density measured for each core at 850~\micron, we can also estimate the
core masses.  Assuming a constant dust temperature and opacity, the conversion between
flux density and mass is given by
\begin{equation}
M = 1.30 \Big( \frac{S_{850}}{1~{\rm Jy}} \Big) 
	\Big( \frac{\kappa_{850}}{0.012~{\rm cm}^2~{\rm g}^{-1}} \Big) 
	\Big(exp\Big(\frac{17~{\rm K}}{T_d}\Big) -1\Big)
	\Big( \frac{D}{450~{\rm pc}}\Big) ^2 M_{\odot}
\end{equation}
where $S_{850}$ is the total flux density at 850~\micron, $\kappa_{850}$ is the dust opacity at 
850~\micron, $T_d$ is the dust temperature, and $D$ is the distance.
For consistency with \citet{salji15a} and \citet{Mairs16}, we assume a 
constant dust grain opacity of 0.012~cm~g$^{-1}$, and a distance to Orion A of
450~pc \citep[e.g.,][]{Muench08}.  Also following \citet{Mairs16}, we assume a 
dust temperature of 15~K.  We note that the assumption of a constant temperature
is necessarily approximate.  For example, \citet{salji15a} and \citet{johnstone06} find 
evidence of variations in temperature across the northern and southern portions of Orion A
respectively, and a comparison of the 450~\micron\ to 850~\micron\ flux density ratio by
Rumble et al. (in prep) suggests that temperatures are generally higher in the north,
as is also seen in the {\it Herschel}-based analysis of \citet{Lombardi14}.  
The cores most poorly represented by a fixed temperature 
are likely to be
some of the protostellar cores, where a higher temperature would be more appropriate.
With the assumption of a constant 15~K temperature, the median core mass is 0.8~M$_{\odot}$
in both the \get-based and \fell-based catalogues, with cores as
small as 0.06~M$_{\odot}$ identified.
In our analysis below, where the flux- or mass-ranking of cores is important to the 
results, we examine both the entire core population and also the starless core
subsample separately.
In Appendix~\ref{app_temp}, we re-run all of the analysis presented in the main
paper, using {\it Herschel}-based temperatures from \citet{Lombardi14} to estimate
the mass-ranking of the cores.  In all cases, we reach similar conclusions as are
presented in the main text.  As we outline in Appendix~\ref{app_temp}, the
\citet{Lombardi14} temperatures were estimated without the inclusion of long wavelength
data, which could lead us to assign artifically high temperature estimates
of some dense cores, since {\it Herschel} is more sensitive to diffuse warm dust than
SCUBA-2 is.  Since diffuse warm dust might not be distributed uniformly across Orion~A,
our main analysis uses the the aforementioned flux-ranking of the dense cores, while
mass-ranking based on {\it Herschel} temperatures is left to Appendix~\ref{app_temp} to
provide an additional test to confirm our results.


\section{CLUSTERING ANALYSIS}
\label{sec_mst}

As seen in Figure~\ref{fig_overview}, the dense cores in Orion~A inhabit a variety of 
environments, ranging from highly clustered in the north around the ISF to more isolated 
in the south. High-mass star formation is ongoing in the north, where the density of gas and 
dust is the highest, while in the south there is more distributed low-mass star 
formation \citep{bally08}. Despite these differences in star forming activity, there are 
visible clusters of dense cores in both the north and south of the cloud, albeit on different 
spatial scales. Clusters in the south are often isolated and more dispersed / extended 
than clusters in the north. \citet{bressert10} showed that protostellar 
clustering, not only in Orion but across the whole Gould Belt, exists along a spectrum of 
spatial densities, suggesting the distinction between clustered and isolated star-formation 
is necessarily relative, rather than absolute. A major difficulty in identifying 
clusters in a cloud as large and varied as Orion A is determining a consistent method that 
works well for all clustering scales that are present. For this reason, we use a minimal 
spanning tree (MST) method following \citetalias{gutermuth09} and \citetalias{kirk11}
\citep[see also][for a completely different application of the MST in analyzing
clustered structures]{cartwright04}.  
The advantage of the minimal spanning 
tree method is that it is able to pick out the relative overdensities that represent 
clusters. The other feature of the MST which is especially advantageous for our analysis is 
that there is no preference imposed on identifying round clusters: since star-formation is 
thought to be strongly associated with filaments \citep[e.g.,][]{andre14}, dense core clusters 
may also show some elongation.

\subsection{Using a Minimal Spanning Tree}
\label{sec_mst_intro}

\begin{figure}[htb]
\begin{tabular}{cc}
	\includegraphics[width=3.2in]{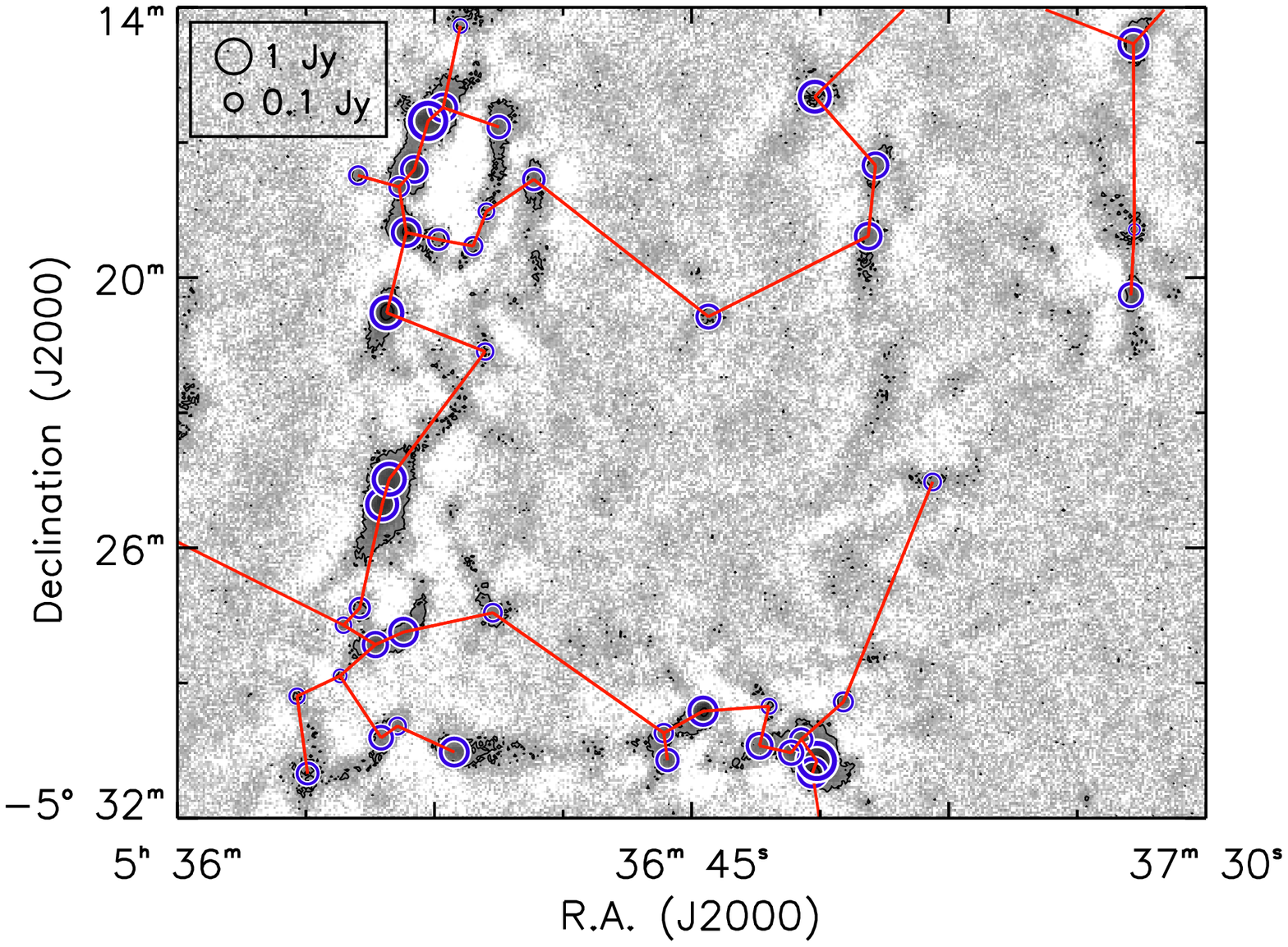} &
	\includegraphics[width=3.2in]{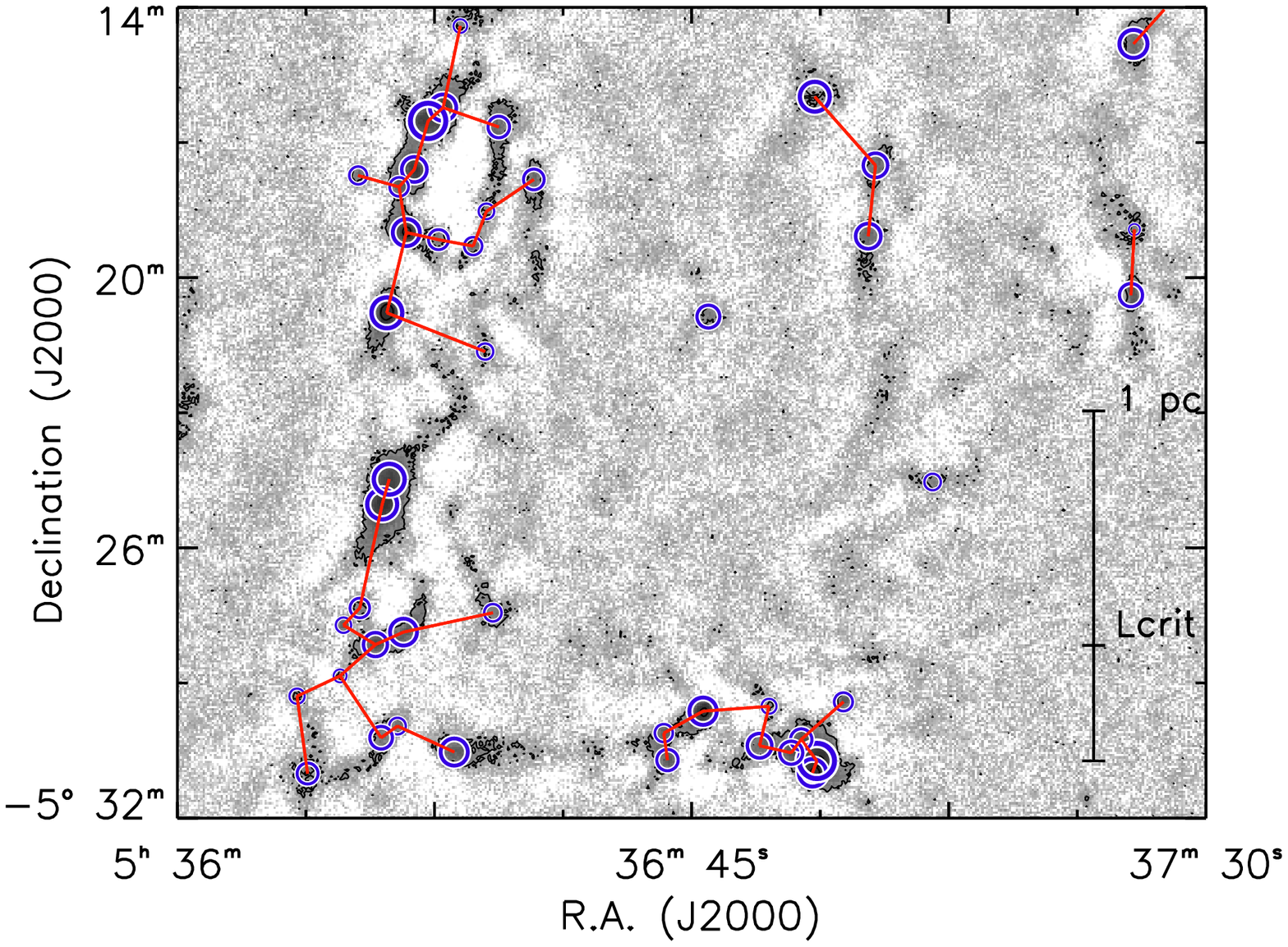} \\
\end{tabular}
\caption{The minimal spanning tree structure for the same region of Orion A shown in the 
bottom inset of Figure~\ref{fig_overview}, also in logarithmic scaling with 
contours at 0.01, 0.1, and 0.5~Jy~arcsec$^{-2}$.
Dense cores identified by \get\ 
are shown as blue circles, with the circle radius scaling with the total flux density measured
for the core.  The branches of the minimal spanning tree are shown in red. 
The left panel shows the full minimal spanning tree structure which connects all cores
to their nearest neighbour (extending beyond the plotted region). 
The right panel shows the potential clusters remaining once branches 
longer than the critical length are removed.  The long scale bar indicates a length of
1~pc assuming a distance of 450~pc, while the shorter segment
shows the value of \Lcrit\ determined, i.e., 0.36~pc.}
\label{fig_msts}
\end{figure}

\begin{figure}[htb]
	\includegraphics[height=3in]{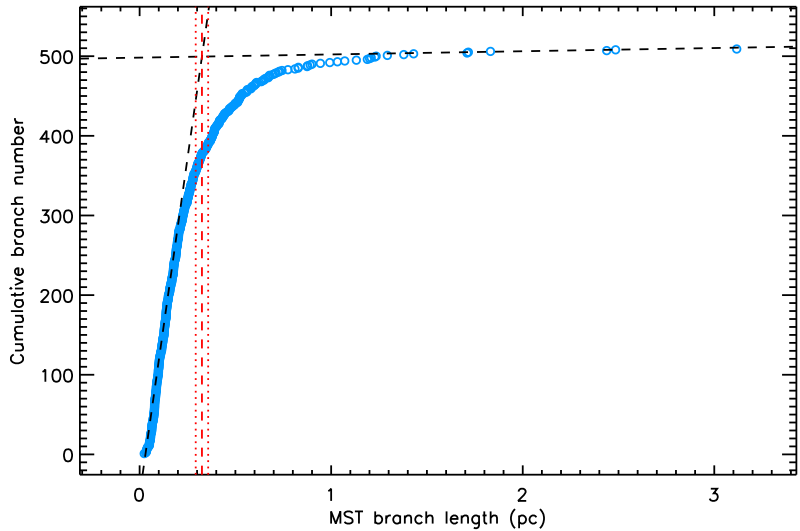}
	\caption{MST branch lengths measured in Orion~A, with ISF cores removed.
	The cumulative number of branches with a given length or smaller are indicated by
	the blue circles.  The dashed black lines show the best linear fits to the 
	two ends of the branch length distribution, while the thick dashed red line
	shows the best-fit \Lcrit\ value (i.e., the branch length at which the two blue
	lines intersect).  The dotted red vertical lines show the range in \Lcrit\ values
	resulting from acceptable linear fits to the distribution.}
	\label{fig_Lcrit}
\end{figure}

An MST is formed by connecting each core to its nearest neighbour and ensuring all cores 
are connected together continuously, as shown in the left panel of Figure~\ref{fig_msts}. The 
connections between cores are known as branches, and the total length of all branches in an 
MST is minimized such that all cores are connected together as efficiently as possible. In an 
MST diagram, clusters are groups of cores connected by shorter branch lengths, a characteristic 
that can be used to extract clusters. \citetalias{gutermuth09} created a criterion for cluster 
membership based on the distribution of MST branch lengths. Figure~\ref{fig_Lcrit} shows 
the cumulative distribution of MST branch lengths for the southern portion of Orion A
(see the following paragraph).  The curve has two distinct 
parts: a steep, initial increase, followed by a turnover to a shallow gradient. This is the 
characteristic shape of the branch length distribution for cluster-forming regions, as shown 
in \citetalias{gutermuth09}. 
The steep part of the distribution at shorter branch lengths represents clustered 
environments, while the longer branch lengths at the flat end of the distribution represent 
isolated cores as well as the cluster-cluster connections. \citetalias{gutermuth09} defines 
\Lcrits as the intersection between linear fits to both ends of the distribution. Removing 
all branches longer than \Lcrits leaves the spatially dense regions connected together, 
which represent clusters, as seen in the right panel of Figure~\ref{fig_msts}. We estimate 
the uncertainty in \Lcrits by finding the steepest and shallowest possible linear fits to 
each of the cumulative distributions and re-calculating \Lcrit. We find, in general, that 
the uncertainty in \Lcrits is small enough that very few cores change their membership status 
when the value is shifted to longer or shorter values.
Appendix~\ref{app_Lcrit} presents a thorough examination of the effect of different
\Lcrits values on the clusters identified and subsequent analysis results.


We initially calculated \Lcrits based on the MST structure across the entire Orion A cloud, 
which gives a value of 0.22$\pm~0.03$~pc. 
A visual examination of the resulting clusters, 
however, showed that this value of \Lcrits did not successfully identify some visually obvious 
clusters, especially in the south where some clusters were fragmented or had too few members 
to be significant. The ISF and its immediate surroundings contain approximately as many dense 
cores as the rest of the cloud combined, which biased the critical length of the entire cloud 
towards a smaller value. We removed the mutual bias that ISF and non-ISF cores had on each 
other by separating cores that were contiguous with a part of the large ISF cluster in the 
initial MST (as defined by \Lcrit=0.22~pc), 
from those that were not. We then 
created a MST and branch length distribution for each sub-population separately. The \Lcrits 
for non-ISF cores was 
0.36~pc~$\pm~0.03$~pc while the \Lcrits for ISF cores was 0.14~$\pm~0.01$~pc.

We identified clusters for the entire cloud using both the ISF and non-ISF \Lcrits values. Using the larger \Lcrits value of 0.36~pc 
produced a set of clusters well-matched to visually-apparent clusters in the southern part of Orion~A as expected, but also resulted in reasonable clustering in the northern part as well. The ISF was classified as one large cluster, which is reasonable given that it is a visually distinct, single structure in the molecular cloud. Applying the smaller, ISF-derived \Lcrits of 0.14~pc 
yielded only the most compact clusters being identified, along with fragments of what are obviously larger structures being picked out as individual clusters. Even the ISF, from which this value was produced, was not divided up in a visually reasonable manner. 
We therefore decided to use the higher \Lcrits value of 0.36~pc 
for the remainder of the analysis on the entire Orion~A cloud.
Appendix~\ref{app_Lcrit} examines the influence of our choice in \Lcrits on the the results
of our analysis and shows minimal variation within a wide range of choices in \Lcrit.
These tests demonstrate that even if we were to find a scheme for determining \Lcrits that
better handled the multiple clustering scales within Orion~A, the results from our analysis
would be largely unchanged.

\begin{figure}[p]
\begin{tabular}{cc}
\includegraphics[width=2.4in]{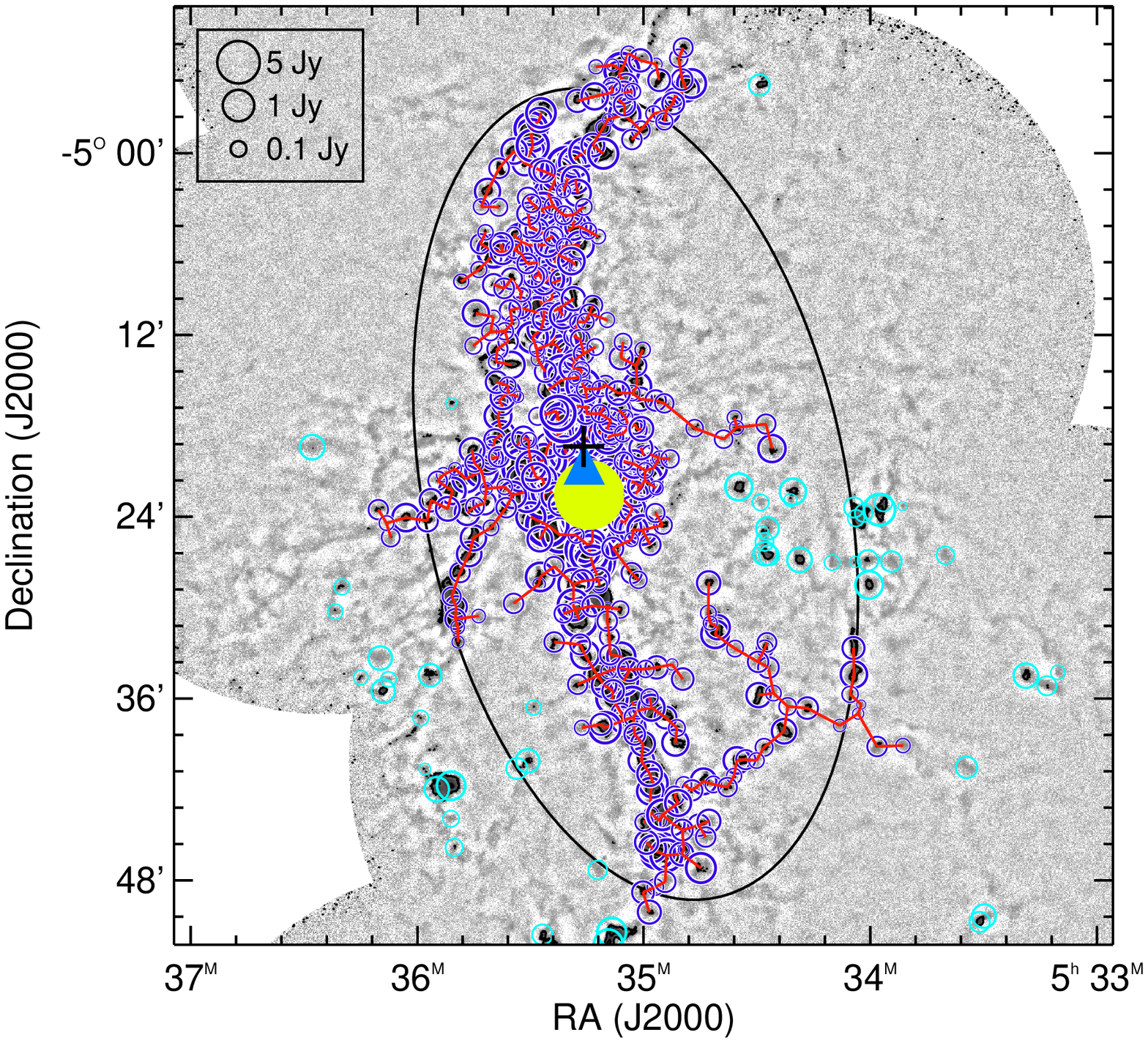} &
\includegraphics[width=2.4in]{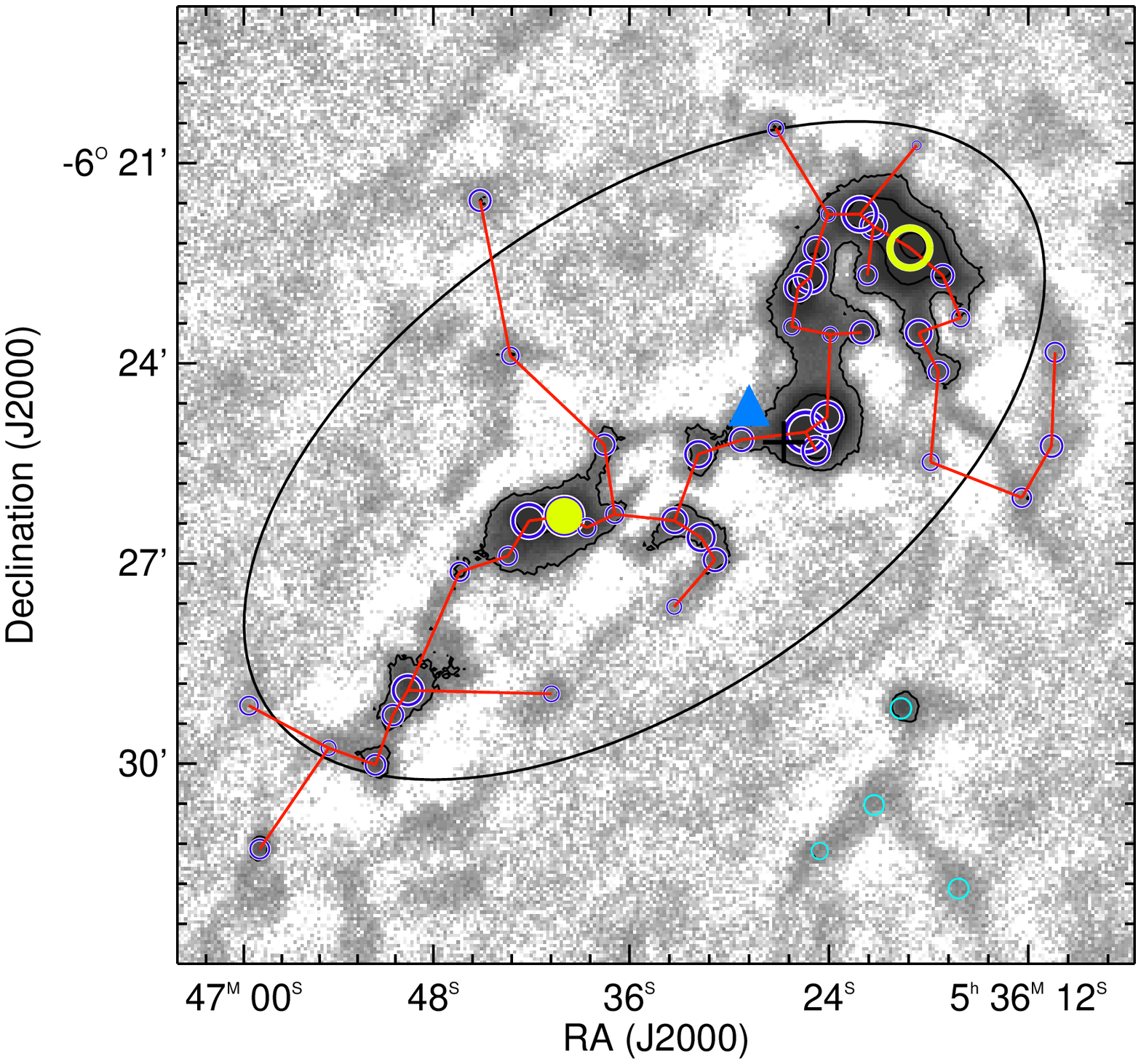}\\
\includegraphics[width=2.4in]{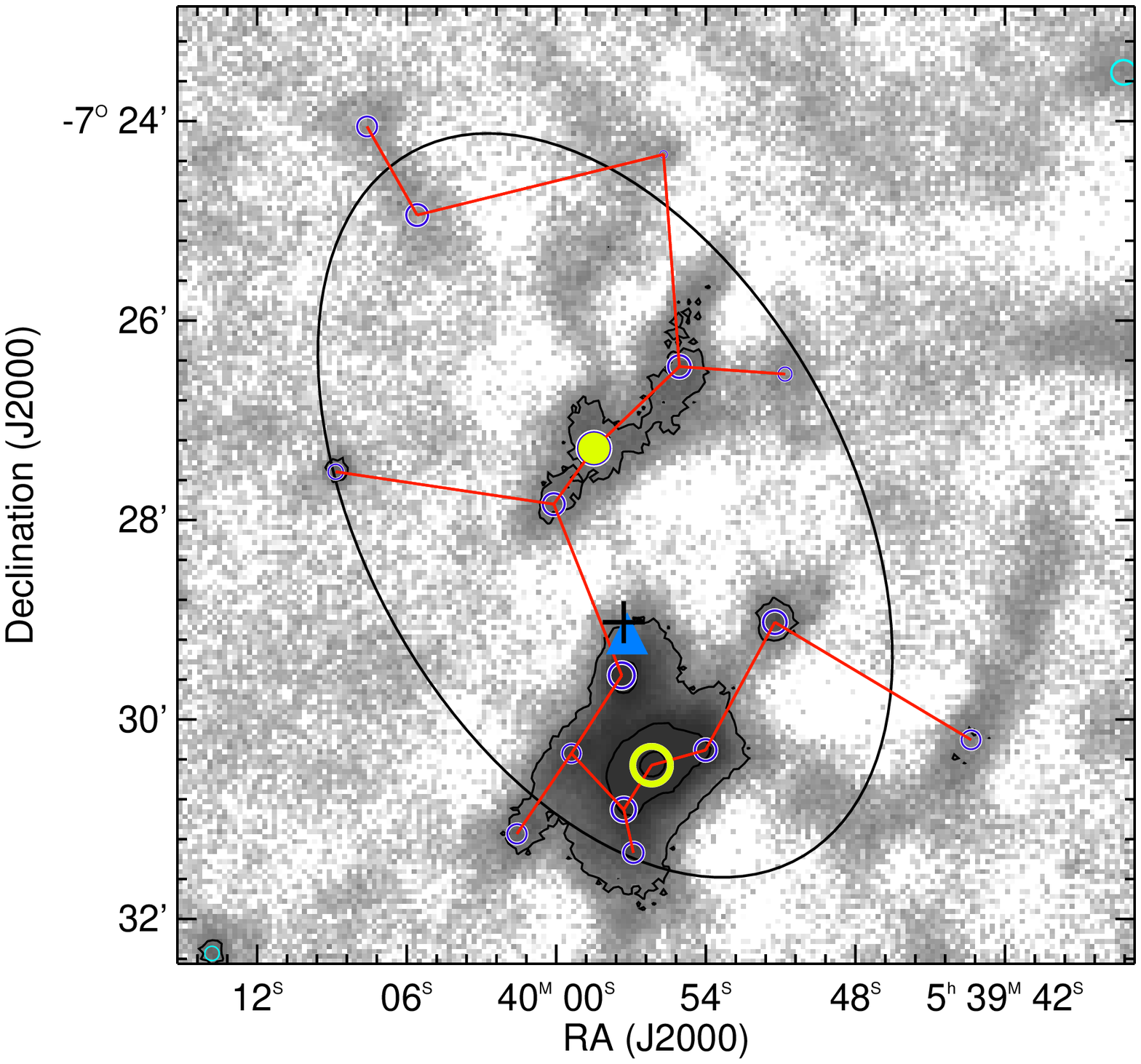}&
\includegraphics[width=2.4in]{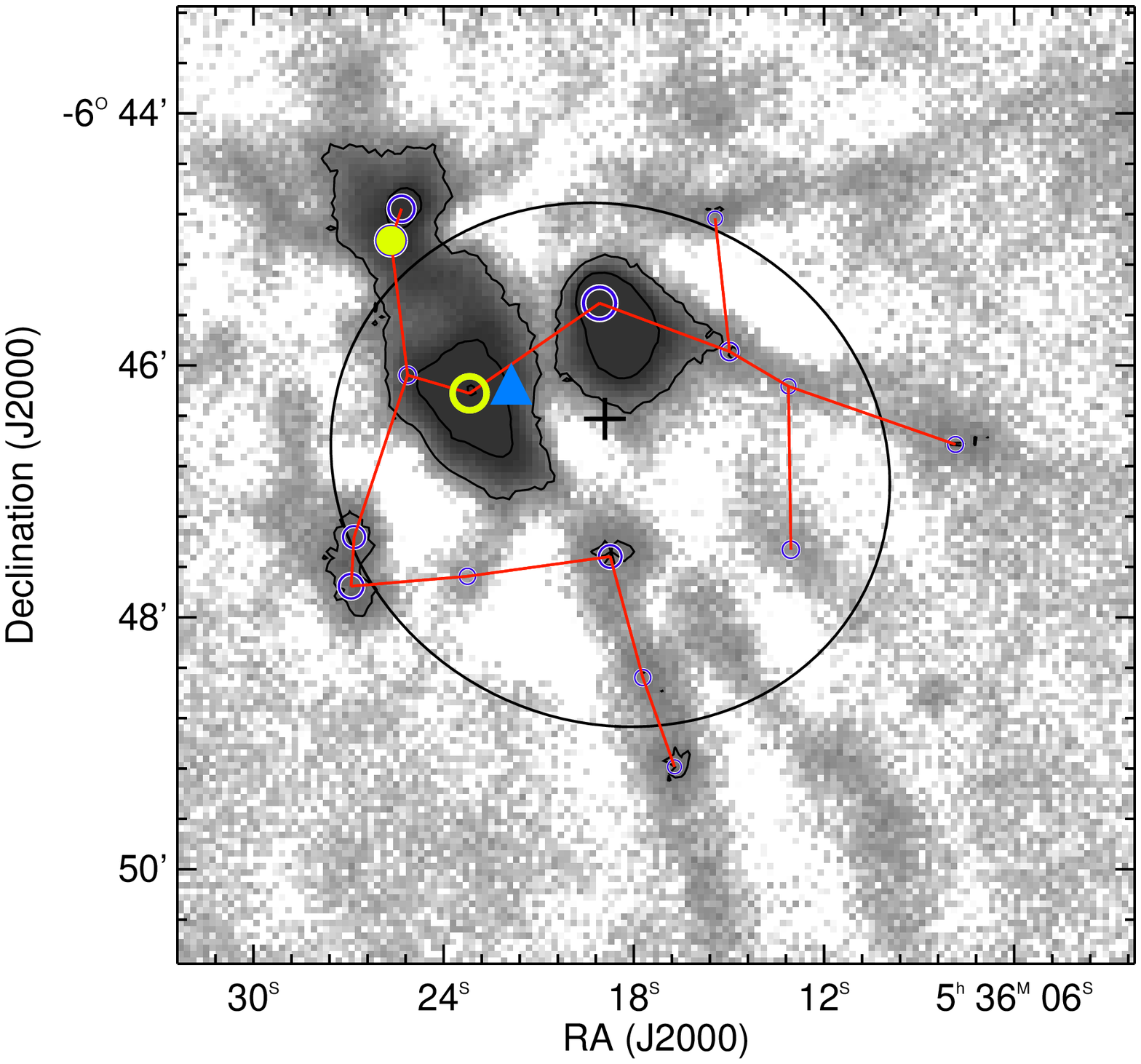}\\
\includegraphics[width=2.4in]{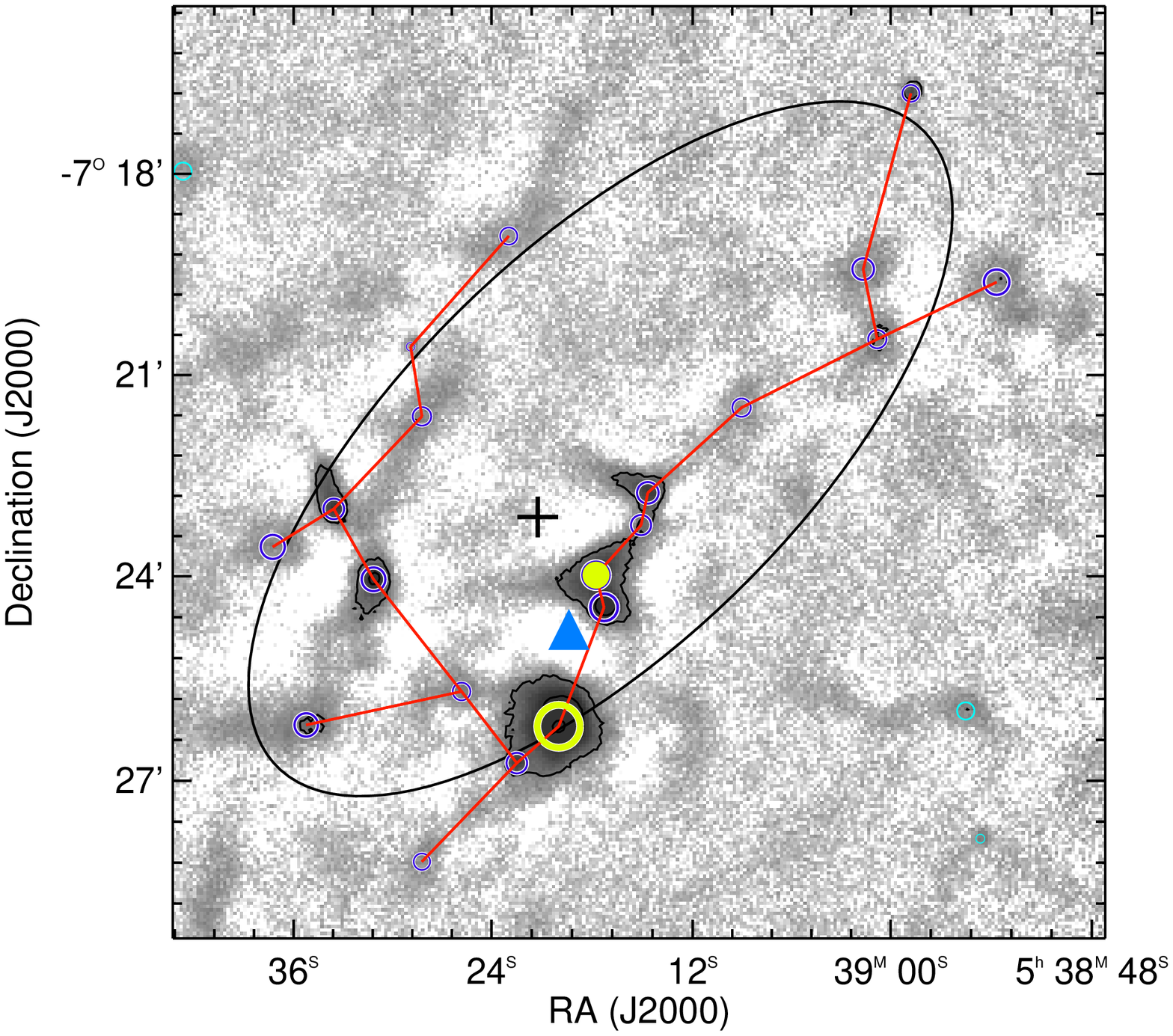} &
\includegraphics[width=2.4in]{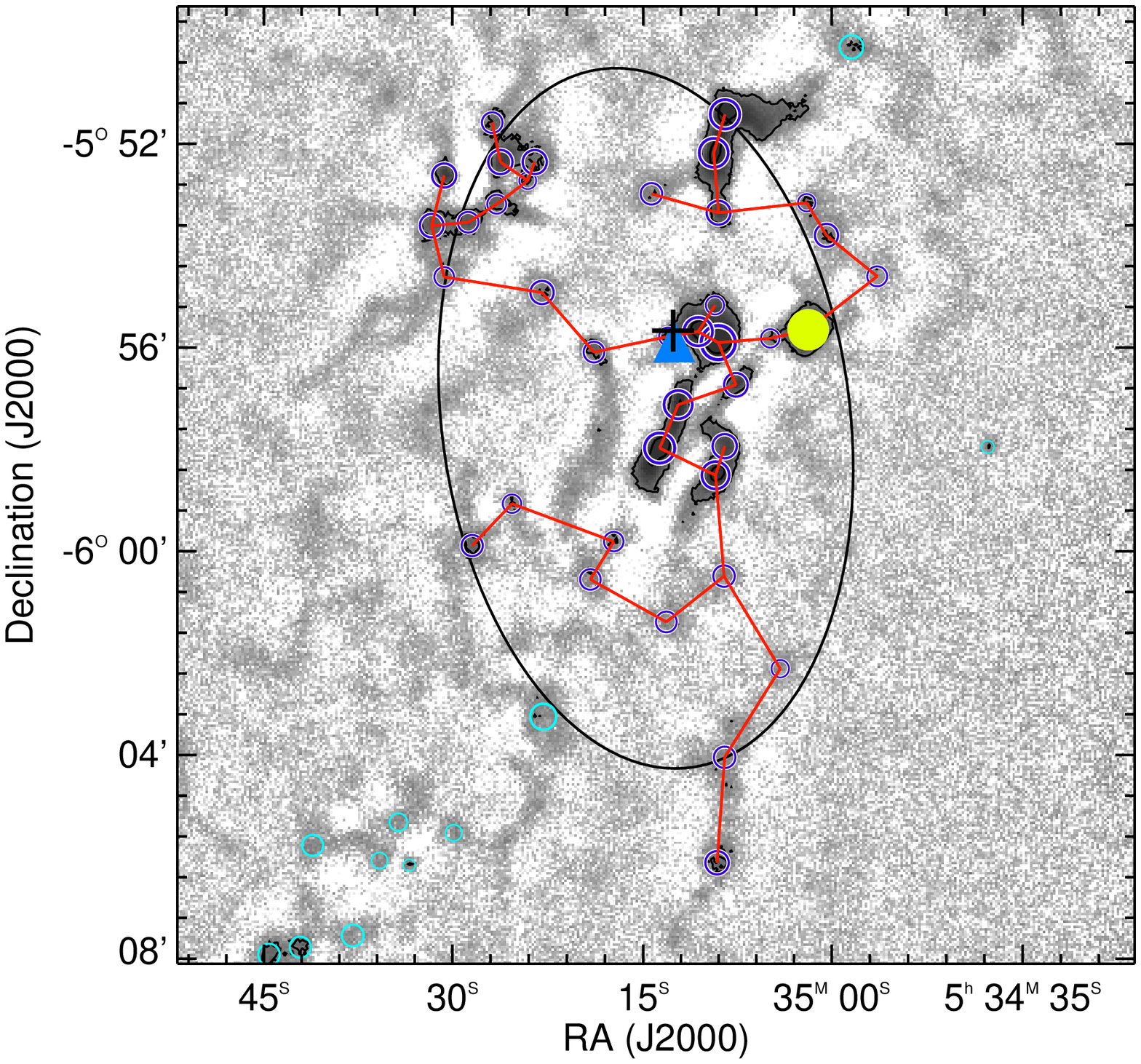} \\
\end{tabular}
\caption{MST-based clusters in Orion~A.  The
	greyscale shows the SCUBA-2 850~\micron\ emission,
	with black contours at 0.02, 0.1, and 0.5~Jy~arcsec$^{-2}$. 
	The dark / light blue circles show  
	cluster / non-cluster members identified using \get, with the circle size
	scaling with the total flux density.  
	Red lines show the MST structure
	and the plus sign shows the cluster centre.
	The black ellipse shows a fit to the cluster perimeter (see Section~\ref{sec_clust_char}).
	The highest flux density cluster member is shown by the open yellow circle,
	and the highest flux density starless core member with the filled yellow circle
	(in some cases, the highest flux density cluster
	member is starless).  The centre of flux density of each cluster is also shown 
	by the large blue triangle, and generally lies closer to the highest flux density cluster
	member than the median position does.
	This figure shows clusters 1 through 6 (top left to bottom right).
	}
\label{fig_all_clusters}
\end{figure}
\begin{figure}[p]
\begin{tabular}{cc}
\includegraphics[width=2.9in]{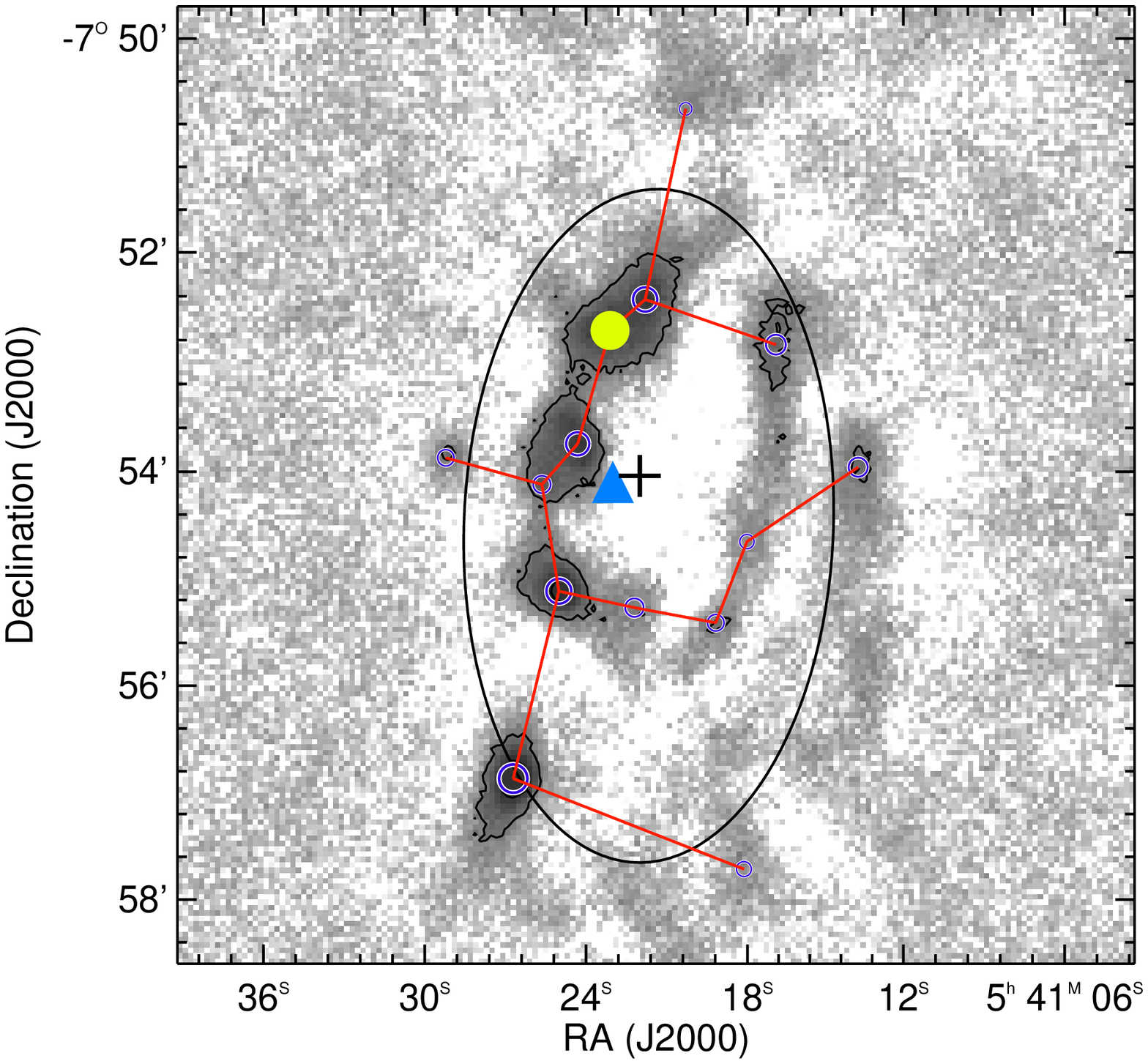} &
\includegraphics[width=2.9in]{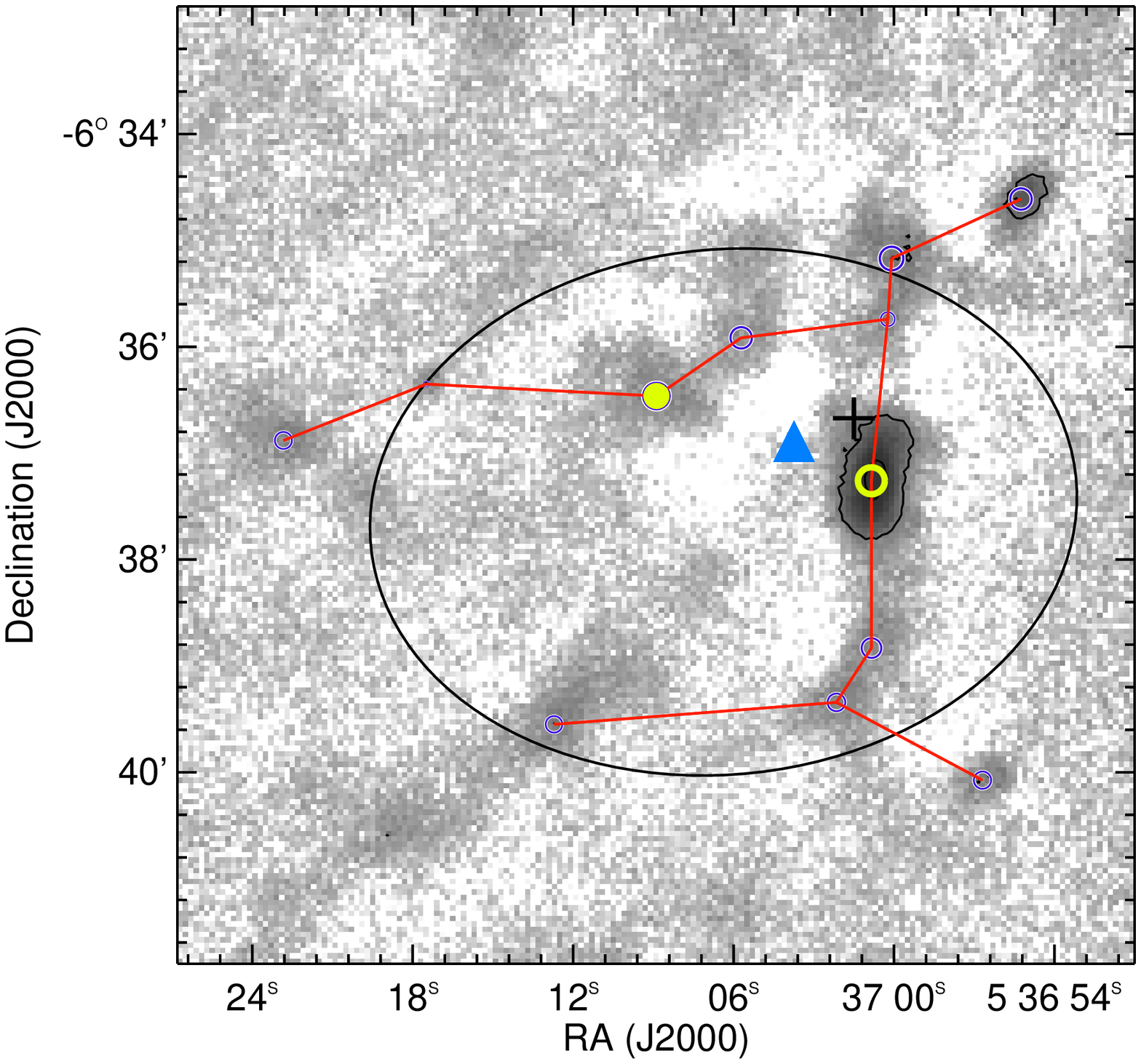} \\
\includegraphics[width=2.9in]{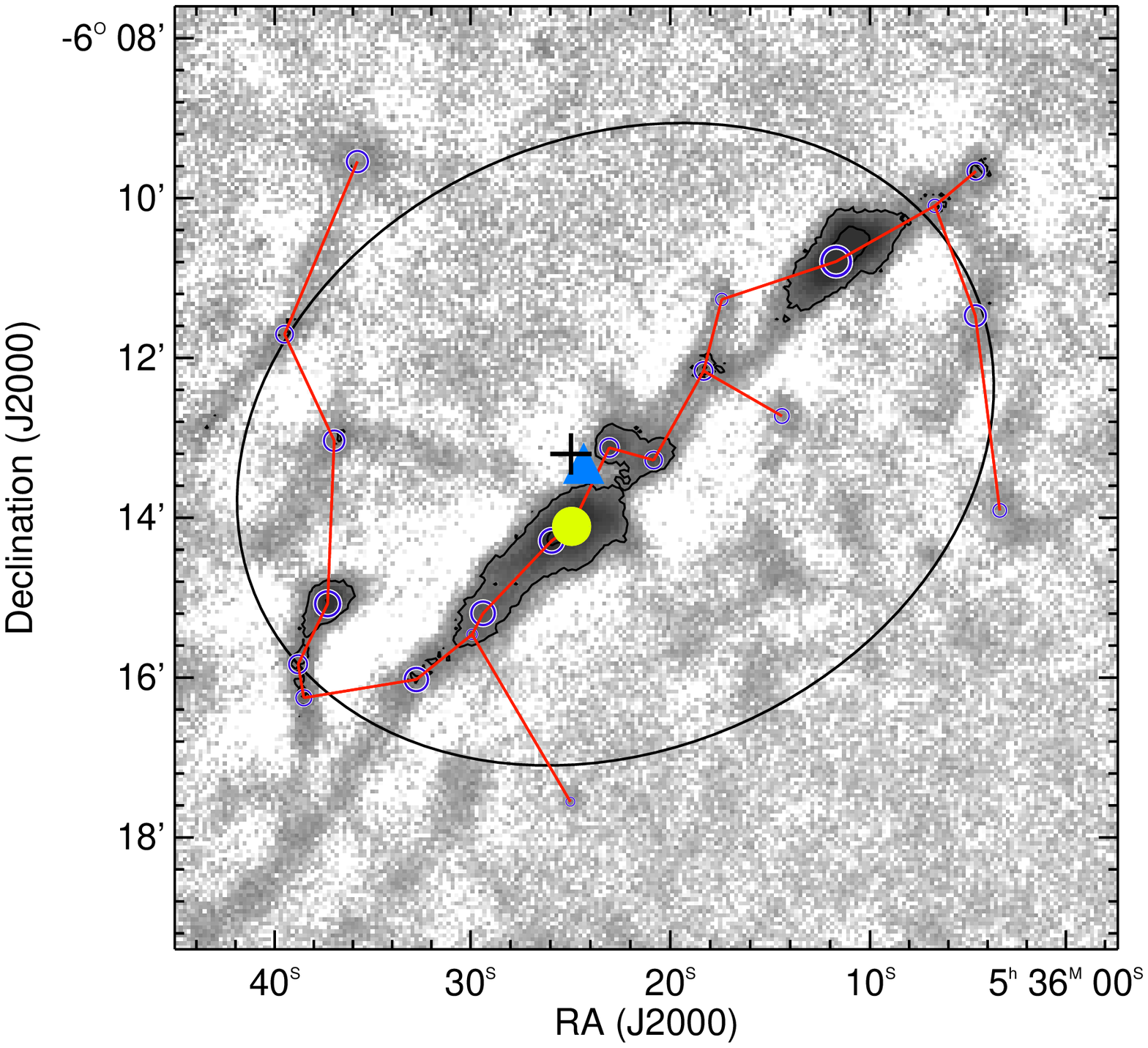} &
\includegraphics[width=2.9in]{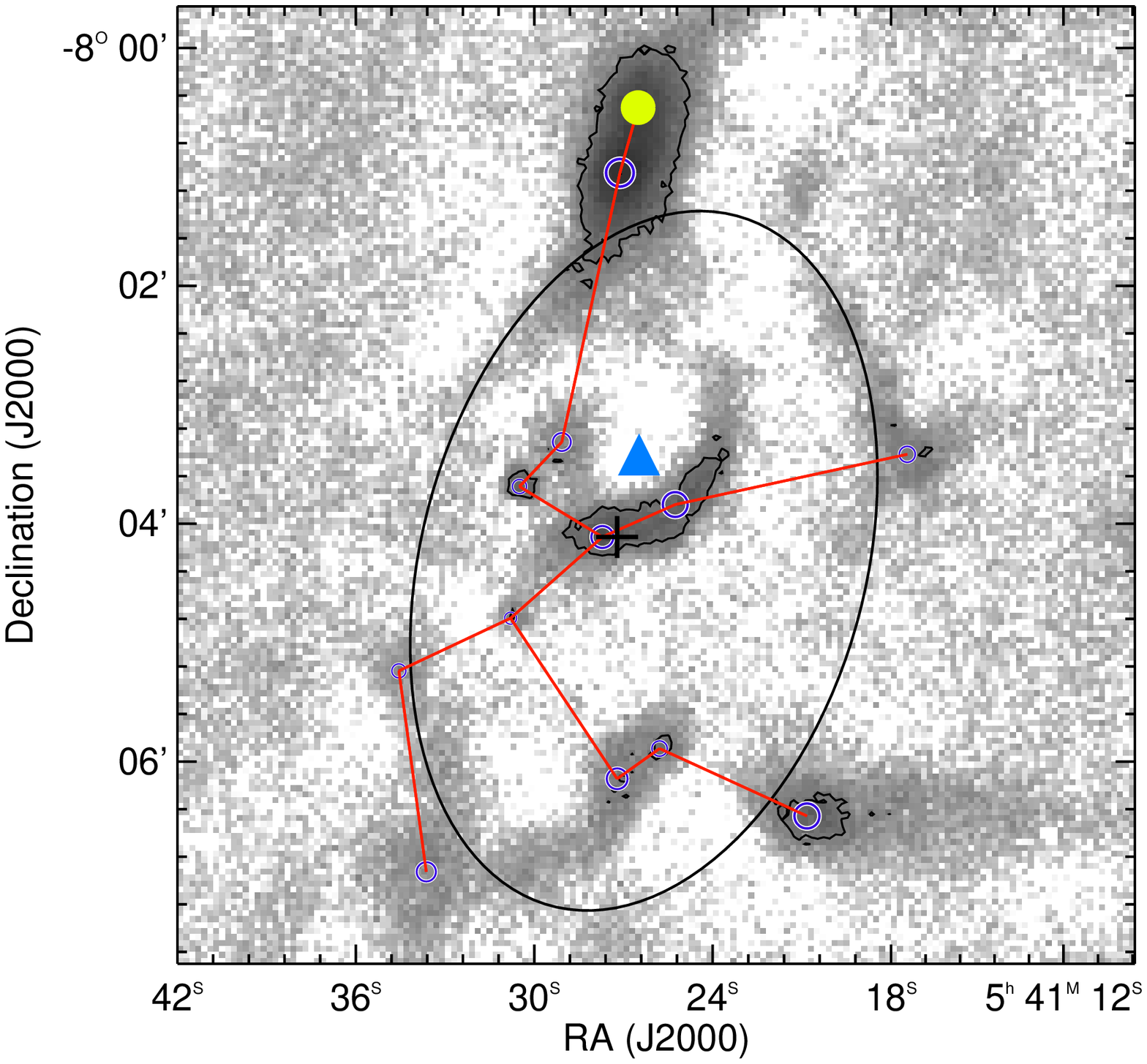} \\
\includegraphics[width=2.9in]{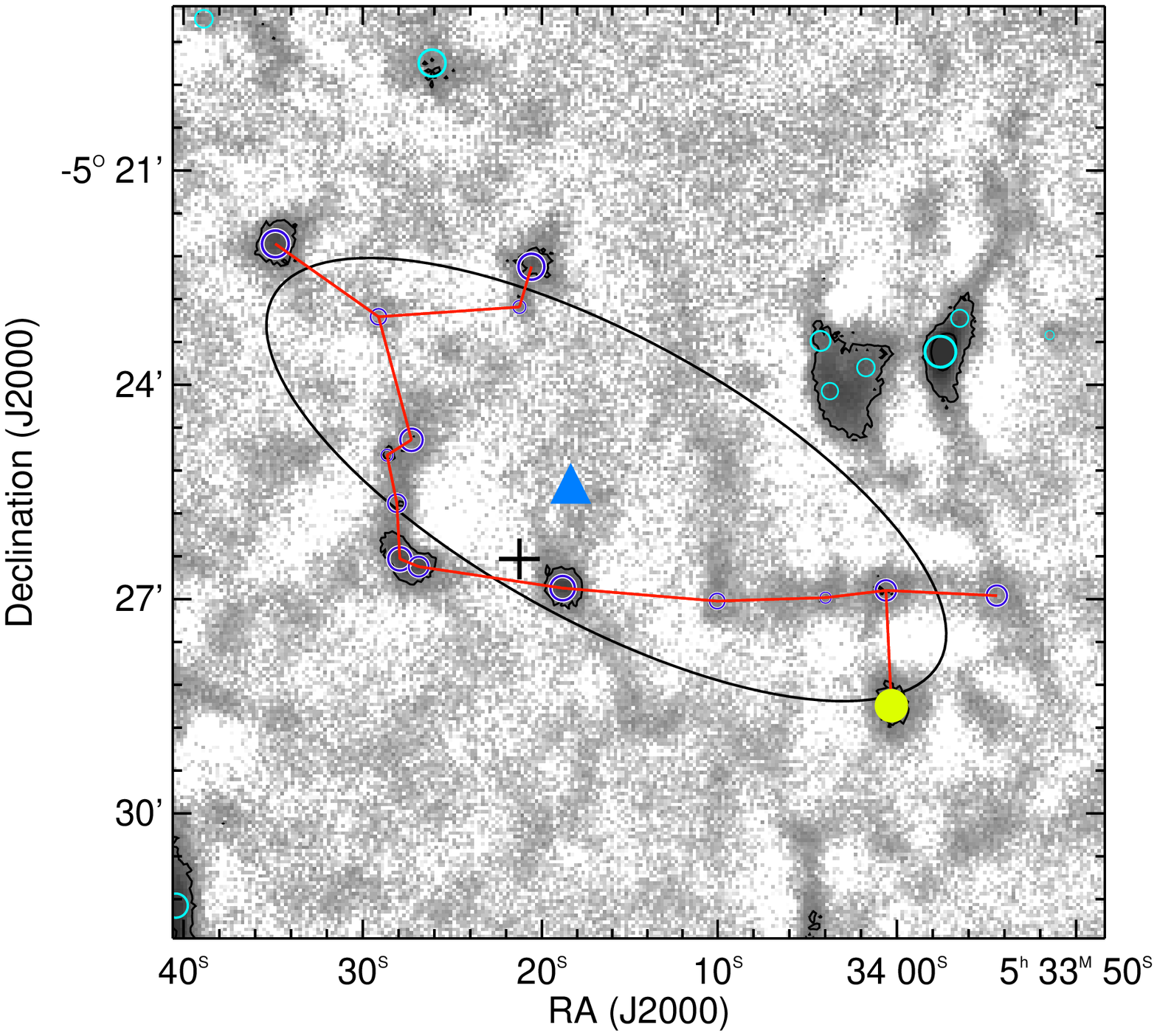} &
\\
\end{tabular}
\caption{The remaining five clusters in Orion~A.  See Figure~\ref{fig_all_clusters} for
	the plotting conventions used.
	 This figure shows clusters 7 through 11 (top left to bottom).
	}
\label{fig_all_clusters2}
\end{figure}

\subsection{Physical Characteristics of Clusters}
\label{sec_clust_char}
After the initial cluster identification, we imposed a minimum membership criterion of 
more than 10 cores 
to be classified as a cluster, which is the same as the threshold applied in \citetalias{kirk11}
and \citetalias{Kirk16b}; Appendix~\ref{app_Nmin} shows our final results vary little
with other minimum cluster sizes. 
In total, 11 clusters were identified.
Individual clusters can be 
seen in Figure~\ref{fig_all_clusters} and \ref{fig_all_clusters2}, while 
their physical properties are listed in Table~\ref{tab_clusters}.
The cluster properties in Table~\ref{tab_clusters} include the number of members,
the location of the cluster centre, the fraction of members classified as protostars,
and the aspect ratio of the cluster, in addition to several quantities introduced in
Section~\ref{sec_mass_seg} below.
One of the clusters (\# 11) is listed as having a protostellar fraction of precisely
zero in Table~\ref{tab_clusters}.  We note that one of the cluster members is in the vicinity
of a protostar, but the protostar-core separation is greater than our 
classification criterion. 

We follow \citetalias{kirk11} and \citetalias{Kirk16b} 
and define the cluster centre as the median position of cluster members.
For few-member systems, the median position is more robust to a single outlying member
influencing the centre than the mean position.  Furthermore, we can better
prevent undue bias in our later measurement of mass segregation by not adopting
the centre of mass (or centre of flux density) position for the cluster centre.
Figures~\ref{fig_all_clusters} and \ref{fig_all_clusters2} also show both the
median position and centre of flux location for each cluster, illustrating that
the latter tends to be closer to the highest flux cluster member. 

\subsection{Elongations}
Following \citetalias{gutermuth09}, 
a two-dimensional perimeter was determined for each cluster by forming 
a convex hull, which is a polygon where all interior angles are less than 180 degrees. 
An ellipse was fitted to this perimeter to estimate the aspect ratio of each 
cluster\footnote{\url{http://www.idlcoyote.com/ip_tips/fit_ellipse.html}}. 
As can be seen in Table~\ref{tab_clusters}, as well as Figures~\ref{fig_all_clusters} and
\ref{fig_all_clusters2}, many of the clusters are noticeably elongated.
We ran some simple checks to determine whether or not these elongations were statistically
significant.  We created 10,000 synthetic clusters each with 10, 15, and 25
members in a random uniform distribution over a 2D circular region.  
We then ran the same convex hull
and ellipse-fitting routine and compared the aspect ratios measured for the observed 
and synthetic clusters.  The two were notably different: a two-sample Kolmogorov-Smirnov (KS) 
test yields 
a probability of 
between 1\% and 2\% that the observed cluster aspect ratios are drawn from the
same population as the synthetic cluster aspect ratios.  We also ran a similar
test using a more centrally
concentrated distribution for the synthetic clusters (positions randomly distributed
equally in radial and angular directions from the cluster centre).  With this second
set of synthetic clusters,
we find an even smaller consistency with the observations, with two-sample 
KS probabilities of the two sets of aspect ratios being drawn from the same 
sample of 0.4\% and lower for clusters of 15 or more members.  Synthetic clusters
with 10 members had a higher probability of consistency than the uniformly random
distribution, but most of the observed clusters have 15 or more members, so
those probabilities are more applicable.  The low probability of consistency
between the observed cluster aspect ratios and those generated by cirularly symmetric
random distributions suggests that
the elongation seen in the observed clusters is real.  

Elongated clusters are 
perhaps not surprising at the dense core stage.  Dense cores appear to be 
strongly associated with filaments \citep[e.g.,][]{andre14} which would tend to
fragment to produce long chains of cores.  Clusters may form preferentially at the
intersection of several filaments \citep[e.g.,][]{Myers09,Schneider12}, but depending on the
geometry and star formation density in each filament, such origins could still easily
lead to initially non-round dense core clusters.

\begin{deluxetable}{cccccccccccccc}
\tablecolumns{14}
\tabletypesize{\tiny}
\tablecaption{Properties of MST-based Clusters\label{tab_clusters}}
\tablewidth{0pt}
\tablehead{
        \colhead{Index}&
        \colhead{R.A.\tablenotemark{a}}&
        \colhead{Dec.\tablenotemark{a}}&
        \colhead{N\tablenotemark{a}}&
	\colhead{S$_{rat}$\tablenotemark{b}}&
	\colhead{S$_{med}$\tablenotemark{b}}&
	\colhead{O$_{rat}$\tablenotemark{c}}&
	\colhead{O$_{med}$\tablenotemark{c}}&
        \colhead{Proto.\tablenotemark{d}}&
        \colhead{Aspect\tablenotemark{e}}&
	\colhead{Maj.\tablenotemark{e}}&
	\colhead{Min.\tablenotemark{e}}&
	\colhead{$\delta$R.A.\tablenotemark{e}}&
	\colhead{$\delta$Dec.\tablenotemark{e}}\\
        \colhead{ }&
        \colhead{(J2000)}&
        \colhead{(J2000)}&
        \colhead{ }&
        \colhead{(Jy)}&
        \colhead{(Jy)}&
        \colhead{(pc)}&
        \colhead{(pc)}&
        \colhead{Frac.}&
        \colhead{Ratio}&
        \colhead{(pc)}&
        \colhead{(pc)}&
        \colhead{(\arcsec)}&
        \colhead{(\arcsec)}
}
\startdata
  1 &    5:35:15.80 &   -5:19:19.20 & 462 & 435.98 &   0.26 &   0.29 &   1.43 &   0.06 &   1.95 &   3.56 &   1.83 &   206.5  &   188.6 \\
  2 &    5:36:26.55 &   -6:25:07.74 &  46 &   9.47 &   0.19 &   1.00 &   0.45 &   0.09 &   1.87 &   0.90 &   0.48 &  -126.6  &     7.4 \\
  3 &    5:39:57.27 &   -7:28:58.92 &  17 &   4.07 &   0.22 &   0.77 &   0.30 &   0.23 &   1.66 &   0.53 &   0.32 &   -11.2  &   -70.2 \\
  4 &    5:36:18.74 &   -6:46:21.70 &  16 &   5.11 &   0.12 &   1.21 &   0.24 &   0.19 &   1.10 &   0.29 &   0.27 &     2.6  &    21.9 \\
  5 &    5:39:21.11 &   -7:23:03.23 &  20 &   2.51 &   0.21 &   0.36 &   0.45 &   0.25 &   2.30 &   0.88 &   0.38 &    56.6  &   -60.7 \\
  6 &    5:35:12.44 &   -5:55:37.56 &  38 &   6.86 &   0.22 &   0.64 &   0.54 &   0.08 &   1.72 &   0.91 &   0.52 &   -32.4  &   103.2 \\
  7 &    5:41:21.65 &   -7:54:01.86 &  14 &   7.24 &   0.15 &   0.84 &   0.22 &   0.43 &   1.83 &   0.41 &   0.22 &     4.9  &    28.0 \\
  8 &    5:37:01.18 &   -6:36:36.39 &  12 &   2.86 &   0.15 &   0.72 &   0.34 &   0.17 &   1.34 &   0.43 &   0.32 &   -73.7  &    52.8 \\
  9 &    5:36:24.78 &   -6:13:09.09 &  22 &   8.43 &   0.14 &   0.25 &   0.47 &   0.05 &   1.28 &   0.64 &   0.50 &    33.6  &    -7.3 \\
 10 &    5:41:27.00 &   -8:04:04.47 &  13 &   3.68 &   0.16 &   1.78 &   0.27 &   0.23 &   1.61 &   0.39 &   0.24 &    13.5  &    12.0 \\
 11 &    5:34:21.02 &   -5:26:23.60 &  15 &   2.78 &   0.20 &   1.58 &   0.46 &   0.00 &   2.67 &   0.70 &   0.26 &    73.0  &   -66.7 \\
\enddata

\tablenotetext{a}{The cluster centre, measured as the median position of all cluster members,
	and the number of cluster members.}
\tablenotetext{b}{The flux ratio and median total flux of cores within the cluster.}
\tablenotetext{c}{The offset ratio (offset of the highest flux core from the cluster centre
	divided by the median offset),
	and the median offset of all cluster members from the cluster centre.}
\tablenotetext{d}{The fraction of protostellar cores in the cluster.}
\tablenotetext{e}{Results from fitting an ellipse to the convex hull of the cluster members:
	the aspect ratio, the major and minor axis lengths, the separation of the centre
	of the ellipse from the cluster centre.}

\end{deluxetable}

\subsection{Offset Ratio}
\label{sec_mass_seg}

\begin{figure}[htb]
\begin{tabular}{cc}
\includegraphics[width=3.0in]{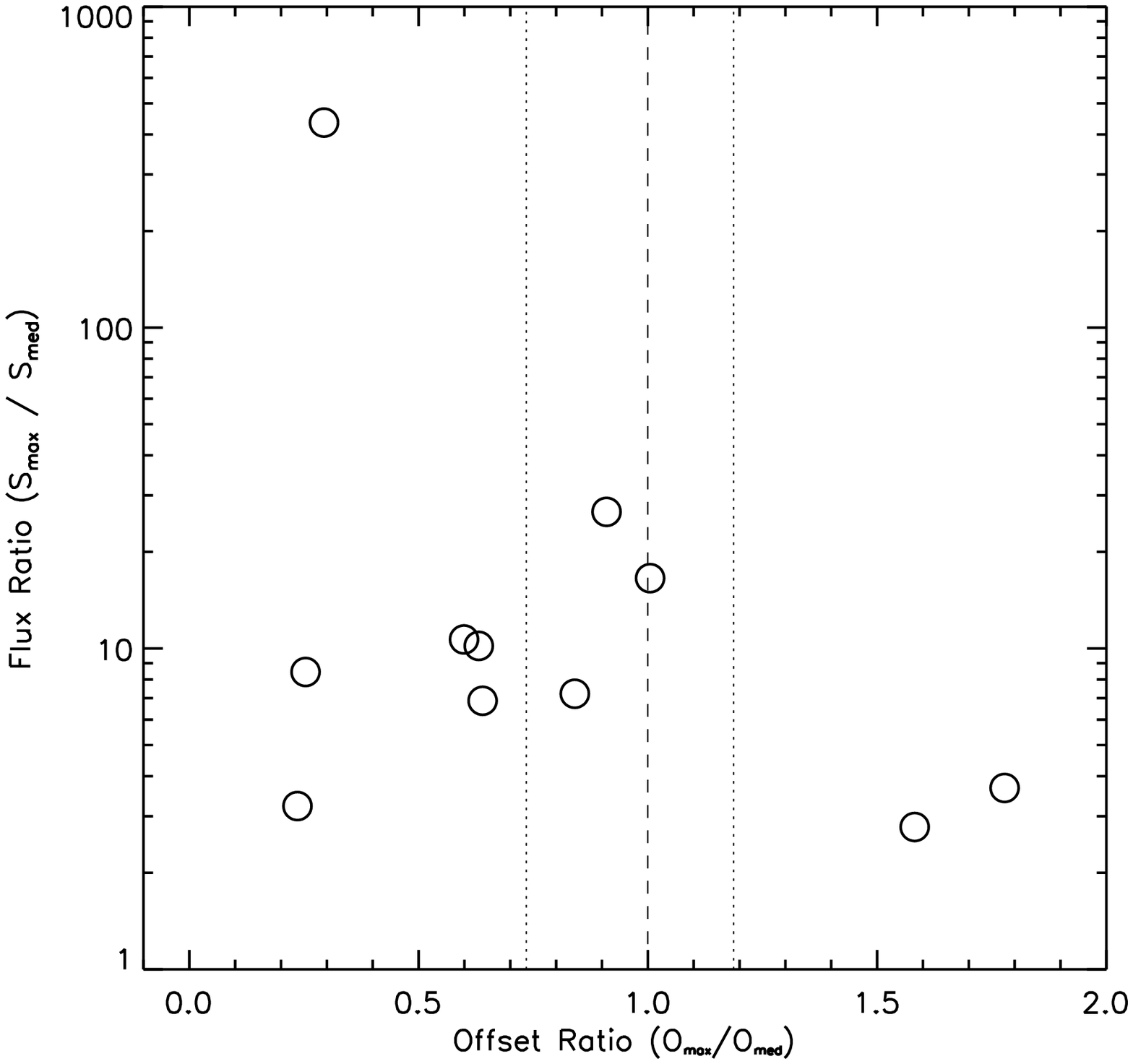} &
\includegraphics[width=3.0in]{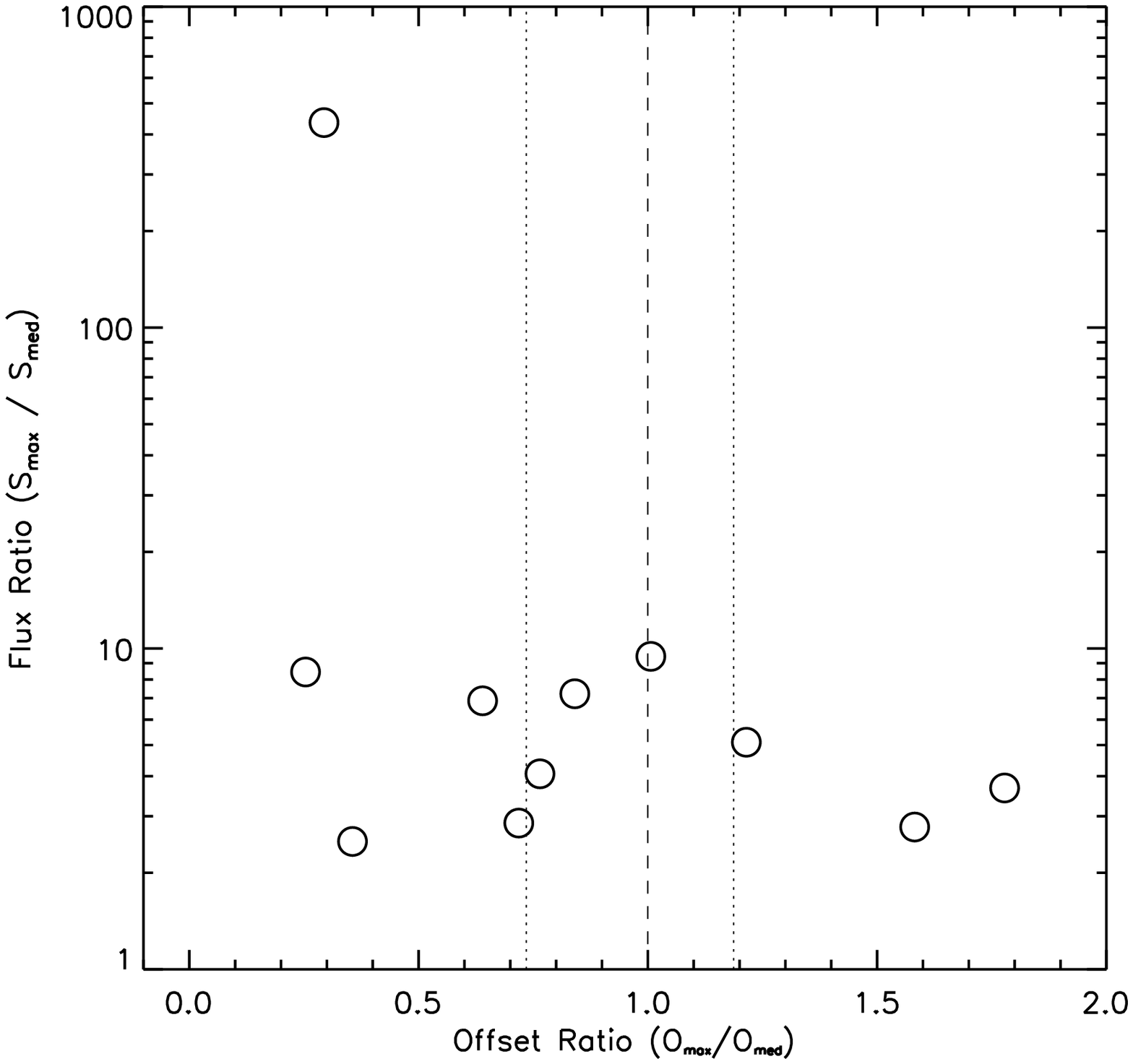} \\
\end{tabular}
\caption{The tendency of a centrally-located highest flux density (most massive) 
member within its cluster.
In the left panel, $S_{max}$ and $O_{max}$ correspond to the highest flux density
core of all cluster members, whereas for the right panel, $S_{max}$ and $O_{max}$ correspond
to the highest flux density core identified as starless (see text).  The median flux
densities and median offset values are calculated for all cluster members, regardless of
classification ($S_{med}$ and $O_{med}$ in Table~\ref{tab_clusters}).
The dotted lines indicate the 25th and 75th percentiles of 
derived for a set of randomly distributed cluster members.  As discussed in
Section~4.4, there is a notable tendency for our cluster sample to have offset ratios
below one (i.e., lying to the left of the dashed line). 
}
\label{fig_offs_ratio}
\end{figure}

Following \citetalias{kirk11} and \citetalias{Kirk16b}, we use the location of the highest flux density 
core in each MST-based cluster to look for indications of whether this core is
located randomly or if it is preferentially located toward the cluster centre.
We first measure the radial offset of each cluster member, i.e., the separation of each core from
the cluster centre.  We then compare the offset of the highest flux density core
to the median offset of all cluster members 
($O_{rat}$ in Table~\ref{tab_clusters}), 
i.e., the offset ratio.  Offset ratio values less than one indicate that 
the most massive member is located close to the center of the cluster, while values larger than 
one indicate the most massive member is in the cluster outskirts.  
For a very rough indication of the relative importance of gravity from the most massive
cluster member, we
calculated the ratio of the flux density of the most massive cluster member
to the 
median cluster member flux density ($S_{rat}$ in Table~\ref{tab_clusters}), 
the flux density ratio.  (Recall that under our assumption of a constant
conversion factor between flux density and mass, the ratio of flux densities and masses are identical.) 
Flux ratios near one indicate that all cluster members have similar flux densities, while clusters with 
flux density ratios much greater than one indicate a large range of flux densities.  
Higher flux density ratios are also usually indicative of the presence of higher flux density cores.
The left panel of Figure~\ref{fig_offs_ratio} shows the flux density and offset ratios of each cluster.

Figure~\ref{fig_offs_ratio} shows that most clusters have centrally-located highest flux density
(most massive) cores.  We ran some tests to check whether or not our observed distribution of
offset ratios are consistent with a randomly-located most massive member.
We created 10,000 synthetic clusters with 15 members (a typical group size) 
placed randomly within a uniform, two
dimensional circular distribution. 
Within these clusters, we randomly selected 
a most massive cluster member and calculated the offset ratio in the same way as the observed
clusters. The 25th and 75th percentiles of this distribution are shown in 
Figure~\ref{fig_offs_ratio}. 
A two-sample KS test was used to compare 
the observed offset ratios with the offset ratios from the random distributions. 
The probability that the random distribution 
and the observed distribution of offset ratios were drawn from the same sample was 7\%, 
a nearly 2-$\sigma$ result.  Other synthetic cluster types (e.g., uniform
spherical distribution, or different number of cluster members) yield similar distributions
of offset ratios, with a roughly equal number of ratios above and below one 
(see discussion in \citetalias{kirk11}), so we would expect similar statistics
for such cases.  The small sample
size of clusters is the driving factor in the statistical significance we find.  Including
offset ratios of dense cores in clusters in Orion~B measured in a similar manner yields
an only 3\% probability of consistency with a random distribution \citepalias[see][]{Kirk16b}.

To interpret the offset ratios derived as relating to {\it mass} segregation, we
would need to assume a one-to-one conversion between the flux- and mass-ranking of cores
within a cluster (or at least that the core with the highest flux density is also the most massive core).
This assumption may not be tenable if cores have a range of temperatures, as discussed in
Section~\ref{sec_core_mass}.  Following \citetalias{Kirk16b}, we additionally search for the location
of the highest flux density {\it starless} core within each cluster, and measure its offset
ratio.  The rationale here is that the if the highest flux density core only has high flux density due
to an elevated temperature, then the highest flux density starless core is the next most 
likely candidate for the highest mass cluster member.  
The right hand panel of Figure~\ref{fig_offs_ratio} shows the offset ratios measured for
the highest flux density starless core in each cluster, and these generally follow a similar
trend, where most have offset ratios less than one.  We note, however, that several of the clusters
have a fairly high protostellar core fraction, which limits the number of starless
cores available for us to perform this test.
In Appendix~\ref{app_temp}, we use {\it Herschel}-based temperatures to further test
the robustness of our analysis to temperature variations, and demonstrate that the most massive
cluster member is nearly always the same as the highest flux cluster member.

\section{MASS SEGREGATION AND SURFACE DENSITY}
\label{sec_m_sigma}
In Section~\ref{sec_mass_seg}, 
we found evidence of a preferential central location of the highest flux density / 
most massive core in each MST-based cluster, suggesting that there is some degree of
mass segregation present in these systems.  We can also search for mass segregation in Orion~A
using a second, completely independent method.
This method is the $S-\Sigma$ method from \citep{maschberger11}, where we have replaced
their mass $M$ by the core total flux density $S$.  The $S-\Sigma$ method uses the local 
core-core surface density
as an indicator of clustering, and compares the mass (total flux density) of cores
to their core-core surface density to look for trends.  In the case of a single cluster,
mass segregation would clearly be visible as higher mass cores being found at higher
core-core surface densities.  In systems with multiple sub-clustered populations, mass segregation
should still be visible as a tendency for higher mass cores at higher core-core surface densities,
although a large scatter in individual values might be expected.
The $S-\Sigma$ measurement is completely
independent of any definition of individual clusters, making it an ideal complement
to the MST-based cluster analysis in Section~\ref{sec_mst}.

\begin{figure}[htb]
\begin{tabular}{cc}
\includegraphics[width=3.3in]{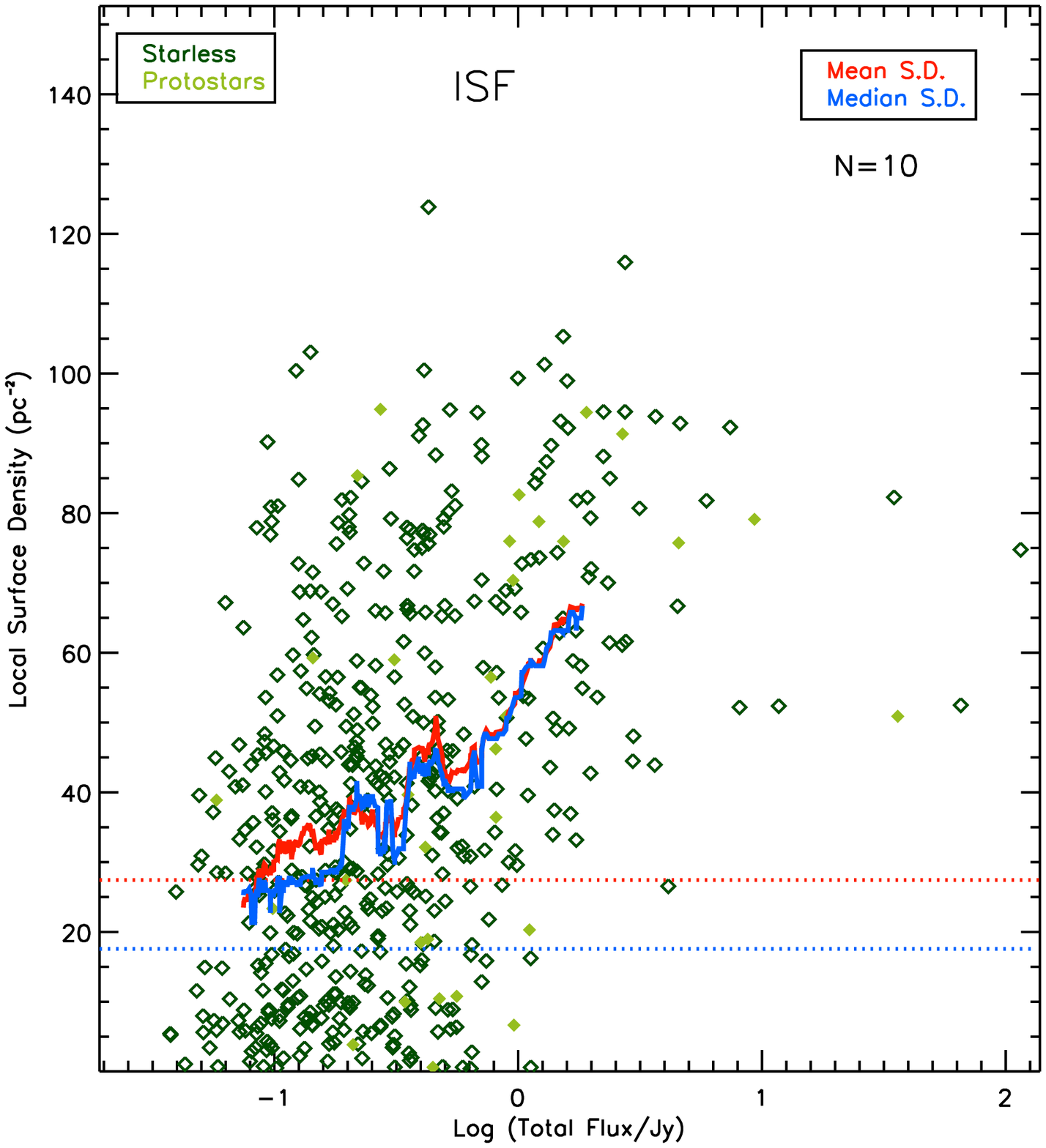} & 
\includegraphics[width=3.3in]{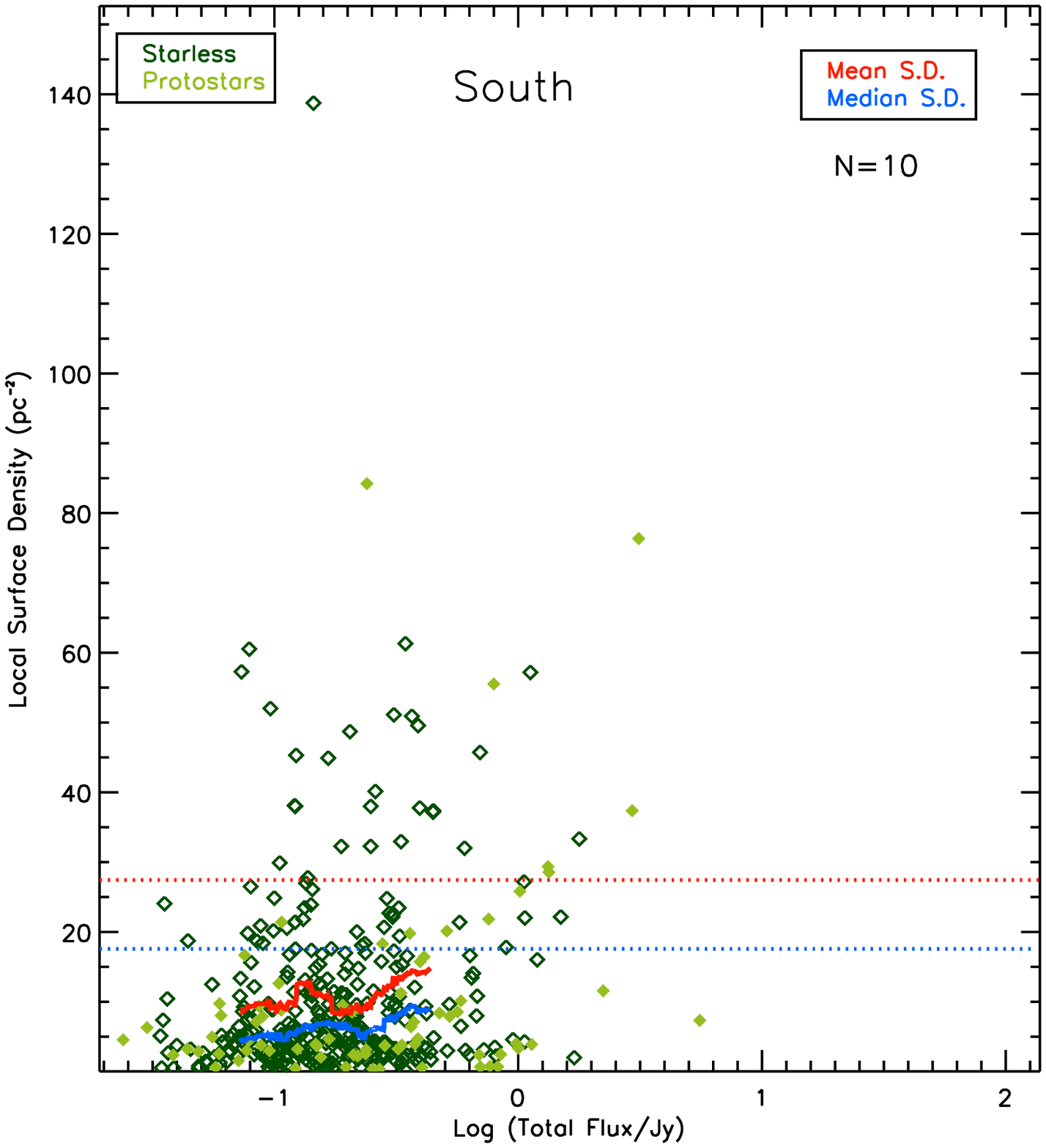}\\
\end{tabular}
\caption{The core-core surface density, as derived from the ten nearest neighbours
(NN10), as a function of core flux density.  
The left hand plot shows dense cores in the ISF (having declinations above 
$\sim -5:50:25$), while the right hand plot shows dense
cores south of the ISF.  Starless cores are shown as dark open diamonds, while protostellar
cores are shown as the light filled diamonds.
The running mean and median 
values for the starless cores are shown by the red and blue lines respectively, 
calculated by binning 30 values on 
either side of each point.  Both lines highlight the general trend of a higher surface 
density of cores around higher flux density cores.  The dotted horizontal lines show the global 
mean (red) and median (blue) values.}
\label{fig_m_sig}
\end{figure}

We calculate the surface density around each core by determining the circular radius
needed to encompass a total of $N$ cores, i.e., 
\begin{equation}
\sigma_N = \frac{N}{\pi r^2_N}
\end{equation}
where $\sigma$ is the core-core surface density, and $r$ is the circular radius encompassing
$N$ cores.  The fractional uncertainty associated with this surface density estimate
scales as $N^{-0.5}$ \citep{Casertano85,gutermuth09}.  In our analysis, similar 
to \citetalias{Kirk16b}, we used both
the closest five cores (NN5) and the closest ten cores (NN10), and find similar
results with both.
In Figure~\ref{fig_m_sig}, we show the NN10 core-core surface densities
plotted against each core's total flux density.  For this figure, we separately
consider cores in the vicinity of the ISF and further south in Orion~A to ensure that
our results are not biased by behaviour present in only one of the two core populations.
Both panels of Figure~\ref{fig_m_sig} show a clear trend
highlighted by the red and blue co-moving mean and median values, that 
the total core flux density tends to be higher in regions with a higher surface density of 
neighbouring cores, although the slope of the co-moving mean and median values
differ between the ISF and the south.  As with our MST analysis in Section~\ref{sec_mst},
we note that protostellar cores may have systematically higher total flux densities
than starless cores for the same intrinsic mass, due to the typically higher temperature
of the former population.  We handle this potential bias in our analysis 
by analyzing the starless cores and protostellar cores separately, as well as jointly.
In our statistical tests below, the protostellar cores represent the smallest population,
and therefore show the poorest statistical significance in the results reported.  
The co-moving mean and median lines shown in Figure~\ref{fig_m_sig} for illustrative
purposes only is based on the starless core population.

We employ a two-sample KS test to evaluate the significance of the apparent
trend of higher flux density cores being found in higher surface density environments. 
First, we compare the core-core surface densities for the higher flux density half of the cores with 
the core-core surface densities for the lower flux density half of the cores to see whether
or not the 
two distributions of surface densities are consistent.  We also run a similar test
using the surface densities associated with the cores with the lowest and highest third of the
flux densities.
In both of these cases, and using either of the 
NN5 and NN10 derived surface densities, 
and running the analysis across all of Orion~A,
or examining the ISF and the south separately, the probability that the 
highest and lowest flux density cores have surface densities drawn from the same parent 
population is less than 
$10^{-7}$\% for either the starless cores alone, or for both the starless and protostellar
cores analyzed together.  For the protostellar cores alone, the smaller population size,
especially when split between the ISF and the south, lead to somewhat less significant results:
the probability is less than
5\% in all cases, with many of the tests giving probabilities of less than 1\%.

We also ran a Mann-Whitney test, which compares the ranking of
values within two samples to determine if one has typically larger values.  As expected
from the clear positive slope seen in the co-moving mean and median lines indicated on 
Figure~\ref{fig_m_sig}, the Mann-Whitney test shows a strong probability of higher flux density
cores inhabiting higher surface density environments.
For the starless cores, or starless and protostellar core samples, 
we find probabilities of $10^{-9}$\% or less that the higher
flux density cores do {\it not} inhabit higher surface density environments, i.e., nearly 100\%
likelihood that higher flux density cores {\it do} inhabit higher surface density environments.
The one exception to this result is the probability measured for the starless core population
in the south, which has a value of 2\% or less, which still implies a 98\% or higher
likelihood of higher flux density cores inhabiting higher surface density environments.  
For the protostellar
core populations, all probabilities are below 7\%, with all but two tests giving 
probabilities below 1\%.
We therefore conclude that the dense cores in Orion~A show strong evidence for a tendency
for high flux density cores to be located in high core-core surface density environments.

In Appendix~\ref{app_temp}, we find similar results using instead estimated masses
based on {\it Herschel}-derived temperatures for the dense cores.

\section{DISCUSSION}

\subsection{Cluster Identification}
Identifying clusters of cores in molecular clouds, which tend to have a hierarchical
structure is difficult at the best of times \citep[see, for example, the discussion of 
hierarchical structure within Orion in][]{Takahashi13}.  This situation is complicated further
in analysis of dense cores, since the identification of cores is also challenging.  We examine the
uncertainty in our main results arising from these complications in a variety of ways. 
In Appendix~\ref{sec_sourcefinding}, 
we examine the impact of differing core identification algorithms.
We first compare the core catalogue used in our main analysis (\get) with another
independent technique (\fell) in Appendix~\ref{sec_compare_gs_fell}, 
and find the main difference is in 
how complex emission structures are divided (rather than whether or not 
emission is associated with
a core at all).  We also compare with the core catalogues from other
SCUBA-2 analysis, namely \citet{Mairs16} in the south (see Appendix~\ref{sec_compare_mairs})
and \citet{salji15a} in the north (see Appendix~\ref{sec_compare_salji}),
both of which use different core identification techniques than our analysis.  
Both of the island and fragment
catalogues of \citet{Mairs16} similarly correspond well with our core catalogue,
while there is poorer correspondence with \citet{salji15a} driven largely by 
constraints in core properties those authors imposed.

To determine whether or not the core identification technique influences our results,
we re-ran all of the analysis presented in the paper using instead the \fell-based
core catalogue.
While the numbers and properties of the cores vary significantly
(919 with \get\ and 773 with \fell), the conclusions drawn from the 
clustering analysis remain largely the same.  

For our results relying on the MST
analysis (Section~\ref{sec_mass_seg}), we additionally test the effects of uncertainty in 
\Lcrit\ and the choice in minimum cluster size on the
clusters identified, as well as the results we derive based on the MST clusters.
These tests are shown in Appendix~\ref{sec_mst_uncert}, and here too, we find that our overall conclusions 
are similar.

These comparisons and tests all suggest that the results of our analysis are
consistent, regardless of the core extraction method and parameters used in the
MST analysis. 

\subsection{Smaller-scale Fragmentation}
Another effect that should be considered is how the dense cores we observe might fragment 
on scales smaller than we are able to resolve with the JCMT.  
The northern portion of the ISF was recently
observed with ALMA by \citet{Kainulainen16}, who identified 40 sources with 3\arcsec\ resolution.
They found evidence of clustering on all scales examined, with no preferential length scale associated
with the clustering; instead, clustering became increasingly evident down to the minimum size probed
by their resolution.  Much of the clustering detected by \citet{Kainulainen16} is on scales below
that observable with our SCUBA-2 data: they report detecting significant clustering at scales of 17000~AU
and smaller ($\le$ 0.08~pc), particularly for their starless sources, while the SCUBA-2 850~$\mu$m beam of 
14.6\arcsec\ corresponds to a scale of $\sim$0.03~pc.

In the protostellar regime, \citet{Kounkel16} searched for binaries and higher order stellar systems
with separations of about 100 to 1000~AU across Orion~A using {\it Hubble} observations.  They find
a companion fraction of about 14\% for the protostellar sources, a similar fraction to that obtained
earlier in Taurus.  They furthermore note that the companion fraction appears to be somewhat higher
for protostars living in high stellar density regions of the cloud, with a roughly 50\% increase in
the companion fraction for protostars in $\Sigma > 45$~pc$^{-2}$ environments versus those in lower
surface density environments.  Based on the \citet{Kounkel16} results, we might therefore expect that
the dense cores in the ISF will fragment more than those further south in L1641.  The degree of extra
fragmentation at higher local source surface densities measured, however, does not appear to be significant
enough to invalidate our assumption that the more massive protostars form out of the more massive 
dense cores, i.e., there is no indication  
that the most massive dense cores form only a large number of low mass protostars. 

Therefore, although small-scale fragmentation certainly exists in Orion~A, the existing data does not suggest
that our results would qualitatively change if our analysis were performed using higher resolution data.

\subsection{Interpretation}

One of the main results of our analysis is that most of the clusters of dense cores in 
Orion~A appear to have some degree of mass segregation already.  We investigated this
using two independent techniques.  First, we used 
an MST to identify clusters which we found have tentative
evidence that the highest flux density (most massive) cluster
member was preferentially found toward the cluster centre.  Second, we applied 
the $S-\Sigma$ technique which uses the core-core surface density  
as a proxy for clustering and found a trend between core flux density and local core-core 
surface density.  
Measurements of mass segregation are often controversial, in part due to differences
in the type of system (e.g., single cluster or multiple subclusters) and the 
size scale that is appropriate to
analyze, as discussed in \citetalias{Kirk16b} and also \citet{Girichidis12}.
The two methods we adopted in our analysis both allow for the possibility of 
multiple clusters or sub-clustered systems within the region analyzed, and both are also
effectively localized measures.

A third popular
method that we did not use is the $\lambda_{MST}$ technique pioneered by \citet{Allison09}.
The $\lambda_{MST}$ method performs very well in tests of searching for
mass segregation within a single cluster \citep{parker15}, but tends instead to measure
features in the large-scale structure when applied to wide areas containing multiple 
sub-clusters, such as the Taurus molecular cloud, as discussed in \citetalias{Kirk16b}.  
Like Taurus, the Orion~A molecular cloud is a multi-parsec structure which is not
best represented as a single star-forming cluster where cluster members have the potential
to interact with each other.
In Orion~A, the typical denser gas velocity dispersion is $\sim$1 to 5~km~s$^{-1}$
based on C$^{18}$O(3--2) HARP observations \citep{Buckle12},
resulting in a crossing time along the narrow axis of Orion~A 
($\sim$5~pc wide in our SCUBA-2 map) of $\>1$~Myr, which is significantly 
larger than the typical estimated dense
core lifetime of several tenths of a Myr 
\citep[e.g.,][]{Kirk05,Hatchell07,Enoch08,Konyves15}.  
The $\lambda_{MST}$ method is therefore expected to trace the large-scale assembly of
lower-density cloud material, while the MST and $S-\Sigma$ methods allow for substructure
within Orion~A to be separately considered.

Using the MST method, we find that most of the clusters identified have a centrally-located
most massive member (i.e., offset ratios less than one).  
The offset ratios measured differ from those expected from a randomly
located most massive member at the 93\% confidence level, with the relatively low statistical
significance driven in part by the small sample size.  Including the dense core clusters
analyzed in Orion~B in a similar manner by \citetalias{Kirk16b}, where the distribution of 
offset ratios are similar (roughly the same proportion of offset ratios below and above one), 
the result becomes more significant (97\% likelihood of non-random locations).  Using
the $S-\Sigma$ technique, we find more massive cores tend to live in higher core-core surface
density environments at well above a 99.99\% significance level.  The 
statistical significance of this measure is much greater in part because every core 
contributes a measurement, rather than using only one measurement per cluster.

Our results, showing that more massive dense cores tend to occupy more clustered
environments, may suggest that the centres of clusters provide a favourable
environment in which more massive protostars can form \citep[see also][]{Myers11}.  
In Orion~A, there is already some evidence that highly clustered environments
favour the formation of more massive stars: \citet{Hsu12} and \citet{Hsu13} compare
the young stellar populations in the ISF and L1641 (southern Orion~A), and conclude that there is
some evidence for a deficit of the most massive stars (O and early B type) in L1641
compared to the ISF.  Our results suggest that within southern Orion~A, there will be a 
relative over-abundance of the most massive protostars within highly clustered zones and
a relative dearth of the most massive protostars in the sparser regions of the south.  
These relative abundances are consistent with the \citeauthor{Hsu12} results, so 
long as the overall fraction
of massive protostars in southern Orion~A is lower than in the ISF, a scenario which seems
reasonable in the context of our results.

Tying our mass segregation results to star formation models
such as turbulent core accretion \citep{mckee03} and competitive accretion
\citep{bonnell01} is not yet possible.
Predictions from each model on the expected spatial distribution of dense core masses
are needed first.  Naively, the competitive accretion model might be expected
to be a better fit, since young {\it protostars} are expected to be mass segregated
in that model.
Unlike the turbulent core accretion model, however,
the majority of material accreted by protostars does not originate in the protostar's
natal core in the competitive accretion model \citep[see, e.g.,][]{Tan14}, therefore 
it is unclear whether 
dense cores should show signs of mass segregation or if segregation should only
be present in the protostars. 
The idea of cluster centres providing a favourable accretion environment is,
however, consistent with 
other lines of evidence suggesting that protostellar clusters form with some degree 
of primordial mass segregation.  As discussed in
\citetalias{Kirk16b}, several recent studies of young protostellar clusters
find evidence of centrally located more massive members 
\citep[e.g.][]{Megeath05,Hunter06,Kryukova12,Elmegreen14}, although at least
one other study comes to a different conclusion \citep{Hunter14}.
Obsevations of slightly older protostellar clusters also often show evidence of
mass segregation, and estimate that the clusters are too young to be able
to explain this through dynamical evolution \citep[e.g.,][]{Carpenter97,
hillenbrand97,Bonnell98,Stolte06,kirk11,Gennaro11,Davidge15}, although this interpretation is
somewhat controversial, with other authors finding a lack of evidence of
primordial (or sometimes present day) mass segregation
\citep[e.g.,][]{Allison09_Orion,Allison10,Parker12,Wright14}.
The observations we analyze here show the clustering behaviour at an early
stage, and it is important to emphasize that the masses, and to a lesser
extent, the locations, of the dense cores will continue to evolve beyond the
snapshot in time that our observations capture.  As noted above,
the centres of clusters are thought to provide an environment favourable to higher
mass accretion \citep[e.g.][]{Myers11}, 
and this has also been observed in some numerical simulations
\citep[e.g.,][]{Smith09}, where the most massive protostar accretes material that
originated from throughout the cluster rather than only its natal core.  
While all of the cores are expected to
continue to accrete material, those inhabiting the centres of clusters may experience
a larger rate of growth than those on the outskirts.  The motion between
dense cores and their surrounding less dense envelopes is generally small 
throughout a cluster \citep[typically less than the sound speed, e.g.,][]{Walsh04,Kirk07,Walsh07}, 
but the motion is larger when considering the dense core versus lower density ambient
cluster material \citep[e.g.,][]{Kirk10}, suggesting accretion of material from larger scales
is important. 

It is interesting to note that the results of our clustering analysis of dense cores
in Orion~A so closely mimics the results seen in Orion~B \citepalias{Kirk16b},
as well as the older, small and sparse protostellar groups analyzed in \citetalias{kirk11}.  
Orion is well-known for having one of the higher density and more turbulent
environments forming a wider range of stellar masses, than is present within other
nearby (Gould Belt) clouds.  
Nonetheless, the southern portion of Orion~A already 
shows a much sparser, looser clustering environment than cores near the ISF, but
both parts of Orion~A exhibit signs of mass segregation.  According
to \citet{Stutz16}, the ISF has a much steeper gravitational potential well 
than the material in L1641 further south.  They postulate that a strong magnetic field around the
ISF is helping to drive cluster-forming instabilities in the gas there, a situation which 
they propose will eventually
propogate further south to L1641.  Other regions with smaller gas reservoirs would likely never
reach an ISF-like stage of cluster formation.
This proposed scenario illustrates the importance of testing core clustering behaviour
under different initial conditions: although we find consistent results across the two varied
environments within Orion~A, other molecular clouds could be different still.
Extension of a clustering analysis to more quiescent,
smaller cluster-forming clouds within the Gould Belt will therefore provide an important
window into how large a role environment plays in clustering properties at these very 
early times, versus the role of dynamical evolution during protostellar accretion.

\section{CONCLUSIONS}
The JCMT Gould Belt Survey mapped $\sim6.1$ square degrees of the Orion~A molecular
cloud down to cores of several hundredths of a solar mass, allowing for dense cores to be
identified in a uniform manner across the highly clustered and active Integral Shaped
Filament region in the north, as well as dispersed, more quiescent zones of star formation
further south.  We analyze the clustering properties of the dense cores across the Orion~A
molecular cloud, and, using two independent techniques, find evidence of mass segregation.
Using Minimal Spanning Trees, we find that the highest flux density (likely most massive) dense core
within each cluster tends to be centrally-located.  We also find that regardless of cluster
definition, dense cores with higher flux densities tend to be located in regions of higher core-core
surface density (the $S-\Sigma$ technique).  
Both of these results have also been found in the complementary analysis
of Orion~B by \citetalias{Kirk16b}.  These two directions of analysis suggest that at 
least in some environments, there may be some level of mass segregation imprinted in 
cluster-forming regions prior to star formation.  Furthermore, the clusters of
dense cores identified using MSTs show statistically significant elongations,
which we speculate may be tied to the filamentary nature of star-forming
gas. 

\acknowledgements{
The authors thank their anonymous referee for a constructive and thorough report which
improved this paper.
The authors wish to recognize and acknowledge the very significant cultural role 
and reverence that the summit of Maunakea has always had within the indigenous 
Hawaiian community.  We are most fortunate to have the opportunity to conduct 
observations from this mountain.
JL and HK thank Herzberg Astrophysics at the National Research Council of Canada
for making this project possible through their co-op program.
The authors thank Peter Martin (U. Toronto) for providing computational support
for the \get\ calculations used in this analysis.
The JCMT has historically been operated by the Joint Astronomy Centre on behalf of the 
Science and Technology Facilities Council of the United Kingdom, the National Research 
Council of Canada and the Netherlands Organisation for Scientific Research. Additional 
funds for the construction of SCUBA-2 were provided by the Canada Foundation for 
Innovation. The identification number for the programme under which the SCUBA-2 data 
used in this paper is MJLSG31\footnote{Note that one scan from Orion~B is presently mislabelled
in CADC with this project code.  Observations
taken during science verification across all GBS regions falls under the project code MJLSG22.}.  
The authors thank the JCMT staff for their support of
the GBS team in data collection and reduction efforts.
The Starlink software \citep{Currie14} is supported by 
the East Asian Observatory.  These data were reduced using a development version from 
December 2014 (version 516b455a).
This research used the services of the Canadian Advanced Network for
Astronomy Research (CANFAR) which in turn is supported by CANARIE,
Compute Canada, University of Victoria, the National Research Council of
Canada, and the Canadian Space Agency.
This research used the facilities of the Canadian Astronomy Data Centre operated by the 
National Research Council of Canada with the support of the Canadian Space Agency.
Figures in this paper were creating using the NASA IDL astronomy library
\citep{idlastro} and the Coyote IDL library\footnote{\url{http://www.idlcoyote.com/index.html}}.
}

\facility{JCMT (SCUBA-2)}
\software{Starlink \citep{Currie14}, CUPID\citep{Berry07}, IDL, Getsources\citep{menshch12} }

\appendix
\section{Source-Finding Algorithms: \get\ versus \fell\ }
\label{sec_sourcefinding}

\subsection{Core Catalogues}

Source identification is a major challenge in submillimetre observations of molecular clouds. 
Sources tend to be irregularly shaped and are often embedded in large-scale structure, which 
makes it difficult for source extraction algorithms that attempt to fit sources with Gaussian 
or power-law profiles. Different extraction algorithms manage these issues in their own way. 
The \get\ algorithm decomposes maps onto a variety of spatial scales and identifies 
structures on each of those scales to characterize best cores and filaments.
In this way, small, faint cores can often be detected alongside
brighter cores even in regions of complex emission structures.  The \fell\ algorithm,
on the other hand, identifies peaks based on local gradients.  This has the advantage
of not assuming any specific shape for a core, but the disadvantage of not allowing for
multiple size-scales of structures to contribute emission to a single pixel.  Our implementation
of each of these algorithms is discussed below.

\subsubsection{Getsources}
\label{sec_sourcefinding_gs}
The \get\ algorithm is a multi-wavelength source extraction algorithm designed for the 
analysis of {\it Herschel} 
observations\footnote{\url{http://www.herschel.fr/cea/gouldbelt/en/getsources/}}\citep{menshch12}.
One of the main difficulties with source finding in star-forming regions is that dense cores may 
have contributions to their observed flux density from any number of sources other than the core itself, 
such as filaments, large-scale overdensities, and noise. Images are analyzed across a range of 
spatial scales to separate emission arising from these sources. The algorithm first decomposes 
an image by convolving it with circular Gaussians of a wide range of sizes (${\sim}$~2-100 
pixels) and subtracting one convolution from the next larger convolution. The Gaussian sizes 
used for convolving at different scales are separated by a small factor (${\sim}$~1.03--1.05) 
to ensure high resolution across the range of decomposed images. Each single-scale decomposition 
is then cleaned of non-relevant signals, such as noise, by using an iterative algorithm that 
finds a unique cutoff threshold for significant flux density based on the standard deviation of the 
single-scale image being cleaned. The clean single-scale images are then combined across all 
wavelengths used in the extraction. Once cores are identified, their properties are extracted 
by fitting a two-dimensional, elliptical Gaussian to the core. For more information on the \get\ 
algorithm see \citet{menshch12}.  We list the few source-finding parameters that were 
not set to the explicit default or single recommended value in Table~\ref{tab_gs}.

\tabletypesize{\small}
\begin{deluxetable}{cccl}
\tablecolumns{4}
\tablecaption{Settings for \get\label{tab_gs}}
\tablewidth{0pt}
\tablehead{
\colhead{Phase\tablenotemark{a}} &
\colhead{Parameter} &
\colhead{Value} &
\colhead{Description}
}
\startdata
P & instrument & scuba8 / scuba4 & Designates data source (for 850 / 450~$\mu$m) \\
P+E & beamsize & 14.6\arcsec\ / 9.8\arcsec\ & Effective beamsize at each wavelength \citep[values from][]{dempsey13} \\
P & mapmaker & other & Specify non-{\it Herschel} software used to create maps \\
P & pixel & 3\arcsec & Pixel size of input map \\
P & rotangle & 0 & Rotation of maps east from north \\
P & cutfact & 0.9 & Used in extraction mask creation; 0.9 is a recommended value \\
\enddata
\tablenotetext{a}{There are two phases in \get\ source-finding: Preparation (P) and Extraction (E)}

\end{deluxetable}

The \get\ catalogue we produced was vetted to remove potentially spurious sources.
We first removed 
sources that had no significant detection at 850~\micron, i.e., sources 
identified by \get\ as having SIG\_MONO\_850 $<7$, which roughly corresponds to a minimum signal
to noise ratio of 7 at 850~\micron, were removed (this threshold 
is the recommended threshold to
distinguish between reliable and tentative sources).  Cores were not 
required to have similar significant detection at 450~\micron. Observations at 450~\microns 
typically have a lower S/N ratio than observations at 850~\micron, so cores that may appear 
distinct and real at 850~\microns may not always have detections at 450~\micron. 
We tested several schemes to eliminate further spurious or suspect sources, and
found that the method described below was the most visually successful in retaining real
sources while eliminating suspect ones.  The approach we adopted was to smooth 
the 850~\microns map using a 5 pixel boxcar kernel and examine
the flux densities of each source.  Peaks were 
excluded from the final catalogue if they fell below 2.1 times the locally averaged (i.e., 
within the extraction ellipse defined by \get) RMS noise per pixel. This 
criterion removed spurious sources 
and ensured that dense cores detected in noisy regions were in fact significant sources.
(We adopt a similar check for the \fell\ catalogue discussed below.) Finally, 
sources which had peak or total flux density errors greater than the respective measurement, which are 
flagged as undetectable by \get, were also removed. 
This vetting reduced the initial \get\ `reliable' catalogue of 1178 sources to the final 
robust core catalogue of 919 sources.

\subsubsection{FellWalker}
\label{sec_sourcefinding_fw}
\fell\ \citep{berry15} is a source extraction algorithm developed for the Starlink CUPID software 
package \citep{Berry07}\footnote{Available at \url{http://starlink.eao.hawaii.edu/starlink/CUPID}}. 
It works analogously to a hiker attempting to find their way to the top of a peak by following 
the route of steepest ascent. The algorithm starts a `walk' from each pixel in an image and 
traces its way to a local maximum by following the steepest increasing flux density gradient. Once the 
program reaches a local maximum it scans a predefined area around the maximum to see if a 
higher-valued pixel can be found, and if it finds one, it jumps to that pixel and continues 
its walk towards the next local maximum. In this manner, \fell\ will eventually find a 
significant local maximum and this point is labeled as a peak. The pixels of each walk 
that end at the same peak are attributed to one core. To limit the physical extent 
of the sources to realistic sizes, parts of a walk that occur along a gradient that is shallower 
than a certain threshold are not included. There are other input parameters that mitigate the 
effects of background noise and sources with excessive substructure, as well as determining how 
significant a peak must be for it to be considered a core.  
Table~\ref{tab_fw} summarizes all of the \fell\ parameters that we set to non-default
values along with a short description of their purpose.
More information about \fell\ and the parameters that can be set is available 
in \citet{berry13}.

\begin{deluxetable}{ccl}
\tablecolumns{3}
\tabletypesize{\footnotesize}
\tablecaption{Settings for \fell\label{tab_fw}}
\tablewidth{0pt}
\tablehead{
\colhead{Parameter} &
\colhead{Value} &
\colhead{Description}
}
\startdata
rms & 0.469 & rms noise per pixel in the image (in mJy/pixel) \\
FellWalker.AllowEdge & 0  & eliminate cores touching map edge \\
FellWalker.CleanIter & 4  & smooth jagged clump edges \\
FellWalker.FlatSlope& 0.05 & increase minimum gradient for pixels to be assigned to a core\\
FellWalker.FwhmBeam & 3 & beamsize in pixels (underestimated to help include small cores)\\
FellWalker.MaxBad & 0 & eliminate cores lying beside bad pixels\\
FellWalker.MinDip & 3*RMS & increase minimum dip required between neighbouring cores\\
FellWalker.MinHeight & 2*RMS & minimum value for peak of core \\
FellWalker.MinPix & 20 & minimum number of pixels for a core \\
FellWalker.Noise & 1*RMS & mininum flux in a pixel assigned to a core \\
\enddata

\end{deluxetable}

In our initial core identification, we experimented with a variety of peak flux density thresholds and 
minimum size requirements.  The final settings were selected to include the maximum 
number of significant cores in the catalogue. \fell\ uses a single global noise value
to determine whether or not sources are real, which we set to the global average RMS noise
per pixel, i.e., 0.0042~Jy~arcsec$^{-2}$ (0.469~mJy~pixel$^{-1}$).  The minimum height
for a core was set to two times this RMS noise level.  Since the noise level
in the mosaic is not constant (in particular, it is higher near the map edges), 
however, a number
of spurious and insignificant sources were also included in the initial catalogue. 
The two main types of spurious sources that we aimed to remove were those present exclusively 
due to 
noise spikes near the edges of the map as well as small, faint sources that were most likely 
artifacts of background fluctuations. 
We eliminated these spurious sources by filtering the initial core list using an
additional signal-to-noise requirement.
We boxcar smoothed the original image using a 5 pixel wide window
and removed sources whose peaks fell below 2 times the locally averaged
RMS noise per pixel. 
The smoothing and RMS cutoff values were found to be the most effective combination after
visually inspecting the results using a wide range of values.  
This process reduced the initial list of 1110 sources down to 773 
robust core detections.

\subsubsection{Comparison of \get\ and \fell\ Catalogues}
\label{sec_compare_gs_fell}
Unlike \get, where a given map pixel may contribute flux density to multiple cores, \fell\
does not allow objects to overlap.  As such, small and faint cores located beside
large and bright cores have a greater chance of being missed, with their flux density
being attributed to the larger core.  This difference explains in part why we find 
fewer cores in the \fell\ catalogue compared to the \get\ extraction, 
despite similar post-extraction cuts being applied.
Additionally, since \get\ uses both 850~$\mu$m and 450~$\mu$m data, the higher
resolution of the latter map can help to de-blend cores which do not appear separate
using the 850~$\mu$m data alone.

\begin{figure}[htb]
\begin{tabular}{cc}
\includegraphics[width=3in]{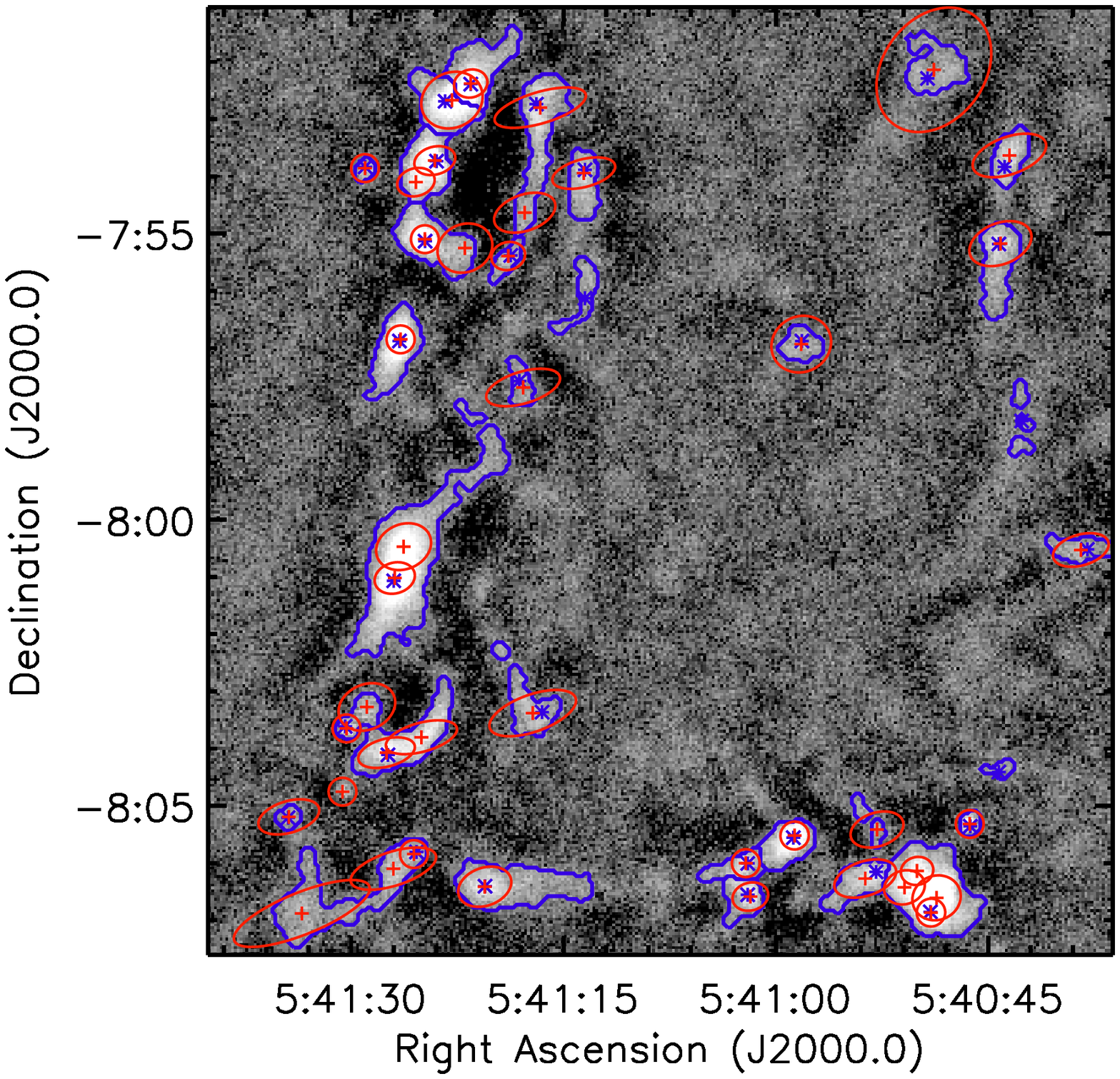} &
\includegraphics[width=3in]{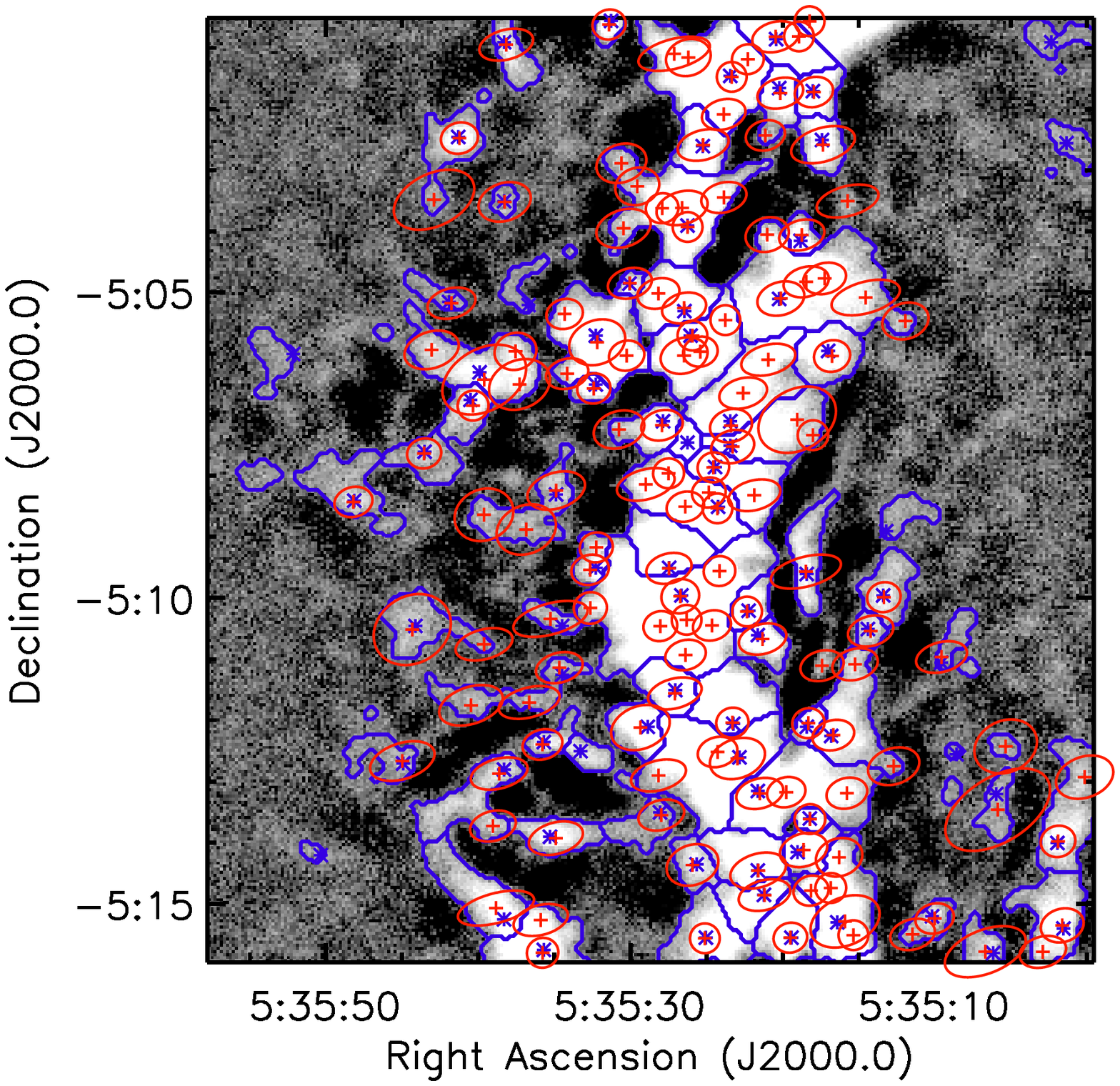} \\
\end{tabular}
\caption{A comparison of \get- and \fell- based core identifications within two small portions
	of Orion~A.  The left panel shows a less crowded field in the southern portion of Orion~A,
	while the right panel shows a field just above the central portion of the ISF.
	In both panels, the greyscale image shows the 850~$\mu$m flux density.  The red plus signs
	and larger ellipses show the \get-based core peak positions and best-fit Gaussian
	FWHM contour.  The blue asterisks and contours show the \fell-based core peak positions
	and full extents.}
\label{fig_compare_gs_fw}
\end{figure}

A direct comparison of some of the dense cores identified using both \get\ and \fell\ is
shown in Figure~\ref{fig_compare_gs_fw}.  This figure highlights two different parts of 
Orion~A: a less crowded field in the south, and a more crowded
region near the ISF.  It is apparent from Figure~\ref{fig_compare_gs_fw} 
that many of the cores have a
reasonable correspondence, with similar peak positions.
Although the sizes of the core contours are not directly comparable (since \get-based
cores would extend beyond their FWHM plotted), it is clear that \get\ typically 
identifies the most peaked part of a structure, while \fell\ allows a much larger,
often irregular, extent of fainter emission to be included.
As discussed above, \get\ also 
has a greater tendency to break up complex emission structures into more pieces.  
An example of this
is the elongated structure at about (5:41:27, -8:00:00) in the left panel of 
Figure~\ref{fig_compare_gs_fw}, where \get\ identified two cores and \fell\ identified
only one.  For the faintest emission structures, there is greater variation between the
two catalogues.  Factors such as the precise core boundaries have a larger impact
on whether the inferred local signal-to-noise level is sufficient for the structure
to be retained in the final catalogue.
An example of a core retained only in the \fell\ catalogue can be seen in the left 
panel of Figure~\ref{fig_compare_gs_fw} at about (5:40:45, -8:04:00).  

Quantitatively comparing the two catalogues, we
find that 902 of 919, or 98\% of \get-based cores, lie within a 
\fell-based core boundary, and 
581 of 773, or 75\% of \fell-based cores, contain at least one \get-based cores.  
These numbers reinforce the visual impression from the right hand panel of 
Figure~\ref{fig_compare_gs_fw} that the majority of the variation between the two
catalogues is caused by how emission is divided into cores, rather than whether
emission is identified as belonging to any core.

\subsubsection{Comparison with Mairs et al.}
\label{sec_compare_mairs}

\begin{figure}[htb]
\begin{tabular}{cc}
\includegraphics[width=3in]{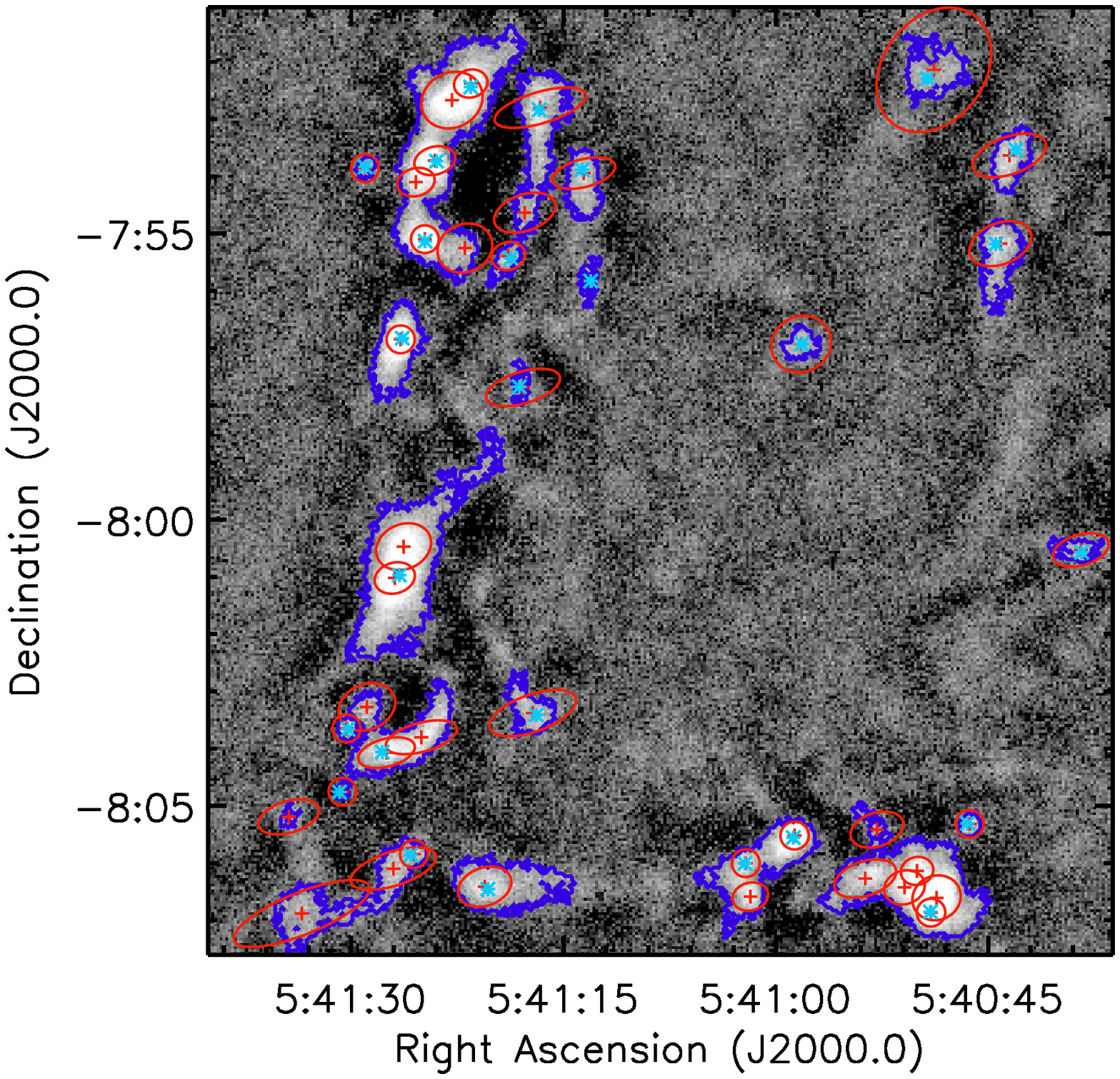} &
\includegraphics[width=3in]{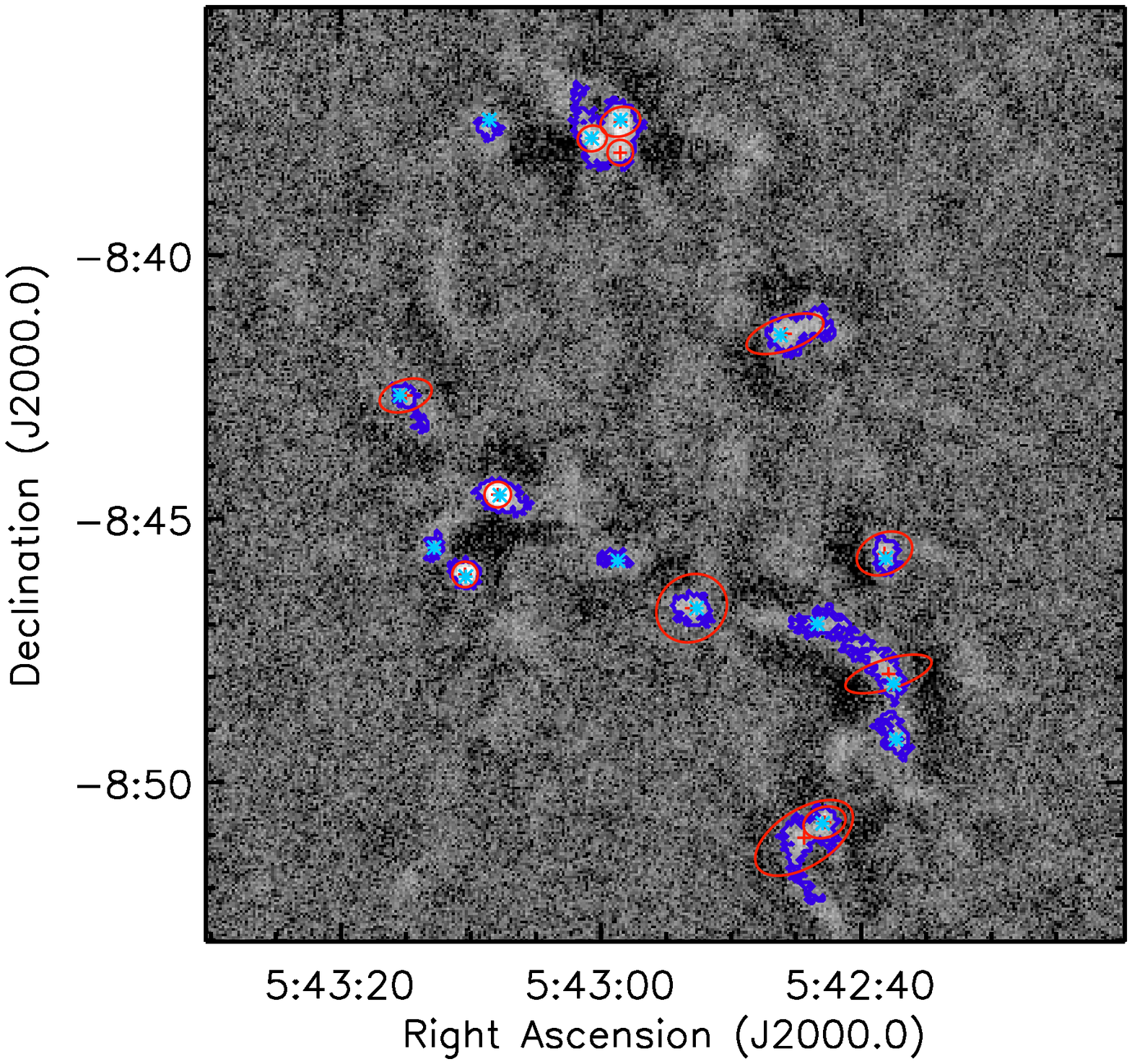} \\
\end{tabular}
\caption{A comparison of \get-based core identification with the islands and fragments from
	\citet{Mairs16} within two small portions
	of Orion~A.  The left panel shows the same field as the left panel of 
	Figure~\ref{fig_compare_gs_fw}, while the right panel shows an even sparser field in the
	very south of Orion~A.
	In both panels, the greyscale image shows the 850~$\mu$m flux density.  The red plus signs
	and larger ellipses show the \get-based core peak positions and best-fit Gaussian
	FWHM contour.  The medium blue asterisks show the peak positions of fragments
	\citep[small scale structures from][]{Mairs16}, while the darker blue contours
	show the full extents of islands \citep[larger scale structures from][]{Mairs16}.}
\label{fig_compare_gs_steve}
\end{figure}

We furthermore compared our \get-based core catalogue with the structures identified in 
\citet{Mairs16} who analyzed a similar SCUBA-2 map of Orion~A, although their
analysis only extended as far north as the bottom of the ISF.
\citet{Mairs16} created two independent lists of structures within southern
Orion~A, large-scale islands and smaller-scale fragments which typically lie within
an island.  The large-scale islands were identified using {\it clumpfind}
\citep{Williams94}, while the smaller-scale fragments were identified using 
a modified implementation of \fell.
We compare our \get-based cores across the same area in southern
Orion~A with both the island and fragment catalogues of \citet{Mairs16} in 
Figure~\ref{fig_compare_gs_steve}.
The agreement between the catalogues is reasonably good.  The left panel
of Figure~\ref{fig_compare_gs_steve} appears to show a better correlation between the
\citet{Mairs16} structures and our \get\ catalogue than the similar comparison of \get\ and \fell\
catalogues in Figure~\ref{fig_compare_gs_steve}, but this trend is less obvious across the
entire area mapped.
For the larger-scale islands, 485 of 518, or 94\% of the \get-based cores in the southern
portion of Orion~A, lie within an island.  Conversely, 244 of 364, or 67\% of the islands,
contain one or more \get-based cores.  The 850~$\mu$m map that \citet{Mairs16} analyzed
had better sensitivity to larger-scale emission structures than the map we used, 
and we find that the \citet{Mairs16} islands which have no match in our \get-based
catalogue tend to be faint diffuse structures in our map with no peaks bright enough
for \get\ to pick up.

Moving to the smaller-scale fragment catalogue of \citet{Mairs16},
we compared the peak positions of the fragments and \get-based cores.  Full coverage
maps were not available for either the \citet{Mairs16} fragments or for our \get\ 
catalogue (in the latter case since each pixel can contribute flux density to multiple cores).
Therefore, we search for evidence of positional coincidence between the peak locations
in each catalogue.  We find that 358 of 436, or 82\% of the 
fragments, lie within one \get-based FWHM\footnote{We used the geometric mean of the
major and minor FWHM values returned by \get.}, and very few additional cores lie
within twice this value.  

\begin{figure}[htb]
\begin{tabular}{cc}
\includegraphics[width=3in]{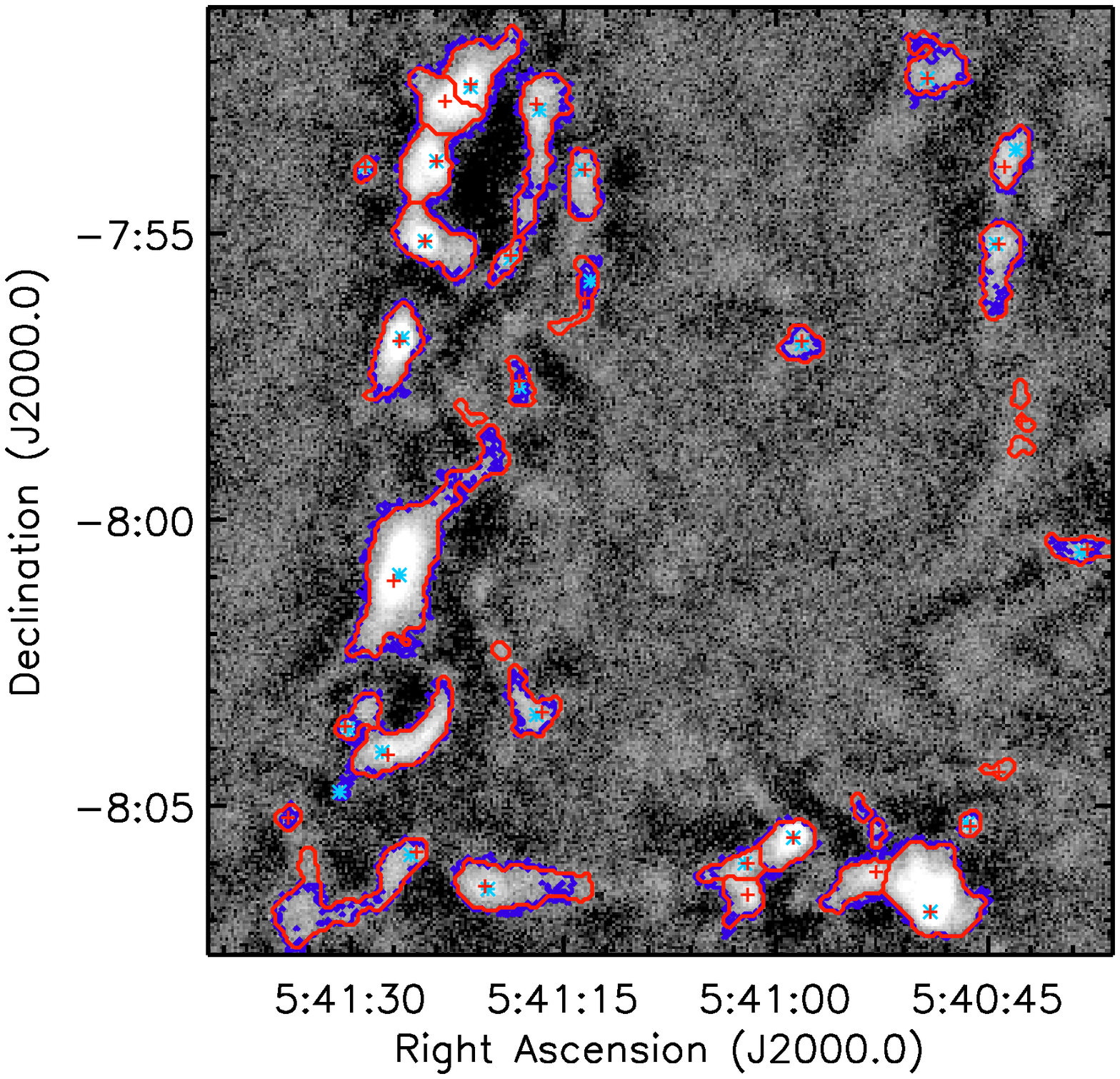} &
\includegraphics[width=3in]{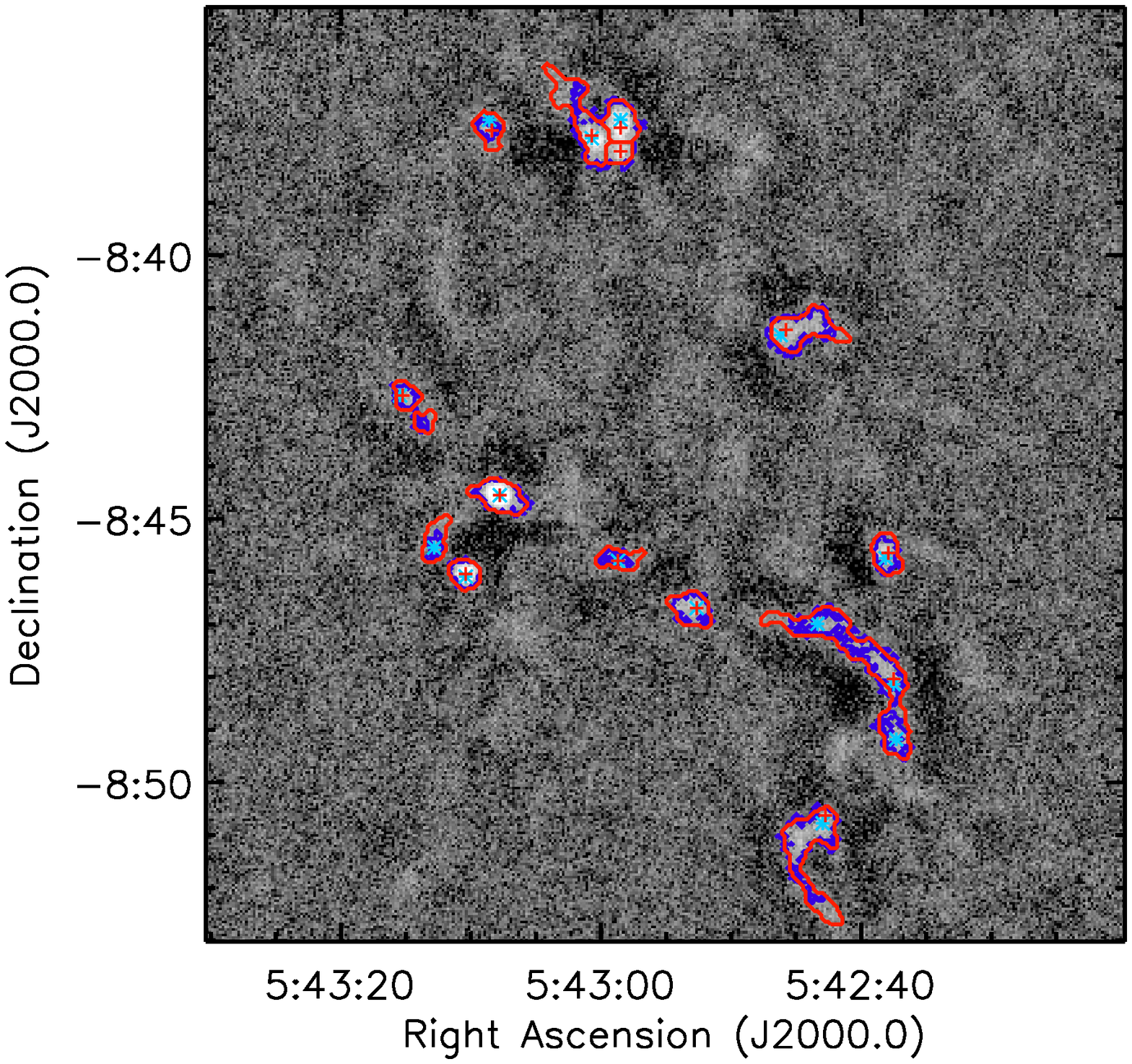} \\
\end{tabular}
\caption{ 
	A comparison of \fell-based core identification with the islands and fragments from
	\citet{Mairs16} within two small portions
	of Orion~A.  The left panel shows the same field as the left panel of 
	Figure~\ref{fig_compare_gs_fw}, while the right panel shows an even sparser field in the
	very south of Orion~A.
	In both panels, the greyscale image shows the 850~$\mu$m flux density.  The red plus signs
	and red contours show the \fell-based core peak positions and boundaries.
	The medium blue asterisks show the peak positions of fragments
	\citep[small scale structures from][]{Mairs16}, while the darker blue contours
	show the full extents of islands \citep[larger scale structures from][]{Mairs16}.}
\label{fig_compare_fw_steve}
\end{figure}

If we instead compare the \citet{Mairs16} catalogues to our \fell-based catalogue,
as shown in Figure~\ref{fig_compare_fw_steve}, 
we find a similar level of agreement.  For the \citet{Mairs16} islands, 408 of 514, or 79\%
of our \fell-based cores, lie within an island, while 290 of 364, or 80\%, of the
\citet{Mairs16} islands, contain at least one \fell-based core.  Meanwhile, all of
the \citet{Mairs16} fragments lie within our \fell-based core boundaries.
\citet{Mairs16} applied \fell\ in a different configuration than our work to identify
their fragments, so it is at first surprising that the agreement with our \fell\ catalogue
is not even better.  A careful visual inspection of the two catalogues and maps used
for identification shows that most of the fragments (and islands) with no correspondence
in our \fell\ (or \get) catalogues are indeed faint extended structures for which
the map we analyze has less sensitivity.  For the \citet{Mairs16} sources that we do
detect with \fell, the agreement is very good.

\subsubsection{Comparison with Salji et al}
\label{sec_compare_salji}
\begin{figure}[htb]
\includegraphics[width=3in]{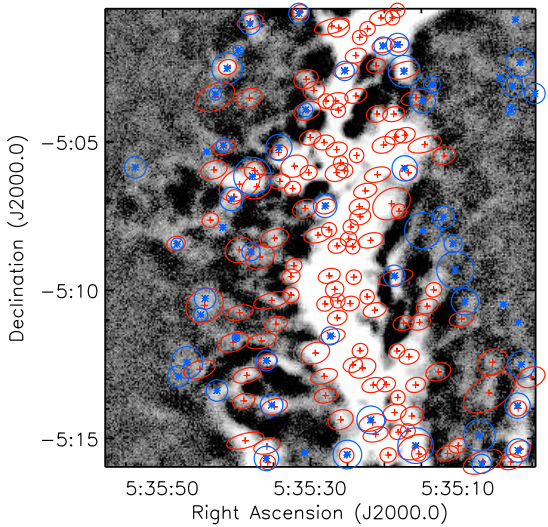}
\caption{A comparison of our \get-based core catalogue with the starless and prestellar 
	cores identified in \citet{salji15a}.  The area displayed is the same as
	the right panel of Figure~\ref{fig_compare_gs_fw}.  The greyscale image
	shows the 850~$\mu$m flux density.  The red plus signs and larger ellipses show the
	\get-based core peak positions and best-fit Gaussian FWHM contour.  The blue
	asterisks and larger circles show the peak positions and sizes of the starless
	and prestellar cores from \citet{salji15a}.
	} 
\label{fig_compare_gs_carl}
\end{figure}

Finally, we compare our \get-based core catalogue with the 432 starless and prestellar cores 
identified in the area around the ISF by \citet{salji15a} using a new core identification
algorithm first presented in that work.
Figure~\ref{fig_compare_gs_carl}
shows a comparison of the two catalogues in the same field as is shown in the right panel
of Figure~\ref{fig_compare_gs_fw}.  The correspondence of cores here is much poorer, with
only 161 of the 432 Salji et al. cores lying within the FWHM extent of a \get-based core.  There
are a variety of factors contributing to this poor match.  The first is related to the
constraints Salji et al. placed on the cores they identified.  Namely, their algorithm was tuned to
explicitly detect only nearly circular cores with sizes ranging between about 0.03~pc
and 0.1~pc \citep[elongated structures were instead analyzed as filaments in][]{salji15b}.
Furthermore, the 48 cores associated with a protostar were removed from Salji et al.'s analysis
and their properties (including location) were not reported.  It seems likely that these 
factors (size and shape constraints and protostar elimination) drive the
poor correspondence between cores along the central ISF spine.  Furthermore, as noted in 
Section~\ref{sec_obs},
\citet{salji15a} analyzed a limited portion of the full dataset across the northern portion
of Orion~A, using an earlier data reduction process which was less sensitive to large-scale
structure.  The combination of these two effects drives a poorer correspondence between
cores identified in the outskirts of the central ISF spine where faint sources are more
prevalent.

\subsection{MST-based Cluster Analysis}
\label{sec_fell_mst}
Recognizing that the differences in source identification could impact the results of our
cluster analysis, we re-ran all of our analysis using the \fell-based catalogue.
Here, we discuss the impact on the MST-based analysis, namely the distribution of offset
ratios.  We follow the same procedure as outlined in Section~\ref{sec_mst} to use an
MST to identify clusters, and find a best-fit value of \Lcrit\ of 0.33~pc 
(with a full range of acceptable fits of 0.31~pc to 0.37~pc).
\fell\ tends to split up
regions of complex emission structure like the ISF into fewer cores than \get, which
resulted in ISF cores having relatively less influence on the distribution of branch
lengths in the region.  Given this difference, a single global fit to the entire Orion~A
distribution of branch lengths provided visually reasonable MST-based clusters across the 
entire field, unlike in our \get-based analysis.   

\begin{figure}[htb]
\includegraphics[width=3.5in]{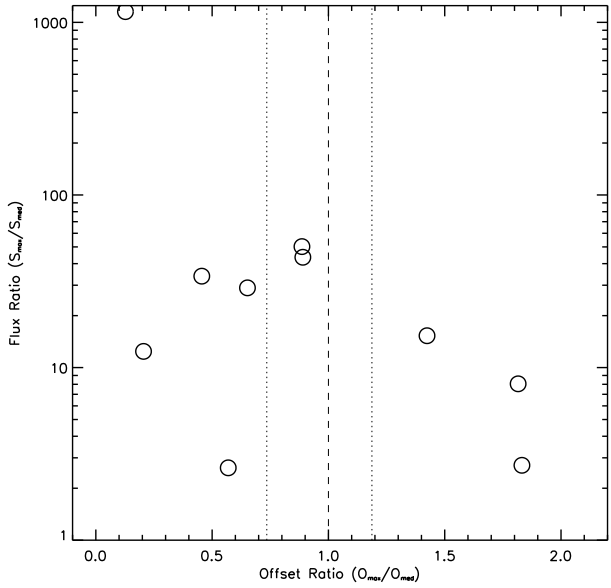}
\caption{Comparison of the offset ratios derived for \fell-based clusters using a membership
	criterion of $N > 10$
	As with the original \get-based analysis, 
	most of the clusters have offset ratios less than one.}
\label{fig_offs_fell}
\end{figure}

Using the standard membership criterion of $N > 10$, we identify ten clusters across 
Orion~A.  Most of these clusters (8 of 10) have a very good correspondence to the 
\get-based clusters.  
In general, the \fell-based clusters have fewer members, due to clustered emission being
divided into fewer cores in \fell. Similar to the \get-based clusters,
most of the \fell-based clusters have offset ratios less than one, with only 3 of 10 having
offset ratios greater than one, as illustrated in Figure~\ref{fig_offs_fell}.

\subsection{$S-\Sigma$ Analysis}
\label{sec_fell_m_sigma}
We next examine the relationship between total flux density and core-core surface density for the
\fell-based cores and compare with the results seen in Section~\ref{sec_m_sigma} for the \get-based
core catalogue.  Figure~\ref{fig_m_sig} shows the $S-\Sigma$ plot for the \fell-based
catalogue, using the ten nearest neighbours to estimate the core-core surface density (NN10) 
as was done for the \get-based analysis shown in Figure~\ref{fig_m_sig}.  
Clearly the same trend for higher total flux densities with increased surface density of
neighbours is seen.  As with the \get-based analysis, the two-sample KS test gives an
extremely low probability that the higher and lower flux density starless or protostellar
cores have a similar distribution of surface densities 
in either the ISF or southern Orion~A, for NN5 or NN10 and either
of the high versus low splits tested in Section~\ref{sec_m_sigma}.  
Similarly, the Mann-Whitney test
shows a strong probability of higher flux density cores inhabiting higher core-core surface density
environments.  Therefore with this metric too, the choice of core identification
method has no effect on our conclusions.

\begin{figure}[htb]
\begin{tabular}{cc}
\includegraphics[width=3.2in]{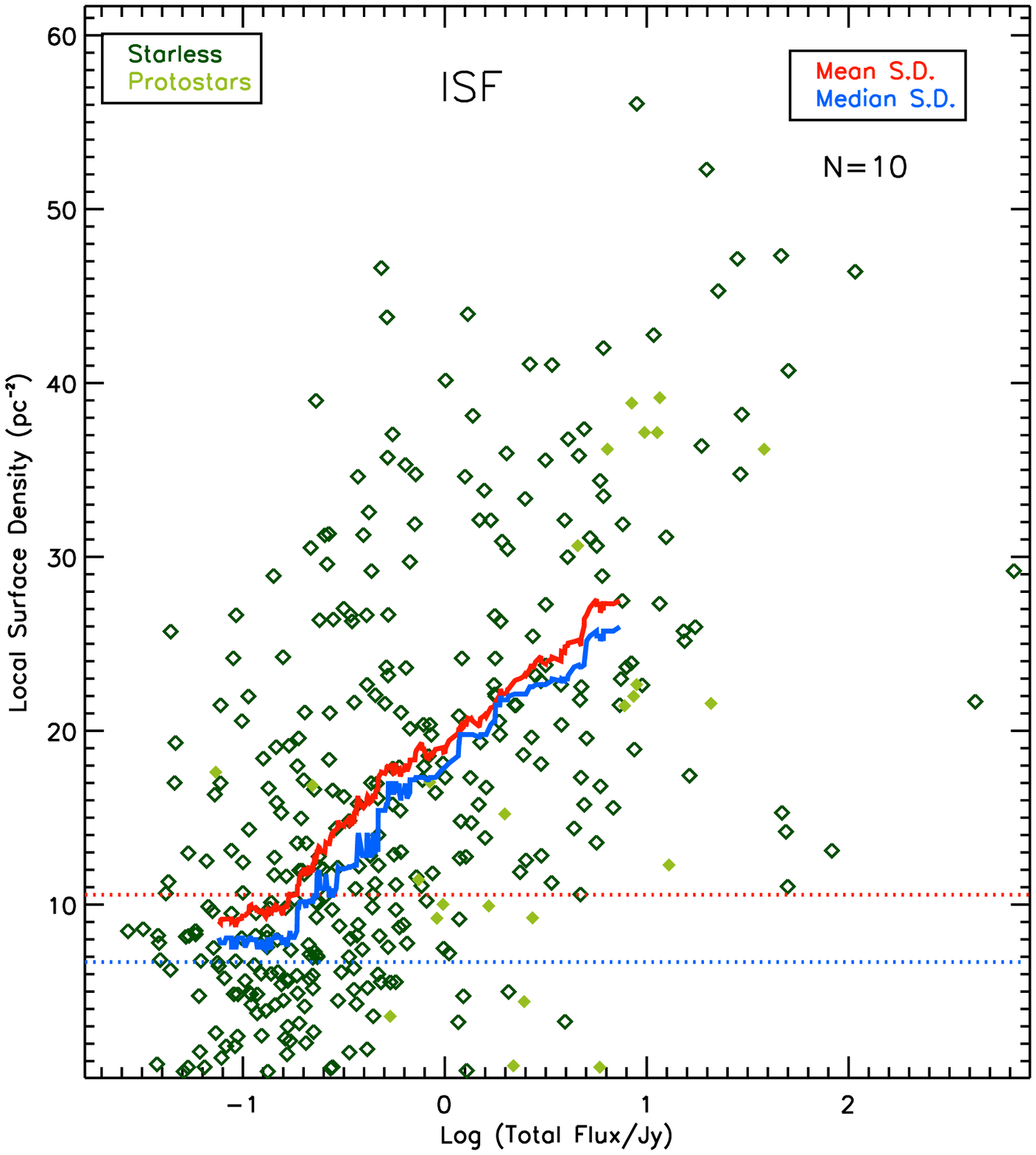} & 
\includegraphics[width=3.2in]{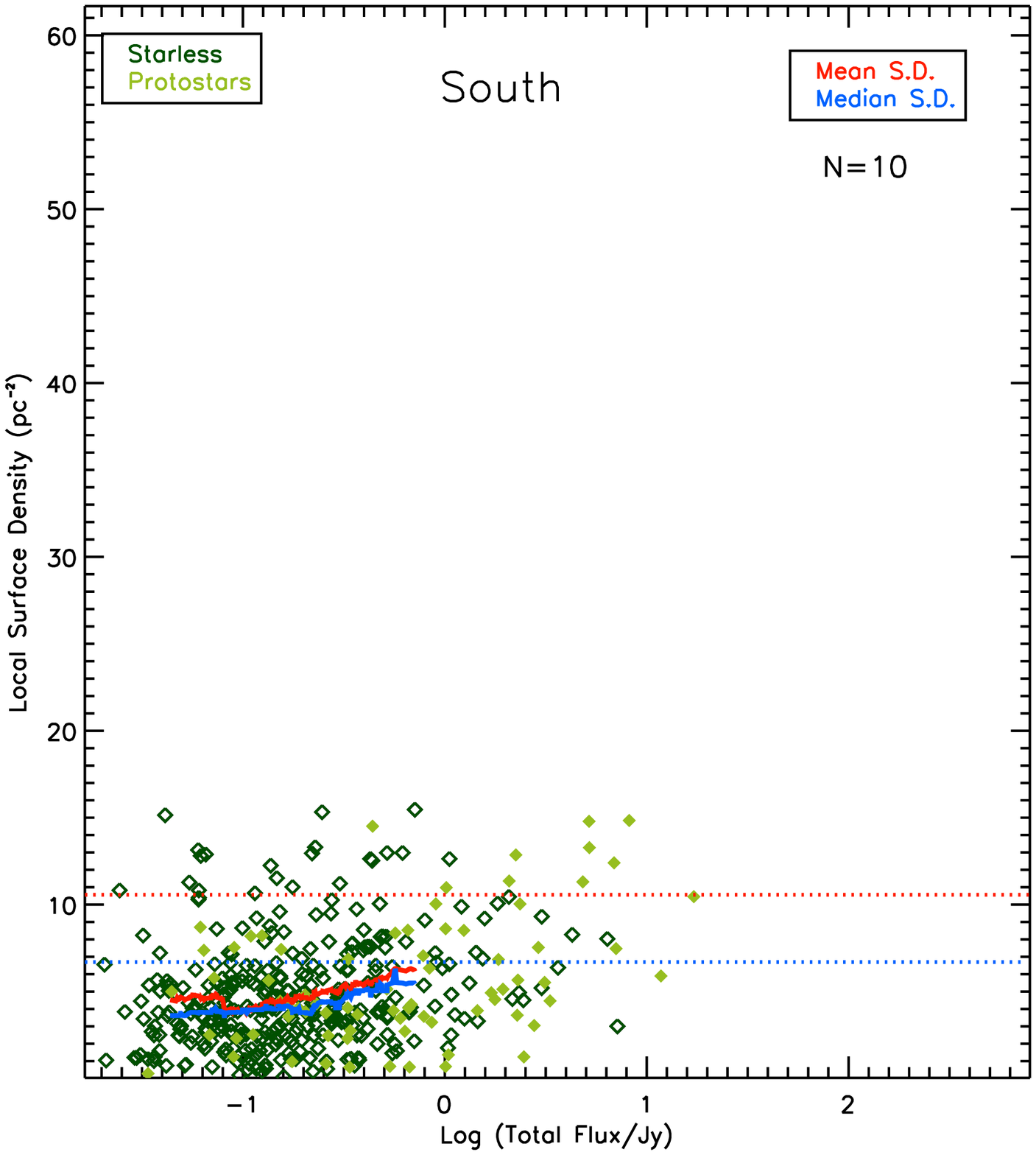} \\
\end{tabular}
\caption{Local core-core surface density versus total core flux density for the 
	dense cores identified
	using \fell.  Similar to the \get-based analysis shown in Figure~\ref{fig_m_sig},
	there is a strong positive trend between total flux density and core-core surface density
	for both the ISF and southern Orion~A.
	See Figure~\ref{fig_m_sig} for the plotting conventions used.}
\label{fig_m_sig_FW}
\end{figure}

\section{Uncertainties in \get-based MST Clusters}
\label{sec_mst_uncert}
As also discussed in \citetalias{Kirk16b}, use of the MST technique requires several user-defined
parameters which have the potential to influence the resulting clusters, and hence,
the analysis.
We examine the influence of varying \Lcrit\ and $N$, the minimum number of cluster members,
in turn.  We find that although the details of the clusters identified vary with changes
to these parameters, the resulting offset ratios show little variation, and hence our
conclusions are robust.

\subsection{Variations in \Lcrit\ }
\label{app_Lcrit}
\begin{figure}[htbp]
\plottwo{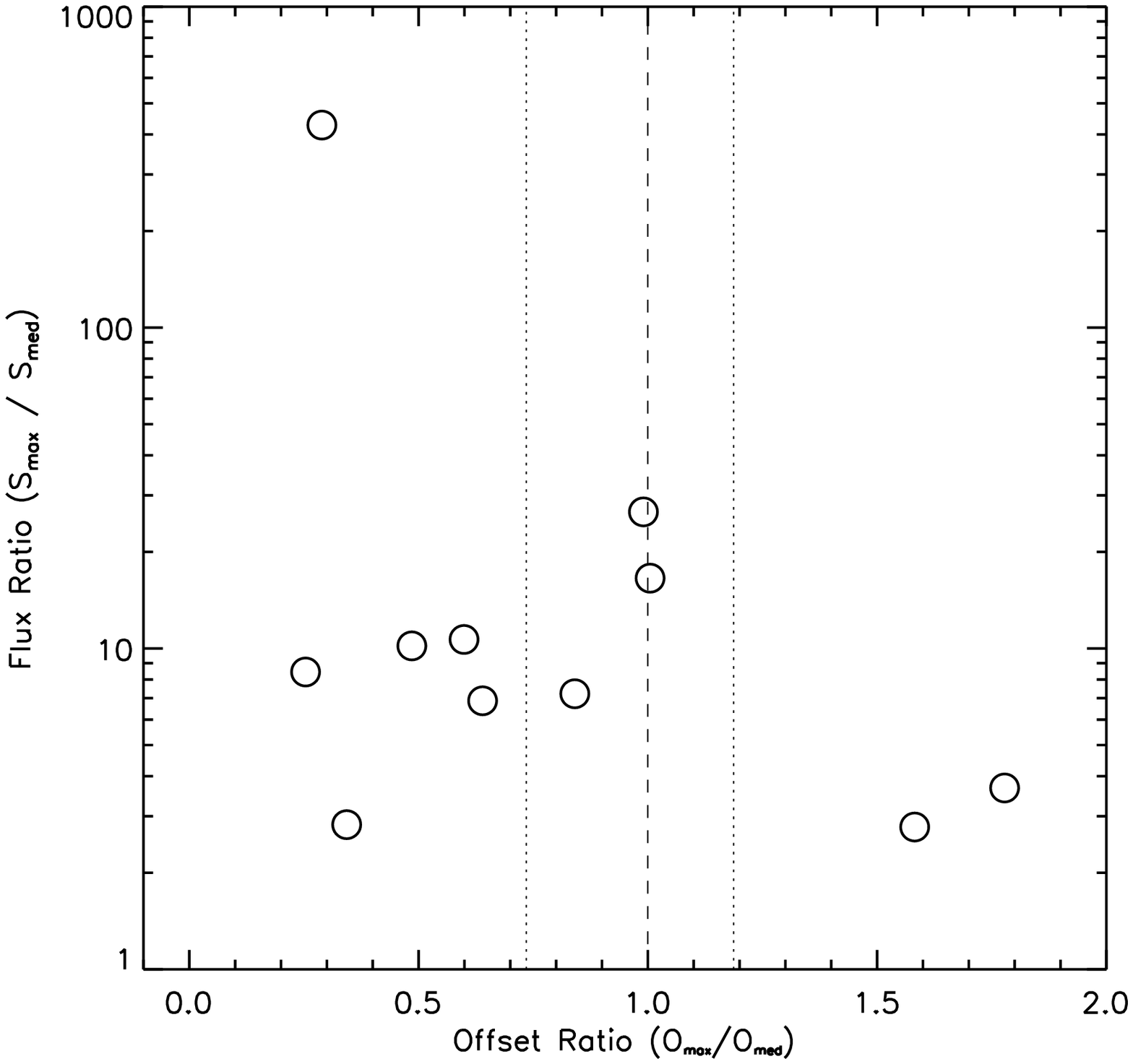}{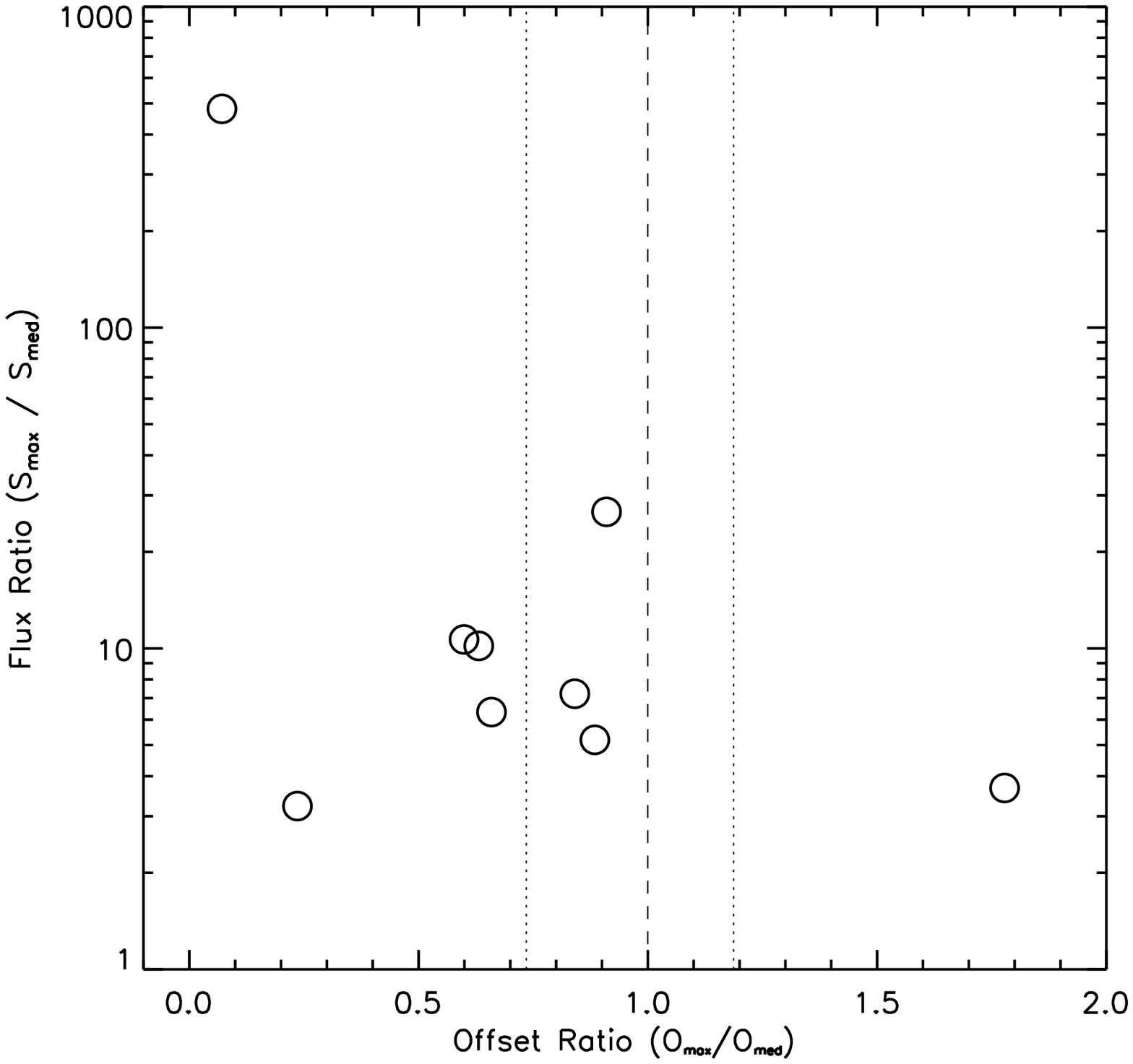}
\caption{Comparison of offset ratios measured for \get-based clusters using the minimum
(left) and maximum (right) values of \Lcrit, 0.32~pc and 0.40~pc 
respectively.  As with
the original analysis, the majority of clusters have offset ratios smaller than one.}
\label{fig_getsources_lcrit_var}
\end{figure}

Here, we examine the effect of the uncertainty in \Lcrit\ on the resulting clusters and
analysis.  Our range of uncertainty in \Lcrit\ based on the cumulative branch length 
distribution measured in the southern part of Orion~A is 0.32~pc to 0.40~pc. 
We re-ran our
analysis using both of these values.  Qualitatively, the clusters appear similar over the
range of \Lcrit\ values tested.  Adopting \Lcrit\ = 0.32~pc, 
only clusters 1, 3, and 5 show
any changes in their membership, and these are restricted to small percentages of members
located at the cluster peripheries.  The changes are slightly larger for \Lcrit\ = 0.40~pc, 
where this increase merges several formerly distinct clusters (2, 6, and 9) into the 
large ISF cluster 1, and a new cluster which barely meets the relaxed criteria is identified.
The remaining clusters (3, 4, 5, 7, 8, and 10), however, still have identical membership.
We ran our offset ratio analysis on the clusters identified using the minimum and maximum
\Lcrit\ found in the south, and show these results in Figure~\ref{fig_getsources_lcrit_var}.
As can be clearly seen in the figure, most clusters have offset ratios less than
one under both scenarios.  

\begin{figure}[htbp]
\includegraphics[width=3in]{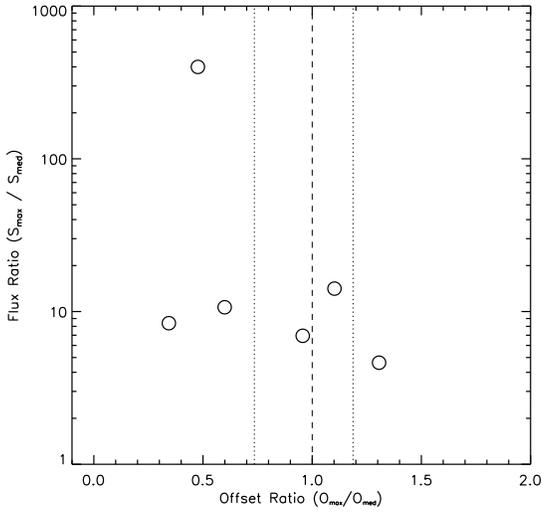}
\caption{A comparison of offset ratios measured for \get-based clusters using the \Lcrit\
value derived for the entire Orion~A core population of 0.22~pc. 
Again, the majority of
clusters have offset ratios less than one.} 
\label{fig_getsources_lcrit_fullcloud}
\end{figure}

We also consider a more extreme variation in \Lcrit.  As discussed in Section~\ref{sec_mst_intro}, 
our initial MST analysis of the entire Orion~A dense core population indicated a smaller
\Lcrit\ value of 0.22~pc, 
which was strongly dominated by cores around the ISF.  Using
this \Lcrit\ value instead noticeably shrinks the membership of clusters 1, 2, 6, 7, and 9,
and reduces clusters 4, 5, 8, 10, and 11 to below the minimum number of cluster members 
required, while only cluster 3 remains identical.  Even with these extreme changes to
the cluster membership, we find the resulting offset ratios tend to be less than 1,
as shown in Figure~\ref{fig_getsources_lcrit_fullcloud}.

It is encouraging to note that a large range of \Lcrit\ values, including the full-cluster
value that excludes most of the clusters in the southern part of Orion~A from the analysis,
provide similar and consistent results.  We conclude that our overall results are therefore
robust to reasonable permutations in cluster definition.

\subsection{Variations in $N$}
\label{app_Nmin}
\begin{figure}[htb]
\plottwo{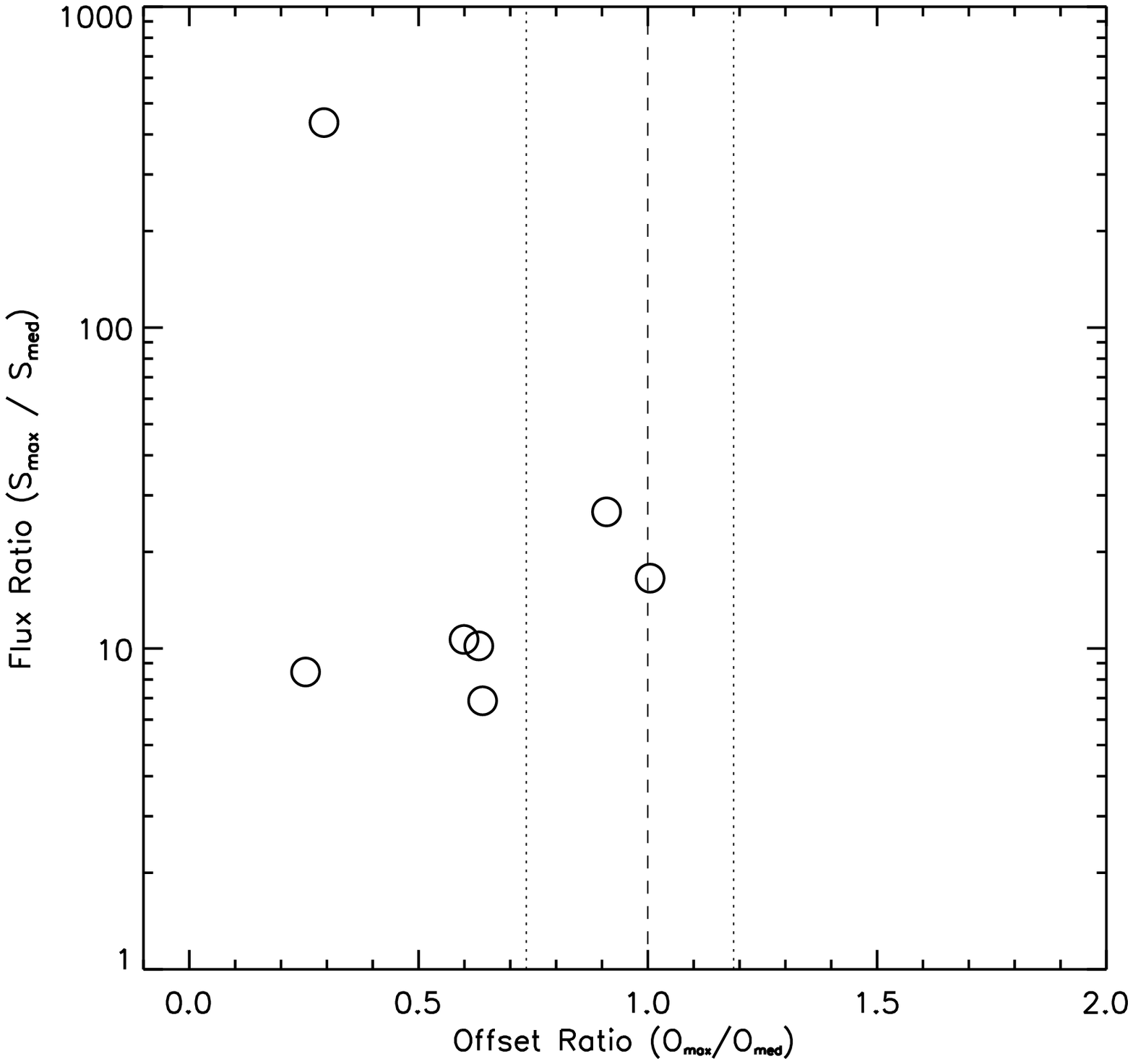}{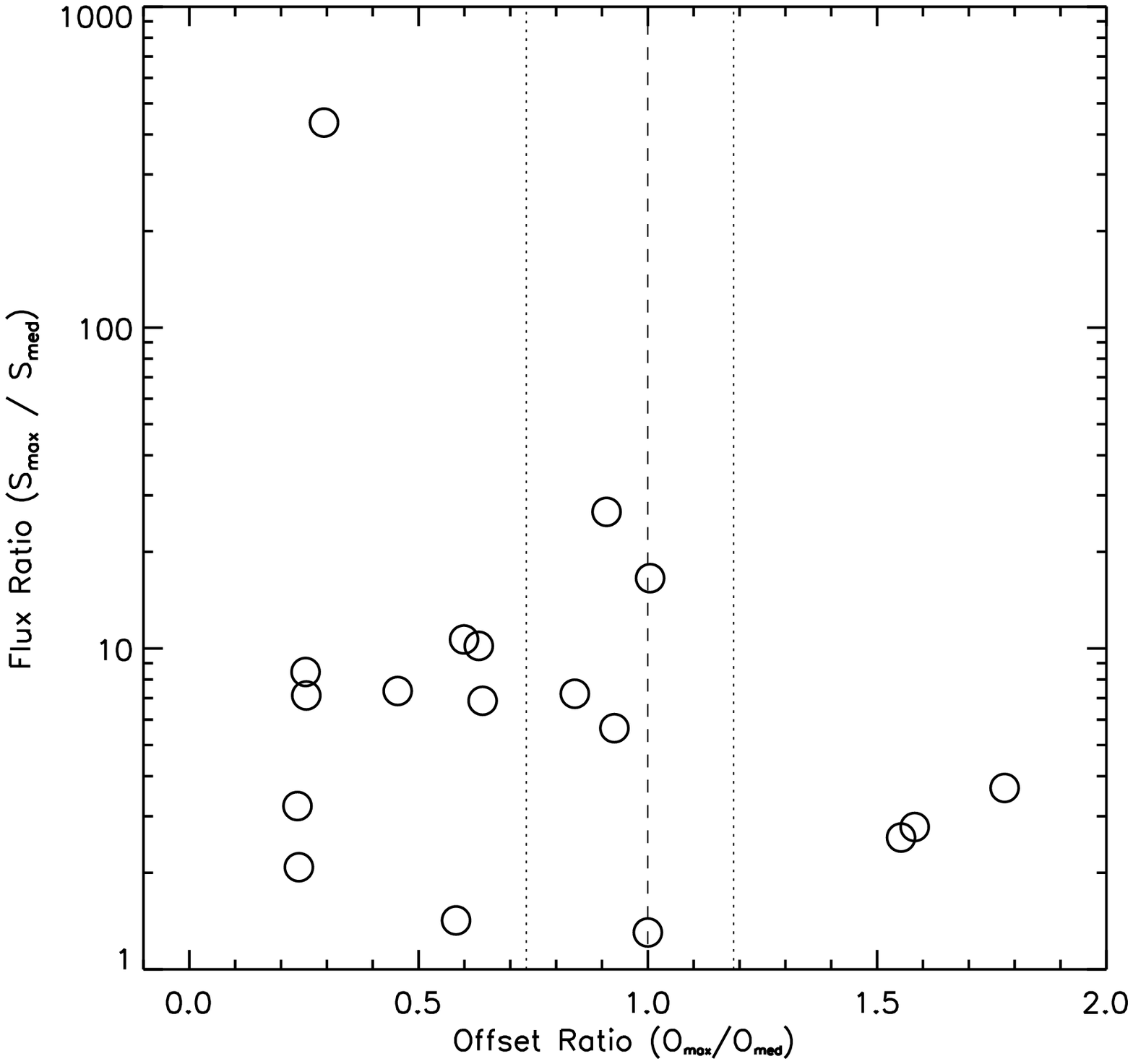}
\caption{Comparison of offset ratios for \get-based clusters using a membership criterion of
	$N > 15$ (left) and $N > 5$ (right).  As with the original analysis, the majority of
	clusters have offset ratios less than one.}
\label{fig_offs_nmin}
\end{figure}

Next, we consider the effect of adopting a different requirement for the minimum number
of cluster members.  Our main analysis requires $N > 10$, but here we present results for
$N > 15$ and $N > 5$.  The $N > 15$ clusters are a subset of the original $N > 10$ clusters,
with four of the eleven initially-identified clusters excluded with this larger membership
criterion.  The $N > 15$ set of clusters show an even stronger tendency than our full
$N > 10$ sample for offset ratios below one, with six of the seven having ratios much
less than one, as can be seen in the left panel of Figure~\ref{fig_offs_nmin}.
Extending the MST-based clusters to those with $N > 5$ members adds a further seven
`clusters' to the original $N > 10$ sample.  We note that some of these additional clusters 
are extremely small, with three containing only six members.  At such a small-N limit,
measuring the offset ratio becomes more uncertain.  The cluster centre, for example,
is much more prone to variations with the inclusion or exclusion of cluster members.
Despite this increased uncertainty, 
most of the additional clusters added when extending the sample
to $N > 5$ do still have offset ratios below one, as shown in the right panel of 
Figure~\ref{fig_offs_nmin}.  We therefore conclude that the offset ratios
measured for the \get-based clusters are robust to variations within the user-defined
parameters of the MST method.

\section{Effect of Using {\it Herschel} Temperatures for Mass Estimates}
\label{app_temp}
We also ran our analysis on the dense core {\it masses}, as estimated using eqn 1 with
dust temperatures measured by \citet{Lombardi14} using {\it Herschel} data.  For 
simplicity, we adopt a constant temperature for each dense core, using the value that
\citet{Lombardi14} measure at each dense core's peak.  The mean and median temperature
measured for the \get-based dense cores are 22.6~K and 19.3~K respectively, while the full
range is 11.7~K to 65.5~K.  It is important to note, however, that the \citet{Lombardi14}
temperature analysis does not extend to wavelengths longer
than those covered by {\it Herschel}.  \citet{Sadavoy13} analyzed {\it Herschel} and 
SCUBA-2 observations in the Perseus molecular cloud, and found that inclusion of the
longer-wavelength SCUBA-2 data improved temperature estimates by 40\% over the best that
could be obtained using {\it Herschel} data alone.  The improved angular resolution at
longer wavelengths and better sensitivity of SCUBA-2 to cold dust are two of the driving
factors in the improvement to the temperature estimate that \citet{Sadavoy13} found.
Many star-forming regions may have dust at multiple temperatures along the same line of
sight, and observations at the shortest {\it Herschel} wavelengths will have greater sensitivity
to any diffuse population of hot dust which is present.
Since the \citet{Lombardi14} temperature estimates did not include
SCUBA-2 data, the temperature estimates for some of our SCUBA-2 dense cores
may be too high, biased by the higher temperatures of surrounding lower-density 
material.  Nonetheless, the {\it Herschel}-based temperature provide us an important
avenue beyond the starless / protostellar core separation employed in the body of the text,
to test the validity of the relative core mass rankings assumed.

\subsection{Offset Ratios for MST-based Clusters}
\begin{figure}[htb]
\begin{tabular}{cc}
\includegraphics[width=3.2in]{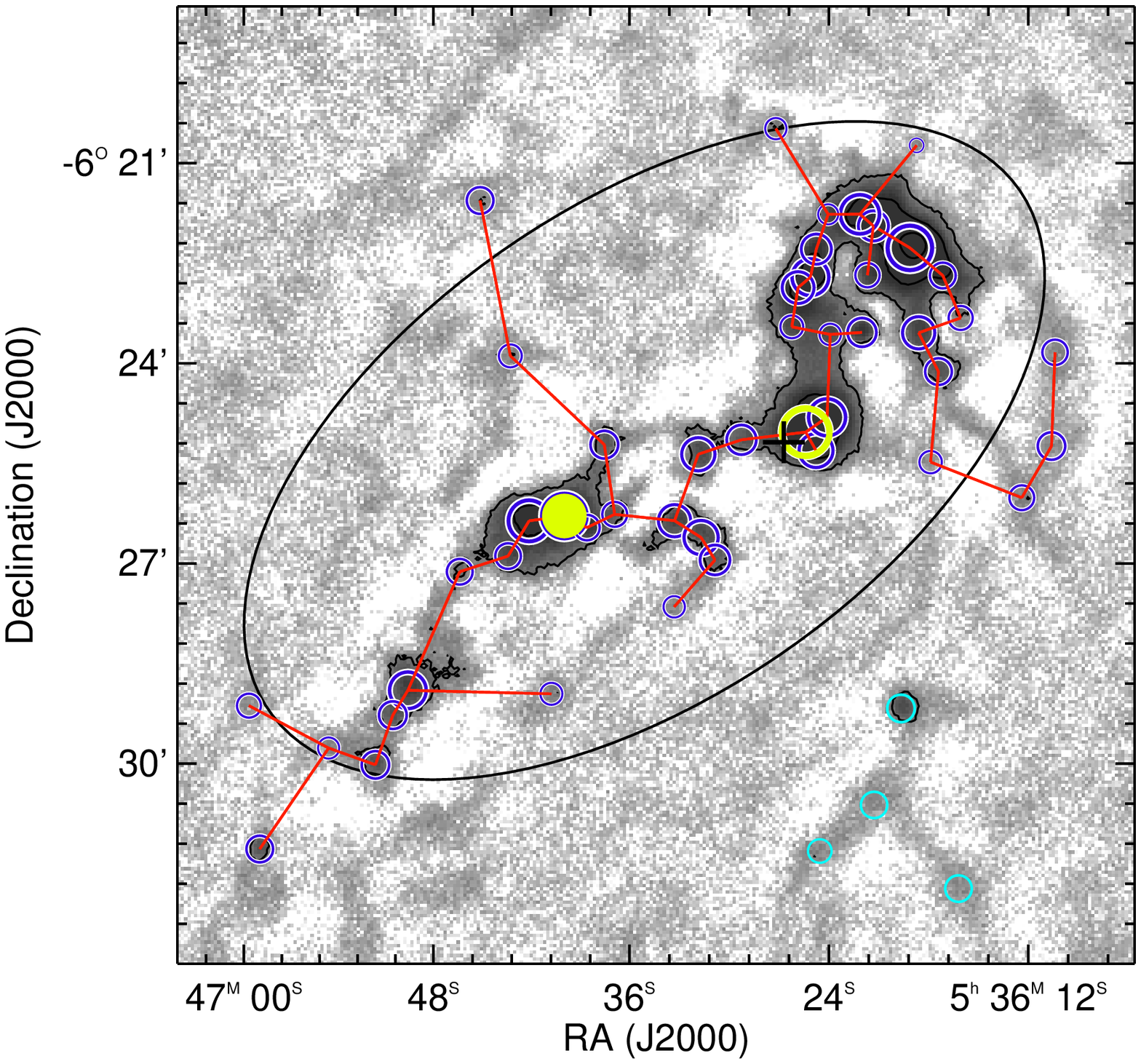} & 
\includegraphics[width=3.2in]{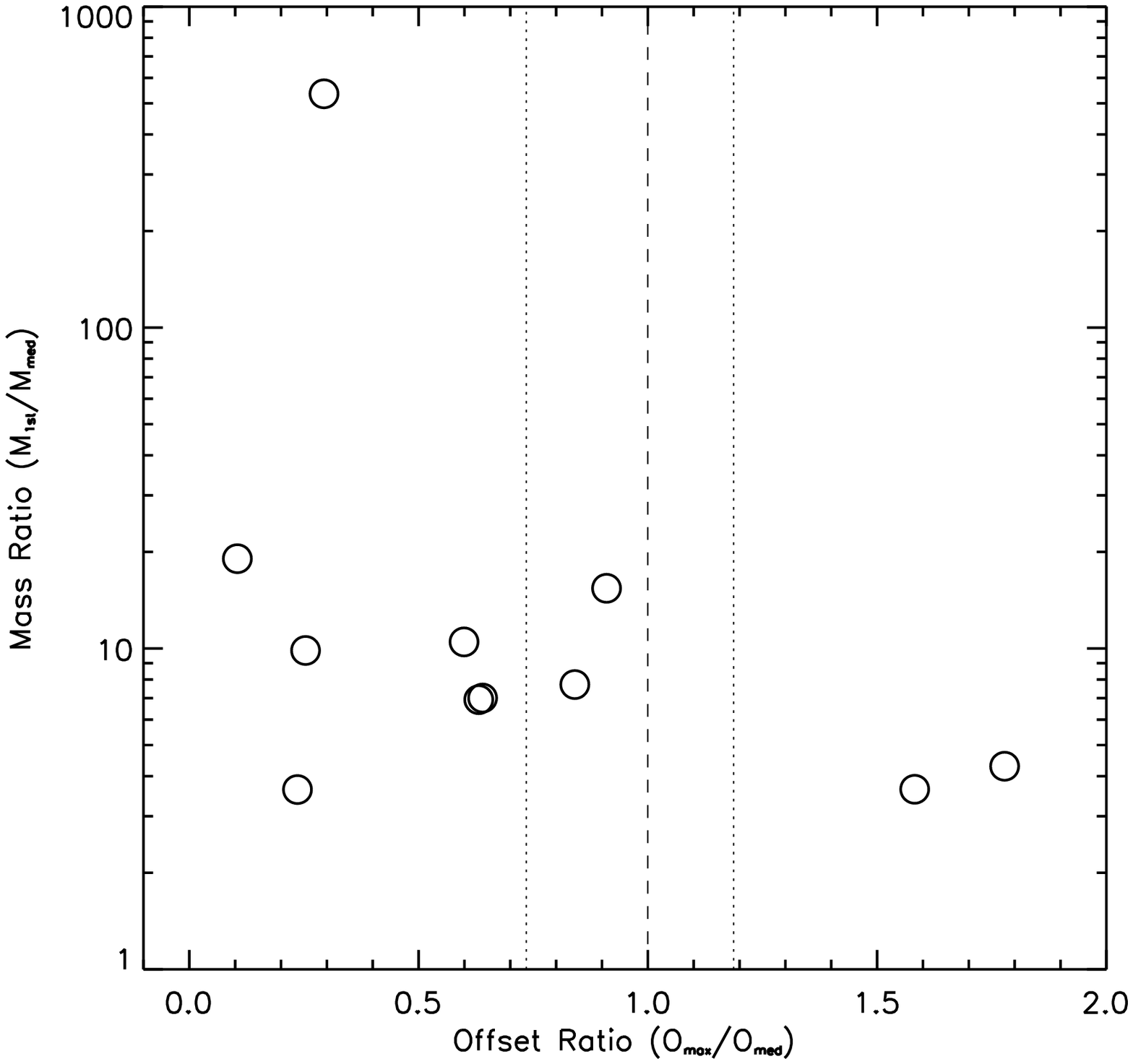} \\
\end{tabular}
\caption{Left: The minimal spanning tree structure for cluster 2, using temperatures from
	\citet{Lombardi14} to estimate the mass rankings of the dense cores.  The most
	massive (protostellar) core is denoted with the thick yellow open circle, and is
	notably different from the position of the highest flux (protostellar) core in 
	the cluster 2 shown in Figure~\ref{fig_all_clusters}. 
	Right: The distribution of offset ratios measured for the \get-based clusters, using
	the most massive dense core in each cluster, showing nearly identical offset ratios
	to those shown in the left panel of Figure~\ref{fig_offs_ratio} based on the highest
	flux cluster member.}
\label{fig_mst_mass}
\end{figure}

We re-examined the offset ratio analysis of Section~\ref{sec_mst} using instead the mass-ranking
of the cores to identify the most massive cluster member.  In all but one cluster (cluster 2),
the most massive cluster member is identical to the highest flux cluster member, and therefore
the offset ratios we measure remain identical.  In cluster 2, the most massive cluster member
lies closer to the cluster centre than the highest flux cluster member did, creating a slightly
larger population of small offset ratio clusters than presented in the main paper.
Figure~\ref{fig_mst_mass} shows the distribution of offset ratios obtained when we use
the mass-ranking of dense cores.  With the revised offset ratio distribution, we find
a less than 2\% probability that the most massive cluster members are randomly located
within their clusters.

\subsection{M-Sigma}
\begin{figure}[htb]
\begin{tabular}{cc}
\includegraphics[width=3.2in]{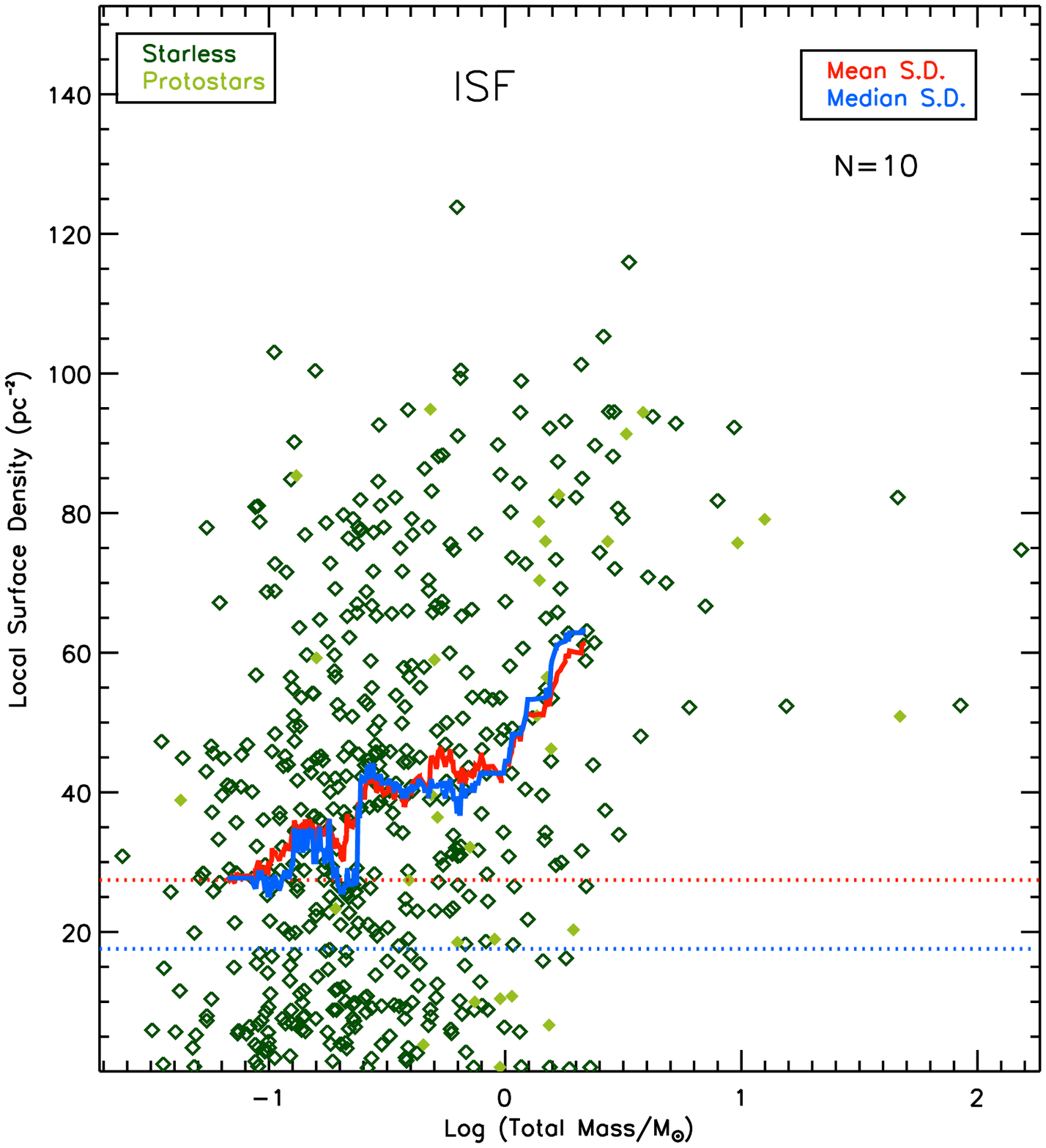} & 
\includegraphics[width=3.2in]{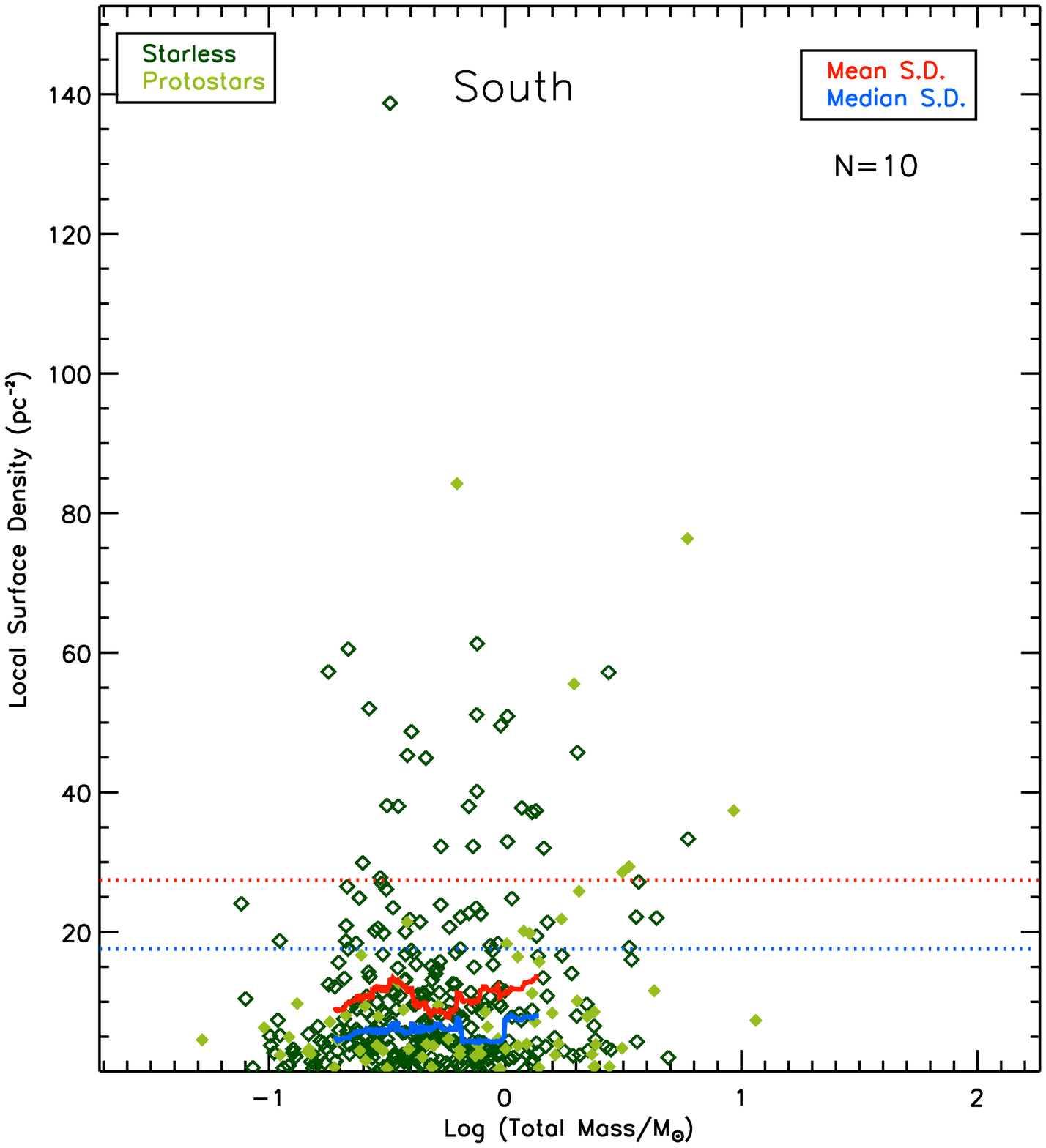} \\
\end{tabular}
\caption{The core-core surface density as a function of their estimated total mass,
	using temperatures from \citet{Lombardi14}.  Similar to Figure~\ref{fig_m_sig},
	higher mass cores tend to be found in more clustered environments, with this
	trend being especially strong in the ISF.}
\label{fig_m_sig_masses}
\end{figure}

We also ran the $S - \Sigma$ analysis of Section~\ref{sec_m_sigma} based instead on the 
estimated masses for all of the dense cores.  Figure~\ref{fig_m_sig_masses} shows the
local core-core surface density as a function of dense core mass for the cores in
the ISF and the South separately.  As we saw in Figure~\ref{fig_m_sig}, both regions
show a significant trend between increasing mass (/flux) and increasing core-core
surface density.  Statistically, we find more significant results for all of the tests
reported here than when using the dense core fluxes reported in Section~\ref{sec_m_sigma}, due to
the increased sample size (since here we can examine starless and
protostellar cores simultaneously).  In other words, we find that the core-core surface densities
are significantly different for the higher and lower mass dense cores, with a strong
tendency for higher core-core surface densities being associated with higher mass
dense cores.

We therefore conclude that our clustering analysis is unaffected by using masses
derived from {\it Herschel}-based temperatures for the cores in this study.

\bibliographystyle{aasjournal}
\bibliography{densecore_15}{}

\end{document}